\documentclass{article}

\usepackage[margin=1in]{geometry}

\usepackage{graphicx}

\newif\ifdraft
\draftfalse
\newif\ifaftersub
\aftersubfalse

\usepackage{setspace}

\usepackage[toc]{appendix}

\usepackage{subcaption}
\usepackage{enumerate,amsthm,amsmath,mathtools,latexsym}
\usepackage[mathscr]{euscript}
\usepackage{bbm}
\usepackage{bbold,dsfont}
\usepackage{listings}
\usepackage{xcolor}
\usepackage{authblk}

\usepackage{hyperref}
\hypersetup{
    colorlinks,
    linkcolor={cyan!50!black},
    citecolor={green!50!black},
    urlcolor={blue!80!black}
}
\usepackage[nameinlink,capitalise]{cleveref}
\usepackage{framed}
\definecolor{shadecolor}{HTML}{F5F5F5}

\usepackage[utf8]{inputenc}
\usepackage{amsfonts}
\usepackage{tikz}
\usepackage{tkz-euclide}
\usepackage{xspace}
\usetikzlibrary{intersections}
\usetikzlibrary{positioning}
\usetikzlibrary{arrows}
\usetikzlibrary{shapes}
\usetikzlibrary{calc}

\usepackage{todonotes}

\newcommand{\omitthis}[1]{}

\usepackage{cancel}

\usepackage{stackengine}
\stackMath

\renewcommand{\epsilon}{\varepsilon}

\title{Nanobot Algorithms for Treatment of Diffuse Cancer}

\author[1]{Noble Harasha\thanks{nharasha@mit.edu}}
\author[1]{Nancy Lynch\thanks{lynch@csail.mit.edu}}

\affil[1]{Massachusetts Institute of Technology (MIT), Cambridge, MA, USA}

\date{\today}

\begin{document}
\maketitle

\begin{abstract}
We consider the problem of a swarm of \textit{nanobots} detecting and treating human cancer that is \textit{diffuse}, that is, dispersed in a region with multiple separate cancer sites in need of treatment. 
We present a mathematical model of nanobots and their colloidal environment that is inspired by actual chemotactic nanoparticles, involving agents noisily following chemical gradients (both attractively and repellently, depending on the chemical). 
We present three incrementally sophisticated algorithms that describe additional chemical payloads that agents carry onboard, beyond the cancer-treating drug K, as well as the rules for when agents drop their payloads: \textit{Algorithm KM}, in which agents simply ascend naturally existing chemical M signals that surround cancer sites; \textit{Algorithm KMA}, in which agents themselves amplify these natural signals by dropping chemical A payloads upon reaching a site; and \textit{Algorithm KMAR}, in which agents choose to either amplify the signal by dropping chemical A or counteract/reduce the signal by dropping chemical R, according to the current unsatisfied \textit{demand} of the site. 
We present simulation results for all of the algorithms, across a set of distinct cancer arrangements, that track both the achieved treatment success as well as the time/duration of the treatment. 
KM has generally successful treatment unless the natural M-signals are weak, in which case the treatment progresses too slowly. 
KMA demonstrates a significant speedup in treatment time (over KM), but also a drop in success except for the most concentrated cancer patterns. 
KMAR has relatively optimal performance across all types of cancer patterns, demonstrating robustness and adaptability in its mechanisms for nanobot coordination. 
\end{abstract}

\section{Introduction}
Motile nanoparticles suspended in a solution, or what are often referred to as ``nanobots'', possess unique potential in various medical applications---as a result of their unique scale---including penetrating biological barriers, navigating within otherwise unreachable regions of the body, and generally increased precision.
Nanobots are being engineered to move within the human body, and more specifically, to navigate to locate target sites of interest, such as a cluster of cancerous cells.
If such nanobots possess a payload of drugs onboard, they can release the drugs once they locate and reach the given target site(s), providing a more selective and precise drug delivery solution that is less toxic to extraneous regions.
In the context of cancer detection and treatment for example, as will be the focus in this paper, this ``targeted drug delivery'' solution presents a particularly dramatic decrease in side effects in comparison to existing treatments such as chemotherapy.
However, as a consequence of nanobots' very small size, their capabilities in sensing, locomotion, memory, and computation are limited, noisy, and/or nonexistent, presenting challenges to this otherwise promising approach.

We investigate the problem of nanobots finding and treating a number of cancerous sites, \emph{distributively} and \emph{autonomously}.
In this work, we consider the process to be stochastic in nature, with nanobots performing a biased random walk, influenced by the chemical makeup and physics of the environment through which they navigate.
The presence of multiple, distinct cancerous sites (as opposed to just a single site, as we investigated in \cite{ourSingleSite}) that are \textit{diffuse} throughout the given environment in varied arrangements, introduces the problem of optimally allocating the treatment; e.g., avoiding sending all of the nanobots to a single, specific cancerous site, leaving the other sites completely untreated.

In Section~\ref{sec:model}, we give a formal, mathematical description of the general model of the nanobot agents and their colloidal environment, including a model of agent locomotion in which agents follow chemical signals by noisily ascending and descending their gradients.
Working within this model, in Section~\ref{sec:algs}, we present three distinct algorithms that describe which additional chemical payloads (beyond a cancer-treating drug) agents carry onboard, as well as the rules for when they release such payloads.
We then present simulation results for all of the algorithms in Section~\ref{sec:results}, investigating the effectiveness and efficiency of the treatment strategies across different  arrangements of cancer sites (e.g., all of the sites spread out versus some sites clustered near each other, two sites both requiring of equal amounts of treatment versus one of the two sites requiring of much more treatment, etc.) as well as for various parameter value settings.

The three algorithms are incrementally sophisticated:
The first algorithm, which we call Algorithm KM, involves agents simply following naturally existing chemical M signals centered around each cancer site.
The second algorithm, KMA, involves agents dropping an additional chemical payload, A, at cancer sites in order to amplify the existing natural signals.
The final algorithm, KMAR, involves agents dropping a different chemical payload, R, if they detect that a given cancer site has already received a sufficient amount of treatment; besides this additional, unique mechanism KMAR is the same as KMA.
The chemical R payload serves to repel agents from moving towards the already treated site, instead encouraging them to explore elsewhere and locate other sites that are in need of treatment.

Simulation results show that Algorithm KM has generally successful treatment but can be slow if the endogenous chemical signals are weak.
Algorithm KMA provides a significant improvement in efficiency over KM due to its amplified chemical signals, but for certain cancer site arrangements has unsuccessful treatment due to the amplification mechanism too strongly favoring a single site, leaving the others untreated.
Algorithm KMAR's added mechanism of dynamically favoring the chemical signals towards cancer sites still needing treatment helps to fix the above issues with KMA, while still preserving much of KMA's efficiency with artificially amplified chemical signals.
Ultimately, KMAR provides an effective and efficient treatment solution that performs well across all settings.
However, we acknowledge that KMAR is the most sophisticated of the algorithms and is thus arguably the most speculative on the level of implementation with current nanotechnologies.

Beyond the comparative advantages and disadvantages between the three algorithms, our results also show the impact of the specific arrangement or pattern of cancer in space on performance.
For the most diffuse cancer (i.e., the least clustered cancer), KMAR's improvements in treatment time over KM and in success over KMA are most significant, particularly for sparser arrangements.
KMA's best performance is seen for concentrated cancer patterns, where its amplified signals cause a successful cascade effect of agents quickly converging on the major site.
For clustered cancer, KM has the highest achieved success, outperforming our more sophisticated algorithms KMA and KMAR in success, but again, KM can be extremely slow in the case of weak chemical M signals.

We conclude the work with discussion and suggestions for future work in Section~\ref{sec:discuss}.
The baseline code used for simulations is available on \href{https://github.com/nobleharasha/nanobots-multiSite-ctsSim/}{GitHub}.

\subsection{Related Work}
Nanoparticles' and nanorobots' uniquely small size makes for great potential in their use in medical applications such as diagnostics, tissue engineering, and drug delivery.
Early works in these areas of nanomedicine are presented in \cite{crandall1996nanotechnology,freitas1999nanomedicine}.
The specific application that is most relevant to us in our work here is that of targeted drug delivery by nanoparticles, specifically for cancer treatment, which is explored in \cite{brigger2012nanoparticles, kostarelos2010nanorobots, zhang2023advanced}.
Using nanoparticles for drug delivery promises reduced side effects as nanoparticles' potential for precision surpasses most other platforms.
Instead of administering a toxic drug to a larger region of cells surrounding those specific cells actually needing the drug, precise targeted drug delivery can limit this exposure of the toxic drug to extraneous cells.
Such treatment by nanobots would ideally have the same level of efficacy as more traditional methods of treatment such as chemotherapy, if not better, in addition to significantly reduced side effects.
Despite the benefits and potential, nanoparticles' extremely small size often yields limited and/or nonexistent capabilities in sensing, locomotion, computation, memory, and communication, bringing along challenges.
While the general problem of having a collective of agents locate an unknown target site when long-distance sensing is impossible is well-studied in the field of swarm robotics, these other limitations in agent capabilities represent a largely novel problem requiring novel algorithmic strategies for agent coordination.
For example, while the work in \cite{strano} shows early promise for basic memory and computation being able to be carried out onboard micro- and nano-scopic particles, much remains to be done for proper medical application of \cite{strano} and other similar technologies.

Regarding the locomotion capabilities of nanoparticles, precise control over the movement of an individual bot is often difficult, if not impossible.
A major exception to this involves nanobots' positions being precisely controlled by external rotating magnetic fields, as in \cite{gwisai2022magnetic, martel2009flagellated}.
However, this represents a centralized perspective, while our study opts for a truly distributed and autonomous framework for nanobots, in an effort to be most scalable and practical.
Restricting ourselves to the distributed framework, on one extreme is the case in which nanoparticles have zero actual locomotion capabilities and are only able to move on account of external forces in their environment.
For example, in \cite{gomez2021markov}, the position of a nanobot changes via ``passive displacement'', simply following the flow of the circulatory system.
There is also the case of standard Brownian motion\footnote{Hereafter, to be clear, ``standard Brownian motion'' refers to there being zero drift.}, in which particles have no tendency to move in a particular direction over time, in expectation.
Our study, though, entails locomotion capabilities beyond this, with agents' movement \textit{not} being fully determined by the external forces of their given environment.
Agents perform some sort of biased random walk in response to the chemical makeup of their environment, inspired by \textit{chemotaxis}.

Chemotaxis is described as the movement of organisms or particles in response to a chemical stimulus, specifically a chemical gradient.
Typically, the environment in which chemotaxis occurs is a colloidal solution, which is a mixture where one substance made up of micro- or nanoscopic insoluble particles is dispersed within another substance termed the medium.
When there is a nontrivial, or more precisely, nonuniform, gradient of the insoluble substance, particles performing chemotaxis indeed perform a biased random walk (as opposed to standard Brownian motion) that has expected behavior in well-defined correspondence with the qualities of the gradient.
The studies in \cite{golestanian2005propulsion, howse2007self, claudiavesicles, sanchez_catalase, sanchez_lm} are relevant for chemotaxis by nanoparticles. 
\cite{golestanian2005propulsion, howse2007self, sanchez_catalase} examine the self-propulsion of nanoparticles where the solution in which they are suspended does not contain a particular global gradient.
The particles indeed move about the space via chemotaxis, but because of the lack of a specific global gradient, the particles do not move in a specific, predictable direction.
In \cite{howse2007self}, a detailed model of the walk being performed by particles is presented, including precise definitions of the drift and rotational velocities.

In contrast to \cite{golestanian2005propulsion, howse2007self, sanchez_catalase}, the authors in \cite{claudiavesicles,sanchez_lm} present novel nanoparticles which are suspended in a solution with a nonuniform chemical gradient.
Via the mechanism of chemotaxis, the nanoparticles here self-propel and move in a \textit{directed} manner, following the gradient.
The study in \cite{claudiavesicles} includes an in-vitro empirical analysis of the nanoparticles' movement following a glucose gradient. 
Their results show that it is precisely the presence of this external, or global, gradient that favors a specific type of random walk from particles in which they tend towards areas of higher (glucose) concentration, in expectation, i.e., they noisily ascend the gradient.
The nanoparticles presented in \cite{claudiavesicles} are asymmetrical, biocompatible vesicles that contain glucose oxidase and catalase. 
In the presence of an external glucose gradient, these encapsulated enzymes react with the glucose, expelling products outwards asymmetrically (because of the vesicles' asymmetrical structure), inducing a slip velocity on the surface of the vesicle in the opposite direction of the gradient \cite{claudiavesicles}.
In turn, the particle is propelled forward in the direction of the gradient.

In \cite{ourSingleSite}, the authors---which include both authors of this work---define a mathematical model which rigorously characterizes the movement of the nanoparticles in \cite{claudiavesicles}, along with some more speculative extensions including a signal amplification mechanism similar to the ideas presented in \cite{doi:10.1073/pnas.0610298104, doi.org/10.1038/nmat3049} as well as our Algorithm KMA here.
In this work, we extend the ideas from  \cite{ourSingleSite} to now consider the case of multiple, distinct cancer sites, or cancerous regions, instead of just a single target site.
With that said, the movement model for nanobot agents in this work is still very similar to the movement model presented in \cite{ourSingleSite}, except for one important distinction.
In \cite{ourSingleSite} (as well as in \cite{claudiavesicles}), agents always tend toward areas of higher chemical concentration, noisily ascending the gradient.
This type of movement is referred to as ``positive chemotaxis''.
However, ``negative chemotaxis'' has also been observed among designed nanoparticle populations, as referenced in \cite{NEGchemotaxEx,posNEGchemotax}, i.e., particles tending towards areas of lower concentration, descending the gradient in expectation.
While the model in \cite{ourSingleSite} did not include negative chemotaxis, we extend the model here to include this type of agent behavior, modeling the noisy gradient descent analogously to its gradient ascent counterpart.

When there are multiple, distinct cancer sites all in need of treatment, the problem of how to best allocate the treatment becomes crucial to success.
As the nanobot swarm does not directly have any knowledge of the cancer sites (how many there are, their locations, their demands, etc.), more nuanced strategies of coordination can become necessary in order for optimal allocation of treatment, and equivalently, optimal success.
This problem is ultimately just a very specific case of the general task allocation problem, in which a number of ``tasks'' must first be located, and then acted upon by agents according to each site's given demand.
Both \cite{adithya,grace} explored this problem, presenting distributed algorithms to solve the task allocation problem (for varying setups).
Many of their high-level strategies for agent coordination were useful inspiration for us here, but importantly, the models in \cite{adithya,grace} assume greater individual agent capabilities than we permit for our nanobot agents, including communication between agents and nontrivial computation.

\section{Model}\label{sec:model}
Building upon the work in \cite{ourSingleSite}, we present a continuous space, discrete time general model for the problem of
\textit{multi-site} cancer detection and treatment by nanobots in the human body. 
Notably we include a precise, feasible model of agent locomotion.

A set of $n$ identical agents---nanobots---move in a bounded portion of two-dimensional Euclidean space $\mathbb{R}^2$. 
Time is discretized.
There are $c$ ``cancer sites''---i.e., distinct clusters of contiguous cancerous cells---each of which is concentrated at a single point in space.
Different cancer sites can require of different amounts of treatment; in the framework of the task allocation problem, this is analogous to tasks having different ``demands''.
Each cancer site naturally produces some unique surface cell marker (unique to cancerous cells versus unaffected cells, not unique to an individual cancer site) which agents can bind to via the appropriate antibody which we assume they possess.
This allows agents to detect the presence of a nearby cancer site once they are within some $\epsilon$ units of distance, with perfect accuracy; we consider them to have ``reached the cancer site'' at this point.
Each agent carries a payload of a drug for cancer treatment that kills cancerous cells; for ease, we hereafter refer to this as \textit{chemical K}.
Once an agent is this $\epsilon$-distance away from the cancer site, it immediately drops its payload of chemical K, delivering its treatment.

We now introduce the notion of a \textit{``signal chemical''}, which is a chemical that all agents are able to sense.
Signal chemicals can also be payloads that are carried and electively dropped by agents.
Signal chemicals diffuse in the colloidal environment, forming gradients which can be followed directionally by agents.
We will consider two main classes of signal chemicals: \textit{attractive} signal chemicals whose gradient is ascended by agents, and \textit{repellent} signal chemicals whose gradient is descended by agents.
That is, if there is a nonnegligible gradient of some attractive signal chemical(s) that is centered at a cancer site, agents will tend to (i.e., noisily) move \textit{towards} that cancer site.
Similarly, if there is a nonnegligible gradient of some repellent signal chemical(s) centered at a cancer site, agents will tend to move \textit{away} from that cancer site.
Note that there could be both attractive and repellent signal chemicals of nonzero concentration present at the same location(s), in which case the signals ``counteract'' each other in some way.
We formalize these mechanisms in the movement model, defined in Section \ref{sec:bots-model}.

No direct interaction or communication occurs between agents.
The movement of an agent is a function of its previous state and the signal chemical(s) that it is currently sensing in the vicinity of its current location.
We investigate the agents' ability to distributively and autonomously locate the cancer sites and drop their drug payloads (i.e., the treatment).

In the appendix, we provide a lookup table for the notation used throughout the paper; see Table~\ref{fig:paramtable} in Section~\ref{sec:appendix-table}.

\subsection{Cancer}
There are $c$ distinct cancer sites, each with location $y_j$ for every $j=0,1,\dots,c-1$, respectively.

We assume that there exist endogenous chemical gradients of a specific attractive signal chemical---hereafter referred to as \textit{chemical M}---centered at each cancer site.
Chemical M can be imagined to
be a type of chemical marker which cancerous cells themselves naturally and continually release, or alternatively, some naturally occurring entity which increases in concentration at closer distances to tumors and cancerous cells.
In either case, we assume that the chemical M gradient(s) is (are) stable and persistent over time. 
As such, for purposes of elegance, we simplify to assume that the concentration of chemical M at all locations is fixed over time.
That is, there initially, and always, exist nonzero chemical M gradients centered at every cancer site.
We assume that the strength of an individual cancer site's surrounding chemical M signal, which is loosely represented by $P_{M_j}$ for each site $j$, respectively, is directly proportional to the amount of treatment that the cancer site requires to be fully killed.
Lastly, regarding the total concentration of chemical M at a certain location, we assume that different cancer sites' M-gradients simply interact additively.

The concentration of chemical M at position $x\in\mathbb{R^2}$ is given by \begin{equation}\label{eqn:M}
    \gamma_M (x) \coloneqq\frac{1}{\pi m} \sum_{j=0}^{c-1} P_{M_j} \text{exp}\left(-\frac{(||y_j - x||_2)^2}{m}\right) \in \mathbb{R}_{\geq 0} \text{,}
\end{equation}
where $m$ and $P_{M_j}$ (the multiset $\{P_{M_j}\}_j$ in total) are parameters.

Let us now formalize the notion of different cancer sites requiring different amounts of treatment.
Recall that the amount of chemical K---or rather, the number of chemical K payloads---needed in order to (first) fully kill a cancer site is dependent on the strength of the chemical M signal surrounding that given cancer site.
More precisely, we fix a (minimum) ratio $r_{K,M}$ of chemical K to chemical M that represents successfully killing all of the cancerous matter at cancer site $j$: where $K_j^{(t)}$ is the number of K-payloads dropped at cancer site $j$ up to time $t$, the condition
\begin{equation}\label{eqn:KtoM}
K_j^{(t)} / P_{M_j} \geq r_{K,M}
\end{equation}
corresponds to the event in which cancer site $j$ has been completely killed/treated by time $t$.

\paragraph{Parameters:}
\begin{itemize}
    \item $c \in \mathbb{Z}^+$ -- number of cancer sites.
    \item For every $j\in \{0,1,\dots,c-1\}$, $y_j \in \mathbb{R}^2$ -- location of cancer site  $j$.
    \item For every $j\in \{0,1,\dots,c-1\}$, $P_{M_j} \in \mathbb{R}_{\geq 0}$ -- parameter in the chemical M concentration function describing the strength of the chemical M signal surrounding cancer site $j$, see Equation \ref{eqn:M}.
    \item $m \in \mathbb{R}_{>0}$ -- parameter in the chemical M concentration function, see Equation \ref{eqn:M}.
    \item $\epsilon \in \mathbb{R}_{> 0}$ -- (maximum) cancer site detection distance.
    \item $r_{\text{K,M}} \in \mathbb{R}_{>0}$ -- ratio of amounts (up to normalization) of chemical K to chemical M needed to kill a cancer site, see Equation \ref{eqn:KtoM}.
\end{itemize}

\paragraph{Other Relevant Notation/Quantities:}
\begin{itemize}
    \item $\gamma_M(x) \in \mathbb{R}_{\geq 0}$ -- concentration of chemical M at position $x\in\mathbb{R}^2$.
    \item $K_j^{(t)} \in \mathbb{N}$ -- number of agents that have dropped their chemical K (drug) payload at cancer site $j$ up to time $t$, where $K_j^{(0)}=0$ for all $j$.
\end{itemize}

\subsection{Signal Chemicals A and R}

The other distinct, specific signal chemicals we will consider are both artificial chemicals that can be carried and electively dropped as payloads by agents: firstly, an attractive signal chemical hereafter referred to as \textit{chemical A}; and secondly, a repellent signal chemical hereafter referred to as \textit{chemical R}.
Unlike for chemical M, chemical A and R are not persistent; they dynamically dissipate and diffuse through the colloidal environment over time.
More specifically, we model the concentration of chemicals A and R across the space over time via instantaneous point-source diffusion.

Beyond chemical K, agents can also carry additional payloads of chemicals A and R which are similarly able to be dropped and released, notably electively (unlike chemical K which is always dropped by agents upon reaching a cancer site), once at the cancer site.
An individual agent can only drop its payload(s) at one cancer site in total.
After an agent has completed dropping all of the payloads it elected to drop, it then for all practical purposes effectively ceases to exist in the environment.

Let $A_j^{(t)}$ be the multiset\footnote{There can be more than one A-payload dropped at the same cancer site during the same timestep by multiple, distinct agents.} containing the timesteps for which all of the unique chemical A payloads dropped at cancer site $j$ \textit{before} time $t$ were indeed dropped, where $A_j^{(0)}$ is the empty set, and the maximum value over all of the elements in $A_j^{(t)}$ is less than $t$.
Similarly, let $R_j^{(t)}$ be the multiset containing the timesteps for which all of the unique chemical R payloads dropped at cancer site $j$ \textit{before} time $t$ were indeed dropped, where $R_j^{(0)}$ is the empty set, and the maximum value over all of the elements in $R_j^{(t)}$ is less than $t$.

Imagine one agent releases its chemical A payload of size $P_A$ at cancer site $j$ at time $t^*$. 
The chemical payload will immediately begin diffusing, with the concentration of this individual payload, at location $x$ at time $t$, being a function of the distance from the cancer site $||x - y_j||_2$ and the time since its release $(t - t^*)$: $\frac{P_A}{4 \pi D_A (t - t^*)} \text{exp}\left(-\frac{(||x - y_j||_2)^2}{4 D_A (t - t^*)}\right)$ \cite{diffusion}, where $D_A$ is the diffusion coefficient for chemical A.
We simplify to assume that the diffusion of each agent's payload is independent---i.e., additive.
Thus, summing over all of the individual A-payloads dropped up until the current timestep, over all cancer sites,

\begin{itemize}
    \item The concentration of chemical A at time $t$ at position $x\in\mathbb{R^2}$ is given by
    \begin{equation}\label{eqn:A}
        \gamma_A^{(t)}(x) \coloneqq\frac{P_A}{4 \pi D_A} \sum_{j=0}^{c-1} \:\: \sum_{t_{j,i}^{*} \in A_j^{(t)}} \left( \frac{1}{t - t_{j,i}^{*}} \cdot \text{exp}\left(-\frac{(||y_j - x||_2)^2}{4 D_A (t - t_{j,i}^{*})}\right) \right) \in \mathbb{R}
    \end{equation}
    where $t_{j,i}^*$ is the time at which the $i$'th A-payload to be dropped at cancer site $j$ was indeed dropped.\footnote{Note that by definition of $A_j^{(t)}$, $(t - t_{j,i}^{*})$ is always positive-valued; thus, $\gamma_A^{(t)}(x)$ is well-defined. $\gamma_R^{(t)}(x)$ is also similarly well-defined.}
\end{itemize}

Similarly,
\begin{itemize}
    \item The concentration of chemical R at time $t$ at position $x\in\mathbb{R^2}$ is given by 
    \begin{equation}\label{eqn:R}
        \gamma_R^{(t)}(x) \coloneqq\frac{P_R}{4 \pi D_R} \sum_{j=0}^{c-1} \:\: \sum_{t_{j,i}^{*'} \in R_j^{(t)}} \left( \frac{1}{t - t_{j,i}^{*'}} \cdot \text{exp}\left(-\frac{(||y_j - x||_2)^2}{4 D_R (t - t_{j,i}^{*'})}\right) \right) \in \mathbb{R}
    \end{equation}
    where $t_{j,i}^{*'}$ is the time at which the $i$'th R-payload to be dropped at cancer site $j$ was indeed dropped.
\end{itemize}

\paragraph{Parameters:}
\begin{itemize}
    \item $P_A\in \mathbb{R}_{\geq 0}$ -- size or amount of an individual chemical A payload, see Equation \ref{eqn:A}.
    \item $D_A\in \mathbb{R}_{>0}$ -- parameter in the chemical A concentration function that gives the diffusion coefficient for chemical A (i.e., how fast chemical A dissipates in the colloidal environment), see Equation \ref{eqn:A}.
    \item $P_R\in \mathbb{R}_{\geq 0}$ -- size or amount of an individual chemical R payload, see Equation \ref{eqn:R}.
    \item $D_R\in \mathbb{R}_{>0}$ -- parameter in the chemical R concentration function that gives the diffusion coefficient for chemical R, see Equation \ref{eqn:R}.
    \item $r_{A,M}\in \mathbb{R}_{\geq 0}$ -- threshold value of the ratio of amounts (up to normalization) of chemical A to chemical M at a specific cancer site used in Algorithm KMAR in the decision of whether an agent drops chemical A or chemical R, see Equations \ref{eqn:AtoM1} and \ref{eqn:AtoM2} in Section \ref{sec:alg-kmar}.
\end{itemize}

\paragraph{Other Relevant Notation/Quantities:}
\begin{itemize}
    \item $A_j^{(t)}$ -- multiset of timesteps when all unique chemical A payloads were dropped at cancer site $j$ before time $t$, respectively.
    \item $R_j^{(t)}$ -- multiset of timesteps when all unique chemical R payloads were dropped at cancer site $j$ before time $t$, respectively.
    \item $\gamma_A^{(t)}(x) \in \mathbb{R}_{\geq 0}$ -- concentration of chemical A at time $t$ at position $x\in\mathbb{R}^2$.
    \item $\gamma_R^{(t)}(x) \in \mathbb{R}_{\geq 0}$ -- concentration of chemical R at time $t$ at position $x\in\mathbb{R}^2$.
\end{itemize}

\subsection{Bots}\label{sec:bots-model}
We now describe the update step for the locomotion of an individual agent, i.e., the movement model.
We clarify that this is a \textit{model} that approximately \textit{reproduces} experimental results of actual nanoparticles following chemical gradients (and thus represents feasible nanobots), rather than being an algorithm carried out directly by the computation power of some nanobot.
Specifically, the actual nanoparticles for which this model is most inspired by is those of \cite{claudiavesicles}.\footnote{Notably, the nanoparticles in \cite{claudiavesicles} only demonstrate the attractive behavior of ascending an external chemical gradient, i.e., positive chemotaxis, while our movement model also includes the repellent behavior of descending a chemical gradient, i.e., negative chemotaxis. This extension to repellent behavior is modeled analogously to the attractive behavior (following the positive or negative gradient vector, respectively, plus added noise) and is only made use of by Algorithm KMAR.}

As a part of an agent's current state, there are two distinct modes of movement, \textit{``Explore''} and \textit{``Follow''}.
In Explore mode, agents perform a simple random walk, i.e., standard Brownian motion with zero drift.
In Follow mode, agents take into account the local makeup of signal chemicals (M, A, and R), ``following'' (either attractively or repellently, depending) the gradient(s) by biasing their orientation accordingly.
In both modes of movement, all steps have the same displacement distance of $\alpha$.
If there are no signal chemicals present at the agent's current location, the Explore and Follow modes are equivalent.

Consider some agent $i$ at time $t$:

The agent's position is given by $x_i^{(t)}$ and its orientation vector is $\theta_i^{(t)}$.

\paragraph{If in \textit{Explore} Mode:}
Let $\beta \sim U(-\pi,\pi)$.
Without loss of generality let $\mu = (1,0) \in \mathbb{R}^2$.

\paragraph{If in \textit{Follow} Mode:}
We first define $$\gamma_{\text{TOT}}^{(t)}\left(x_i^{(t)}\right) \coloneqq \gamma_M\left(x_i^{(t)}\right) + \gamma_A^{(t)}\left(x_i^{(t)}\right) - \gamma_R^{(t)}\left(x_i^{(t)}\right) \:.$$
If $\gamma_{\text{TOT}}\left(x_i^{(t)}\right)$ is identically zero, follow the step for Explore mode.
Otherwise, let $$\mu \coloneqq \nabla \gamma_{\text{TOT}}^{(t)}\left(x_i^{(t)}\right) \in \mathbb{R}^2 .$$
Let 
\begin{equation}\label{eqn:beta}
\beta \sim \mathcal{N}(0, \sigma^2), \:\: \sigma^2 = \left ( b \cdot || \mu ||\right)^{-1},
\end{equation}
which we then force to be within $[-\pi, \pi]$ by updating $$\beta = [(\beta + \pi) \:\:\text{mod}\:\: (2\pi)] - \pi \:.$$
Note that larger $b$ values yield movement that is more biased toward following the given chemical signal(s) (i.e., farther from standard Brownian motion), or greater ``orientation-bias''.

\paragraph{Update Position (For Both Modes):}
Representing $\mu$ as a column vector, we then update agent $i$'s orientation vector as 
\[\theta_i^{(t+1)} = 
\begin{bmatrix}
\cos{\beta} & -\sin{\beta} \\
\sin{\beta} & \cos{\beta}
\end{bmatrix} \mu\]
such that $\beta$ is the angle formed between $\mu$ and $\theta_i^{(t+1)}$.
Agent $i$'s position is then updated as follows, taking a step of length $\alpha$ in the direction of its orientation vector.
\[ x^{(t+1)}_i = x^{(t)}_i + \alpha \cdot \frac{\theta_i^{(t+1)}}{||\theta_i^{(t+1)}||_2}.\]

To summarize, if the attractive chemical signal (M plus A) dominates the repellent chemical signal (R) in steepness, then the agent is biased to move roughly \textit{towards} the nearest local maximum of the attractive chemical gradient (e.g., the nearest cancer site, often).
If instead, the repellent chemical signal dominates the attractive signal, then the agent is biased to move roughly \textit{away} from the nearest local maximum of the repellent chemical gradient, which is, again, often the nearest cancer site.
The attractive behavior of nanobots ascending a chemical gradient in the way described by the above movement model is known to be feasible behavior as was demonstrated by the model validation carried out in \cite{ourSingleSite}.
However, the repellent behavior aspect of the model remains to be formally validated via direct comparison with experimental results from actual nanoparticles performing negative chemotaxis.

For the scope of this work we will assume a bounded space $[0,\phi_{\text{max}}] \times [0,\phi_{\text{max}}] \subseteq \mathbb{R}^2$.
Each timestep, each agent follows the above update step (for its given current movement mode) repeatedly until its new location is indeed within the given boundary.\footnote{For sufficiently small $\alpha$ a valid new location will eventually be achieved with probability $1$.  
We assume that all of these attempts begin from the same starting location, and happen within one timestep.}
Despite space being bounded, note that we model diffusion of chemicals in an unbounded space; imposing this boundary is a simplification we choose to make in order to aid convergence/termination in simulations.

There are $n$ total nanobot agents, but within a given single time step, all agents' movements are independent as no direct interaction nor communication occurs between them.
If chemical payloads are dropped during the current timestep, this will indeed change the given concentration functions of chemicals M, A, and R, but this will not affect the update step for the positions of other agents until the next timestep; hence the aforementioned notion of independence.

We also define a maximum runtime cutoff $T^*$ (in the number of discrete timesteps).
The reality of nanobots moving within the human body forces a finite timeline: after some amount of time, the nanobots will naturally dissolve or disintegrate, becoming useless for our purposes of delivering their drug treatment.
This is often referred to as the ``clearance time'' and is a possible intuitive interpretation of the value $T^*$.

To be concrete, the values of all parameters which involve distance and/or time hereafter are in SI units unless otherwise specified.
To be consistent with \cite{claudiavesicles}, we can think of one timestep in our model to be equivalent to one real-time second.
For example, parameter settings of $\alpha=2\cdot 10^{(-5)}$ and $\phi_{\text{max}}=0.005$ mean that agents travel roughly\footnote{This statement would be a precise equality assuming that agents move in a perfectly straight line during the duration of one timestep.} twenty micrometers per second, within a bounded square space which is half of a centimeter large in both directions.

\paragraph{Parameters:}
\begin{itemize}
    \item $n\in \mathbb{Z}^+$ -- total number of nanobot agents.
    \item $\phi_{\text{max}} \in \mathbb{R}_{> 0}$ -- length and width of the total space of activity/operation; total area is $(\phi_{\text{max}})^2$.
    \item $\alpha \in \mathbb{R}_{> 0}$ -- displacement length of each step in agent motion, i.e., ``step size''.
    \item $b\in \mathbb{R}_{>0}$ -- parameter in the movement model loosely describing how biased agent movement is towards following the gradient(s) of the chemical signal(s), i.e., the amount of ``orientation-bias'', see Equation \ref{eqn:beta}.
    \item $T^* \in \mathbb{N}$ -- maximum runtime cutoff in number of discrete timesteps, or clearance time.
\end{itemize}

\paragraph{Other Relevant Notation/Quantities:}
\begin{itemize}
    \item For every $i\in\{0,1,\dots,n-1\}$, $x_i^{(t)}\in\mathbb{R}^2$ -- position of agent $i$ (arbitrary index\footnote{We assume that each agent has its own unique identifier or index $i\in\{0,1,\dots,n-1\}$; i.e., $x_i^{(t)}$ and $\theta_i^{(t)}$ refer to the same agent. Since all nanobot agents are identical and do not interact or communicate directly with each other, these indices are completely arbitrary and only for our outside use here; no agent actually needs to store in memory nor know these indices' values in implementation.}) at time $t$.
    \item For every $i\in\{0,1,\dots,n-1\}$, $\theta_i^{(t)}\in\mathbb{R}^2$ -- orientation vector of agent $i$ at time $t$.
\end{itemize}

\subsection{Metrics}
We define two metrics of performance, a \textit{success} metric that measures the effectivity of the treatment and a \textit{treatment time} metric that measures efficiency:

Our first metric of performance $S$, measuring what proportion of the total cancer has been killed by time $T^*$ (i.e., how successful or effective the treatment is), is given by
\begin{equation}
    S(T^*)\coloneqq \frac{\sum_{j=0}^{c-1} \min\left\{\frac{K_j^{(T^*)}}{r_{K,M}}, P_{M_j}\right\}}{\sum_{j=0}^{c-1} P_{M_j}} \in [0,1] \text{ .}
\end{equation}
That is, the success metric is a function of $T^*$.
Note that after the first $\lceil P_{M_j} \cdot r_{K,M} \rceil$ K-payloads are dropped by agents at a given cancer site $j$, all additional K-payloads dropped there from that point on in time are useless; once a site's demand is satisfied, further treatment is unnecessary.
These ``useless'' K-payloads are not only not contributing to the treatment of the given cancer site where they were dropped, but they are also not contributing to the treatment of other sites that could have actually used the extra drug.
A successful treatment in which all cancer sites are treated does so by optimizing the allocation of K-payloads such that this issue of wasted treatment does not occur often.
The metric $S$ accurately captures this.
Given individual agents' random, noisy behavior and considering how the non-toxicity of such targeted cancer treatment by nanobots allows for repeated, even regular, treatment, our goal will be for just some significant portion of the cancerous cells to be killed, i.e., $S(T^*)$ close to, but not necessarily equal to, one.
Hereafter, the ``success'' of a given treatment refers to the metric $S$.

Our second metric of performance $T_{\text{fin}}$ measures how long it takes for the treatment to stabilize and (mostly) finish (i.e., how efficient the treatment is, ignoring its success).
Let $S'(t) \coloneqq \frac{S(t+\delta) - S(t)}{\delta}$, where $\delta \in \mathbb{Z}^+$ is a parameter.\footnote{Note that the calculations for $S'(t)$ require knowledge of the value $S(t+\delta)$, where $(t+\delta)>t$. We assume $T_{\text{fin}}$ is calculated after the entire (finite) simulation run has already completed; $S(t)$ is thus known for all $t$ at this point.}
That is, $S'$ is the average rate of change in $S$ over a window of $\delta$ timesteps.
We then define $T_{\text{fin}} \coloneqq \min\left\{ t \:|\: S'(t) \leq D \right\}$, where $D \in \mathbb{R}_{\geq 0}$ is a parameter with some very small value.\footnote{Note that for sufficiently small $D$, $T_{\text{fin}}$ actually becomes equivalent to $\min\{t \:|\: S(t+\delta)=S(t)\}$, i.e., the first point in time for which $S$ in constant for $\delta$ consecutive timesteps.}
That is, $T_{\text{fin}}$ is the first point in time for which the average rate of change in $S$ (given by our defined $S'$) is less than or equal to $D$.
When the success metric flattens out and stops increasing any further, we can assume that all of the agents are either (1) already planted at their final site or (2) ``stuck'', unable to find a site in reasonable time; it is at this point in time in which we deem the treatment to have (mostly) finished.
Hereafter, the term ``treatment time'' will refer to the quantity $T_{\text{fin}}$.

As will be made clear by our results in Section~\ref{sec:results}, for a given fixed algorithm and environment setting we run $x\in\mathbb{Z}^+$ repeated simulations, i.e., there are $x$ trials for each experiment.
We define $S_{\text{avg}}(t)$ to be the average value of $S(t)$ over all $x$ trials, as its own separate, well-defined function over $t$/$T^*$.

Unfortunately, when the number of nanobot agents $n$ is not extremely large (as will indeed be the case for our simulations), $S(t)$ as a curve over the domain $t$/$T^*$ is not particularly ``smooth''.
For these types of piecewise functions with very limited codomains, the notion of its rate of change becomes difficult and misleading; e.g., $S$ can stay constant for many (more than $\delta$) timesteps well before the treatment has actually stabilized and finished.
As a result of this, in place of using sufficiently large $n$, in this work we use the smoother $S_{\text{avg}}$ when calculating $T_{\text{fin}}$ in order for the metric to more accurately capture the true ``treatment time''.
Analogously to $S'$, let
\begin{equation}
    S_{\text{avg}}'(t) \coloneqq \frac{S_{\text{avg}}(t+\delta) - S_{\text{avg}}(t)}{\delta} \:.
\end{equation}
We then redefine
\begin{equation}
    T_{\text{fin}} \coloneqq \min\{ t \:|\: S_{\text{avg}}'(t) \leq K\} \:.\footnote{Technically, in simulations, we will only track $S$ (and thus $S_{\text{avg}}$) every $\delta'$ timesteps (for purposes of the efficiency of running all of the experiments), i.e., for $t=0,\delta',2\delta',\dots$, where we ensure that $\delta\geq\delta'$ and $\delta \text{ mod } \delta' \equiv 0$. Thus, more precisely, $T_{\text{fin}} \coloneqq \min\{ t \:|\: t \text{ mod } \delta' \equiv 0, \: S_{\text{avg}}'(t) \leq D\}$.}
\end{equation}
For this work, we fix $\delta=30000$ and $D=3\cdot 10^{(-7)}$.

Note that $T_{\text{fin}}$ is not guaranteed to exist/be defined, since our simulation runs are finite and some algorithms such as RW have slow progressing treatment.

\section{Algorithms}\label{sec:algs}
We now present the three distinct algorithms which describe which chemical payloads agents carry and when these payloads are dropped.
The least sophisticated Algorithm KM, in which agents simply follow the natural chemical M signals, is a feasible nanobot algorithm aligned with current technologies, as existing experimental works have explored nanobots ascending a stable chemical gradient.
From Algorithm KM to KMA, in an effort mostly to improve treatment time, we add chemical A payloads which allow for the natural chemical M-signals to be amplified.
From KMA to KMAR, in an effort to improve success (via improving the optimality of treatment allocation), we add chemical R payloads which uniquely allow for agents to \textit{descend} chemical gradients and thus, be driven away from sites that are already treated to explore elsewhere.
We also consider the simple random walk as a baseline of comparison, to more accurately gauge the improvements in overall performance (both metrics of success and treatment time) that these incrementally sophisticated algorithms produce.

\paragraph{Bot Initialization and Execution:}
The following applies to all of our algorithms.

At time $t=0$, each agent $i$ of the total $n$ agents is initialized by having its position chosen uniformly at random (independently between all other agents) within the bounded space $[0,\phi_{\text{max}}] \times [0,\phi_{\text{max}}]$: i.e., for all $i$, $x_i^{(0)} = (x,y)$ where $x,y \sim \mathcal{U}(0,\phi_{\text{max}})$.
Further, for all $j$, $K_j^{(0)}$ is initialized to be equal to zero, and $A_j^{(t)}$ and $R_j^{(t)}$ are both initialized to be the empty set.

Then, at each subsequent timestep $t$, the following execution is carried out for each agent $i$: 
The agent's new position $x_i^{(t)}$ is given by following the update step as defined in the movement model in Section~\ref{sec:bots-model}.
At this point, if the agent's new, updated position is within $\epsilon$ units of distance of any cancer site $j$,\footnote{For all of our experiments, all cancer sites are guaranteed to have pairwise distances between each other which are greater than $\epsilon$. Thus, if an agent is within $\epsilon$ units of distance of some cancer site, it cannot be within $\epsilon$ units of distance of any other site.} the rules defined by the given algorithm---defined in this section below---outline which chemical payloads beyond K, if any, are to be dropped.
According to this, if an A-payload is dropped, then $A_j^{(t)}$ is updated to be equal to $(A_j^{(t)} + \{t\})$ (with multiplicity as it is a multiset), and similarly, if an R-payload is dropped, then $R_j^{(t)}$ is updated to be equal to $(R_j^{(t)} + \{t\})$.
Also, the value of $K_j^{(t)}$ is always incremented by one.
Lastly, this given agent is deemed to be ``terminated'', with no further payloads to drop and its position fixed indefinitely.
If agent $i$ did not reach a cancer site yet, nothing beyond updating its position occurs.

This continues for every timestep (and for every agent) until either all agents have reached a cancer site and thus terminated, and/or the clearance time is reached, i.e., until $t$ is greater than $T^*$.

\subsection{Algorithm KM}

Each agent has a payload of chemical K only.
There is no chemical A or R anywhere---i.e., $\gamma_A^{(t)}$ and $\gamma_A^{(t)}$ are both identically zero, or equivalently, $A_j^{(t)}$ and $R_j^{(t)}$ are the empty sets for all $j$ and $t$.
Though there still exists the endogenous, time-constant chemical M gradients.
When an agents reaches a cancer site, it always, and immediately, releases its K payload to deliver its marginal treatment.

\subsection{Algorithm KMA}\label{sec:alg-kma}

Each agent has a payload of chemical K and a payload of chemical A.
When an agent reaches a cancer site, it always, and immediately, releases both K- and A-payloads to deliver its marginal treatment and amplify the current/existing chemical signal, respectively.
That is, there are chemical M gradients centered at each cancer site, with this attractive signal being amplified and boosted over time whenever an agent plants at the respective site and releases chemical A.
There is no chemical R anywhere---i.e., $\gamma_R^{(t)}$ is identically zero, or equivalently, $R_j^{(t)}$ is the empty set for all $j$ and $t$.

\subsection{Algorithm KMAR}\label{sec:alg-kmar}
Each agent has a payload of chemical K, a payload of chemical A, and a payload of chemical R.
When an agent reaches some cancer site $j$ at time $t^*$, it always, and immediately, releases its K-payload to deliver its marginal treatment.
Then, if 
\begin{equation}\label{eqn:AtoM1}
\gamma_A^{(t^*)}(y_j) / P_{M_j} < r_{A,M}
\end{equation}
the agent also releases its A-payload, but does not release its R-payload.
Otherwise, if 
\begin{equation}\label{eqn:AtoM2}
\gamma_A^{(t^*)}(y_j) / P_{M_j} \geq r_{A,M}
\end{equation}
the agent instead releases its R-payload, but does not release its A-payload.
That is, when the chemical A signal at a cancer site is too strong, agents will release chemical R (instead of chemical A) at that site in an effort to encourage agents to explore and administer treatment elsewhere.
Note that $r_{A,M}$ is a parameter of the model specific to this algorithm.

Intuitively, or rather, ideally, we want to have agents find the cancer sites efficiently by leveraging chemical A just as in Algorithm KMA, but once a given cancer site is fully treated, we want to immediately stop having agents come there.
More reasonably within the constraints of our model, once a given cancer site is fully treated, we want chemical R to be released there by any further agents that happen to still reach that site in order to \textit{start gradually} encouraging agents to move away from the already treated site.
However, this requires agents being able to sense the current amount of treatment already administered---or equivalently, to sense the current amount of remaining treatment still required---which we assume to be impossible.
As such, we must use chemical A as a proxy to \textit{estimate} the current amount of treatment (K) already administered since K and A payloads are often dropped in conjunction with each other.

We define $r_{A,M} \coloneqq k \cdot r_{K,M} \cdot P_A$, where $k$ is a positive real.
Recall that before \eqref{eqn:AtoM2} ever holds true, agents will always drop both chemicals K and A when they reach a cancer site, so $A_j^{(t)}=K_j^{(t)}$ holds in this case.
Up to significant error---due to diffusion, different A-payloads being dropped at different points in time, and different sites' payloads interacting additively---$\gamma_A^{(t)}(y_j)$ is roughly proportional to $A_j^{(t)} \cdot P_A = K_j^{(t)} \cdot P_A$.
In order for $\gamma_A^{(t)}(y_j) / P_{M_j} \geq r_{A,M}$ to be somewhat estimating/coinciding with the event $K_j^{(t)} / P_{M_j} \geq r_{K,M}$, then $r_{A,M}$ should be proportional to $r_{K,M} \cdot P_A$; hence our above formulation for $r_{A,M}$.
We will vary the multiplicative factor $k$ to both tune our estimation as well as vary how early we shift from dropping A-payloads to dropping R-payloads.

\subsection{Simple Random Walk, RW}
Lastly, as a baseline of comparison, we will also consider the simple random walk, ``RW'', in which agents are always in Explore mode (as defined by the movement model in Section \ref{sec:bots-model}), i.e., agents are completely blind to any chemical signals/gradients including chemical M.
All agents have a payload of chemical K which they immediately drop upon reaching a cancer site.

\section{Results}\label{sec:results}

All agents start at independently random initial locations, sampled uniformly within $[0,0.005] \times [0,0.005] \subseteq \mathbb{R}^2$, i.e., we fix $\phi_{\text{max}}=0.005$.
We consider a set of five handpicked, distinct cancer site and demand (i.e., $\{P_{M_j}\}_j$) arrangements, which we enumerate in Subsection~\ref{sec:arrangements}.
We provide simulation results for the three algorithms individually (in Subsections \ref{sec:res-km}, \ref{sec:res-kma}, and \ref{sec:res-kmar}, respectively), as well as for comparison between the algorithms (in Subsection~\ref{sec:res-comp}), with all experiments being carried out for all of the different site and demand arrangements.

For simplicity (and without loss of generality), we fix $r_{K,M}=1$, i.e., (at least) $P_{M_j}$ agents are needed in order to fully treat/kill cancer site $j$.
We also fix the total demand summed over all sites to be $50$ and the total number of nanobot agents to be $55$, i.e., $\sum_{j=0}^{c-1} P_{M_j} = 50$ and $n=55$.
Consequently, we have a small margin of error of five (ten percent) extra agents which, if there were to be perfect allocation of agents between sites according to the respective $P_{M_j}$ demands, would not actually be needed for treatment.
We also fix $\alpha=\epsilon=2\cdot 10^{-5}$ and $m=10^{-6}$ for all experiments.

Instead of running separate simulations for many unique clearance times, we run all simulations until a maximum cutoff of $t=200000$ (seconds)\footnote{Note that a treatment or clearance time of $200000$ seconds is over two full days. This amount of time may not be realistic in practice, but nonetheless this maximum runtime cutoff provides us with a large window of time to more thoroughly track how a given treatment progresses over time.} and imagine the success metric $S$ for every $t$ along the way to be the total achieved success if the clearance time $T^*$ were to have been equal to $t$, respectively.

\subsection{Cancer Site and Demand Arrangements}\label{sec:arrangements}
Specifically, the site and demand arrangements we consider in our experiments are the following:\footnote{While the \textit{scale} of the $P_{M_j}$ values were arbitrarily chosen, i.e., $\{15,35\}$ instead of $\{1.5,3.5\}$, this is without loss of generality as varying other parameters such as $b$, $P_A$, and $P_R$ allows us to equivalently (though indirectly) model varying strengths of chemical M signals. The value of $r_{K,M}$ can also easily be adjusted such that $P_{M_j}$ values of $15$ and $1.5$ both correspond to ``demands'' of $15$ agents, for example.}
\begin{align*}
    \textbf{(a) -- }&\text{Two well spaced-out sites with equal demand: } \\
    &c=2, \: \{y_j\}_j = \{(0.0015,0.0025), (0.0035,0.0025)\}, \: \{P_{M_j}\}_j = \{25,25\} \text{ ;} \\
    \textbf{(b) -- }&\text{Two well spaced-out sites with unequal demand: } \\
    &c=2, \: \{y_j\}_j = \{(0.0015,0.0025), (0.0035,0.0025)\}, \: \{P_{M_j}\}_j = \{15,35\} \text{ ;} \\
    \textbf{(c) -- }&\text{Five well spaced-out sites with equal demand: } \\
    &c=5, \: \{y_j\}_j = \{(0.0025,0.0025),(0.001,0.004),(0.001,0.001),(0.004,0.004),(0.004,0.001)\}, \\
    &\{P_{M_j}\}_j = \{10,10,10,10,10\} \text{ ;} \\
    \textbf{(d) -- }&\text{Four sites clustered together and one other site faraway, all with equal demand: } \\
    &c=5, \: \{y_j\}_j = \{(0.0008,0.0027),(0.0012,0.0023),(0.0008,0.0023),(0.0012,0.0027),(0.004,0.0025)\}, \\
    &\{P_{M_j}\}_j = \{10,10,10,10,10\} \text{ ;} \\
    \textbf{(e) -- }&\text{One major site with nearly all of the total demand and two outliers, all sites well spaced-out: } \\
    &c=3, \: \{y_j\}_j = \{(0.0025,0.0025),(0.001,0.002),(0.004,0.003))\}, \: \{P_{M_j}\}_j = \{46,2,2\} \text{ .}
\end{align*}
While this selection of specific arrangements is clearly inexhaustive with regards to all possibilities, they are intended to be representative of larger classes of common cancer patterns which are all mutually distinct.
Arrangements (a) and (c) are representative of typical diffuse cancer, where (a) and (c) are the sparse and dense versions of this, respectively.
On the other hand, (e) is representative of typical concentrated cancer.
Arrangement (b) is somewhere in between these diffuse and concentrated categories, with the discrepancy in demand between the sites in (b) being less extreme than for (e).
Similarly, arrangement (d) is a much less extreme version of concentrated cancer with the area of concentration being larger for (d) than for (e), spread out over a cluster of multiple sites.

Within all Figures \ref{fig:KM} through \ref{fig:compareAlgs}, the site/demand arrangements (a)--(e) above correspond to subfigures (a)--(e), respectively.

\subsection{Results for Algorithm KM}\label{sec:res-km}

For our simulation results, we begin with Algorithm KM, the basic case of only having the time-constant, endogenous chemical M signals for agents to follow.

\begin{figure}
    \centering
    \begin{subfigure}{.48\textwidth}
        \centering
        \includegraphics[width=.85\textwidth]{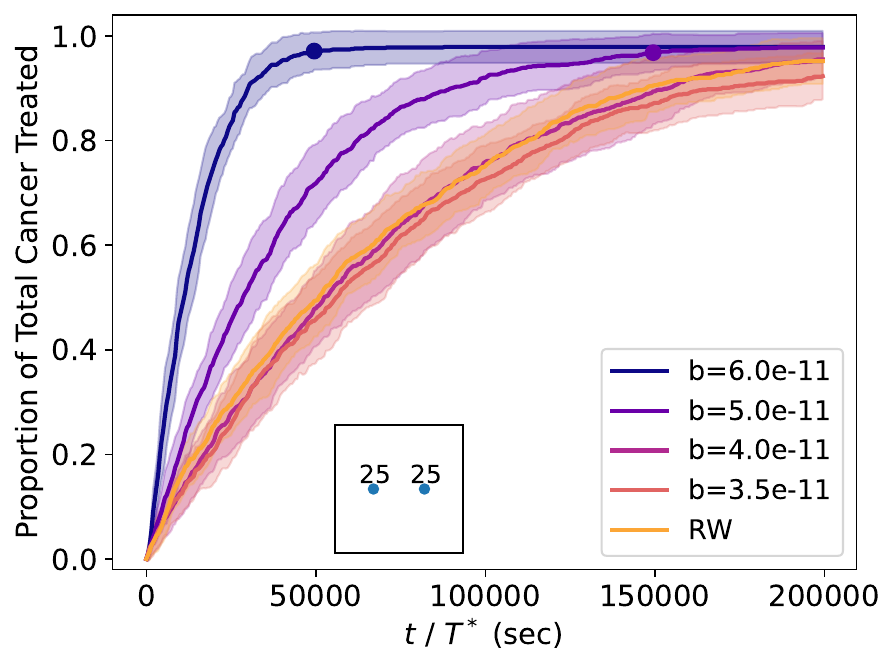}
        \caption{For $b=6\cdot 10^{(-11)}$: $(T_{\text{fin}}, S_{\text{avg}}(T_{\text{fin}}))=(49500,0.971)$; for $b=5\cdot 10^{(-11)}$: $(149500,0.968)$.}
        \label{fig:KM-2Even}
    \end{subfigure}
    \hfill
    \begin{subfigure}{.48\textwidth}
        \centering
        \includegraphics[width=.85\textwidth]{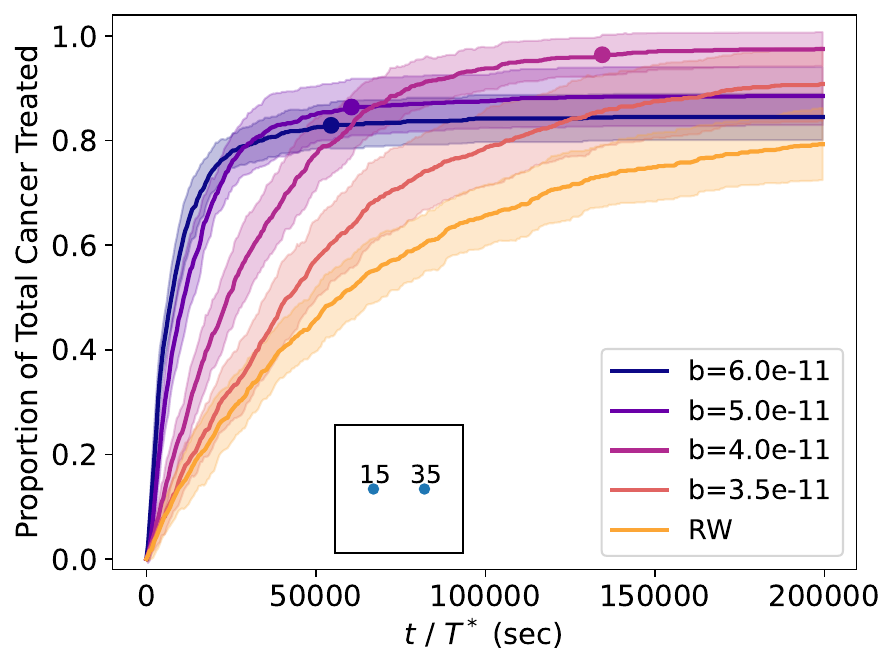}
        \caption{For $b=6\cdot 10^{(-11)}$: $(T_{\text{fin}}, S_{\text{avg}}(T_{\text{fin}}))=(54500,0.829)$; for $b=5\cdot 10^{(-11)}$: $(60500,0.864)$; for $b=4\cdot 10^{(-11)}$: $(134500,0.964)$.}
        \label{fig:KM-2UNeven}
    \end{subfigure}
    \hfill
    
    \centering
    \begin{subfigure}{.48\textwidth}
        \centering
        \includegraphics[width=.85\textwidth]{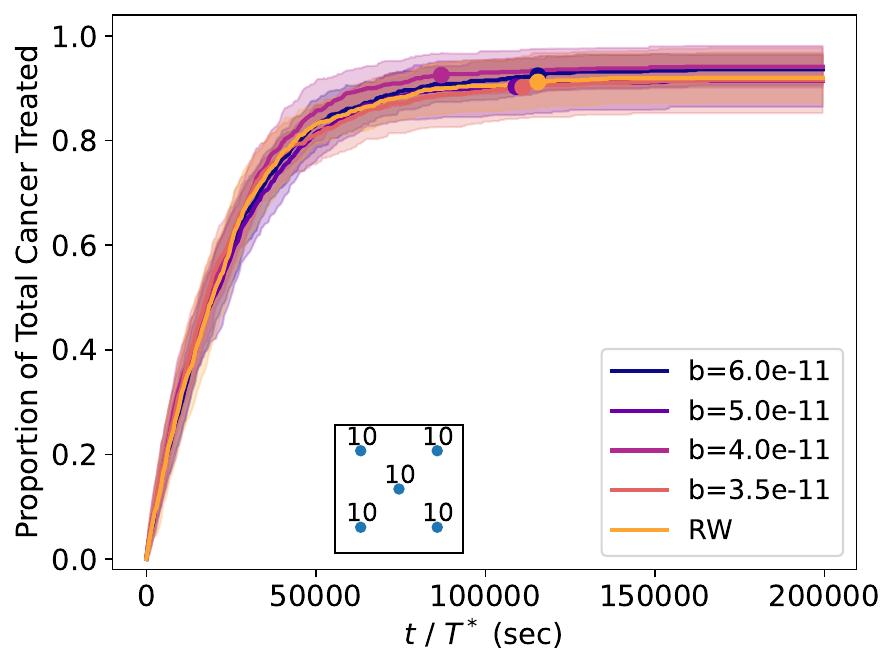}
        \caption{For $b=6\cdot 10^{(-11)}$: $(T_{\text{fin}}, S_{\text{avg}}(T_{\text{fin}}))=(115500,0.924)$; for $b=5\cdot 10^{(-11)}$: $(109000,0.903)$; for $b=4\cdot 10^{(-11)}$: $(87000,0.925)$; for $b=3.5\cdot 10^{(-11)}$: $(111000,0.902)$; for RW: $(115500,0.912)$.}
        \label{fig:KM-spreadout}
    \end{subfigure}
    \hfill
    \begin{subfigure}{.48\textwidth}
        \centering
        \includegraphics[width=.85\textwidth]{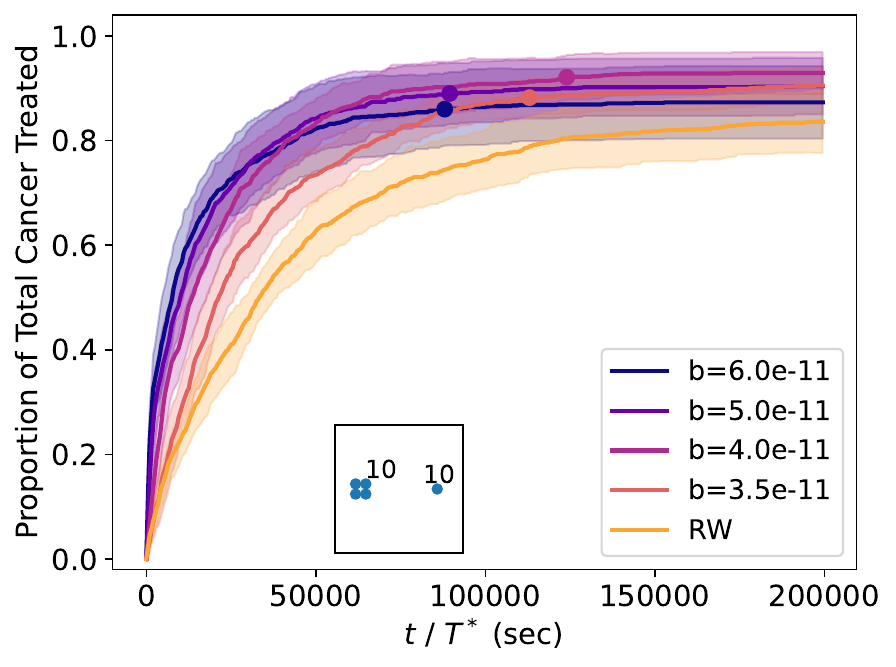}
        \caption{For $b=6\cdot 10^{(-11)}$: $(T_{\text{fin}}, S_{\text{avg}}(T_{\text{fin}}))=(88000,0.86)$; for $b=5\cdot 10^{(-11)}$: $(89500,0.89)$; for $b=4\cdot 10^{(-11)}$: $(124000,0.921)$; for $b=3.5\cdot 10^{(-11)}$: $(113000,0.882)$.}
        \label{fig:KM-clusterplusone}
    \end{subfigure}
    \hfill

    \centering
    \begin{subfigure}{.48\textwidth}
        \centering
        \includegraphics[width=.85\textwidth]{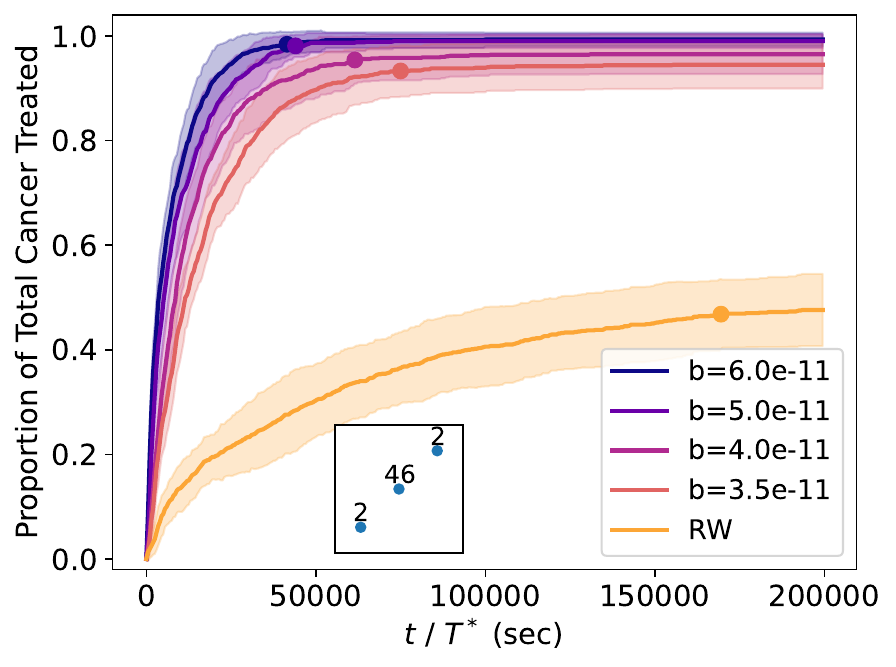}
        \caption{For $b=6\cdot 10^{(-11)}$: $(T_{\text{fin}}, S_{\text{avg}}(T_{\text{fin}}))=(41500,0.984)$; for $b=5\cdot 10^{(-11)}$: $(44000,0.981)$; for $b=4\cdot 10^{(-11)}$: $(61500,0.954)$; for $b=3.5\cdot 10^{(-11)}$: $(75000,0.933)$; for RW: $(169500,0.468)$.}
        \label{fig:KM-onebig}
    \end{subfigure}
    \hfill

    \caption{
    \textit{Alg. KM:   } 
    Simulation results for Algorithm KM with $n=55$ agents, across varying cancer site and $\{P_{M_j}\}_j$ arrangements (depicted by small scatterplots) and for different amounts of orientation-bias. RW serves as baseline of comparison. $\phi_{\text{max}}=0.005$, $\alpha=\epsilon=2\cdot 10^{-5}$, $m=10^{-6}$, $r_{\text{K,M}}=1$. Plotted lines show average success $S$ (at that given point in time) over $20$ trials, and shaded regions are standard deviations.
    Points on main plot have coordinates $(T_{\text{fin}}, S_{\text{avg}}(T_{\text{fin}}))$ for the given $b$ setting whose color they match with.
    For $b$ settings with no such point, $T_{\text{fin}}$ is undefined, i.e., $T_{\text{fin}} > 200000$.
    }
    \label{fig:KM}
\end{figure}

For our five chosen cancer site and demand arrangements (as listed in Subsection~\ref{sec:arrangements}) as well for different orientation-bias parameter $b$ values, we plot the success metric $S$ (total proportion of cancerous mass treated) over time ($t$/$T^*$).
See Figure~\ref{fig:KM}, in which cancer site arrangements are depicted in the small scatterplots where each cancer site $j$ is a point labeled with its $P_{M_j}$ value.

Our results show that with greater orientation-bias ($b$ value), the treatment progresses faster but, for certain site arrangements, has suboptimal success---following the M-signals more closely gets agents to some final site faster, but not always not the ideal site for optimal allocation.
For reasonable orientation-bias and sufficiently long clearance time $T^*$, Algorithm KM has high success with approximately $85$ to $100$ percent of the total cancer being treated, significantly outperforming the simple random walk RW for all site arrangements other than (c).
Again, for all arrangements other than (c), KM also significantly outperforms RW in treatment time $T_{\text{fin}}$.
For arrangement (c) which represents dense diffuse cancer, RW has just as good, if not better, success than KM since every site already initially has some agents close to it, in expectation, by means of agents' uniformly random initial locations.

The strength of the chemical M signal for each individual cancer site is directly proportional to its given demand.
As such, the observed high success of Algorithm KM is reasonable and intuitive. Having agents, which start at random initial locations, follow the chemical M gradients yields reasonably good allocation of treatment in expectation, because more agents tend towards those sites that require more agents as those sites' M-signals are stronger.

The effect of the arrangement of cancer sites on performance is extremely significant.
For Figure~\ref{fig:KM-2Even} (arrangement (a)) in which there are two cancer sites with equal demand, we observe that greater orientation-bias (i.e., greater values of $b$) corresponds to significantly faster progressing treatment---even just from $b=5\cdot 10^{(-11)}$ to $b=6\cdot 10^{(-11)}$, the treatment time $T_{\text{fin}}$ is nearly three times faster.
Under greater orientation-bias, agents follow the M-signal(s) more closely and thus reach a final site faster.
The average maximum success of treatment is approximately equivalent and preserved across all levels of orientation-bias for arrangement (a).\footnote{The ``maximum success of treatment'' refers to the global maximum success metric value over all $t/T^*$; i.e., if the nanobots have infinite time to perform their treatment before dissolving/dying, what proportion of the total cancerous mass could they kill. Note that this is not much different from $S_{\text{avg}}(T_{\text{fin}})$.}
With that said, the speed at which treatment progresses is also relevant.
Say, for example, that the nanobots become useless after six hours, dissolving in the body and/or being cleared away; i.e., the clearance time $T^*$ is equal to $21,600$.
Then the treatment for the setting $b=6\cdot 10^{(-11)}$ has approximately $75$ percent success, $b=5 \cdot 10^{(-11)}$ has approximately $40$ percent success, and the other lower orientation-bias settings have less than $25$ percent success.
This is a completely different picture than for $T^*=200000$, where all levels of orientation-bias have between $90$ and $100$ percent success.
This is to stress how vital the clearance time $T^*$ can be to our discussion of algorithm success.
For this arrangement in which both sites have equal demand, the optimal allocation of treatment is to send half of the agents to each site.
Then, with agents starting at uniformly random initial locations (and with the sites being symmetrically placed in the environment), the simple random walk running for a sufficiently long time should yield this optimal even allocation of agents/treatment and thus $100$ percent success, in expectation.
In fact, notice that all of the plots in Figure~\ref{fig:KM-2Even} besides the ones for the two greatest $b$ values are still slightly increasing in their success metric  $S$ value over time.

For Figure~\ref{fig:KM-2UNeven}, in which the cancer sites now have uneven demand, the effect of the orientation-bias parameter on performance is different.
While greater orientation-bias still yields faster treatment time by the same reasoning that agents will reach \textit{some} site faster in expectation, here, when the orientation-bias is too strong, the maximum success of treatment achieved decreases.
With an uneven demand arrangement, as the orientation-bias increases, the M-signal belonging to the site with greater demand will begin to dominate the weaker M-signal; consequently, agents will converge towards the site with greater demand, leaving the other site untreated.
For this uneven demand arrangement across cancer sites, sending \textit{all} of the agents to the site that requires more treatment (which is what often happens under the greatest orientation-bias setting) still ends up yielding a reasonably high success metric value.
But note that this treatment scenario is completely ignoring the cancer site with lower demand; this could arguably be unacceptable in application.
There is a point, though, before which increasing $b$ does not reduce the maximum success achieved; as such, the greatest $b$ value before success decreases is the optimal parameter setting as higher $b$ settings speed up treatment time.
Here, that is $b\approx 4 \cdot 10^{(-11)}$, though this optimal $b$ value is likely to be dependent on the specific cancer site and demand arrangement, among other things.
Further experiments would be needed to investigate this, testing algorithm performance for the same site locations as (b) but with a different demand arrangement of $\{5,45\}$, for example.

When there are several cancer sites spread throughout the environment, as for Figure~\ref{fig:KM-spreadout}, treatment is fast and successful across all amounts of orientation-bias.
However, our algorithm is not actually introducing any improvement at all over the simple random walk algorithm RW, as RW already performs well.
This is due to the fact that agents, with initial locations distributed uniformly at random, are already close to some cancer site from the start.

For Figure~\ref{fig:KM-clusterplusone} where there is a cluster of several cancer sites close together in addition to one outlier cancer site far away from the cluster, we see a similar trend relative to orientation-bias as for Figure~\ref{fig:KM-2UNeven} in which greater orientation-bias can sometimes yield suboptimal success.
Compared to Figure~\ref{fig:KM-2UNeven} though, the drop in success under greater orientation-bias is less extreme here, as all of the sites have equal demand.
Greater orientation-bias still again yields faster treatment time.

Lastly, for Figure~\ref{fig:KM-onebig} in which there is one site with almost all of the total demand as well as two other sites elsewhere with low demand, greater orientation-bias corresponds with faster as well as slightly more successful treatment.
Ignoring RW, the difference in total average success $S_{\text{avg}}(T_{\text{fin}})$ between the various $b$ settings is minimal.
Under weaker orientation-bias there is more random/noisy behavior, so there will be more trials in which the treatment happens to have poor success, bringing the average value success metric down.
On the other hand, under greater orientation-bias, agents' movement is more deterministic, or predictable, yielding treatment with optimal $100$ percent success more often, and thus a slightly higher average maximum success metric with lower variance.\footnote{Recall that we have $n=55$ agents, which is greater than the total demand $\sum_j P_{M_j} =50$.}
Ultimately, this is the site and demand arrangement with the best performance, generally, for all orientation-bias settings.
Here, one site has an extremely strong M-signal resulting in agents, even under weak orientation-bias, ascending that site's gradient efficiently; this leads to fast treatment time $T_{\text{fin}}$.
Further, since that site has most of the demand, even if the other two sites are completely untreated, the total success of the treatment $S(T^*)$ will be close to one.
This is also the arrangement for which Algorithm KM outperforms the simple random walk most drastically, as many agents, completely blind to the chemical M signals, waste their K-payloads at one of the sites with low demand after it has already been fully treated.

For weak orientation-bias, our results show that Algorithm KM has generally slow treatment time.
As weak orientation-bias is essentially equivalent to weak chemical M signals, a characteristic of the natural environment which is out of our control when considering actual implementation of such nanobot treatment, it is worth considering further strategies for improving treatment time such as Algorithm KMA.

Ultimately, through the results in this section, we show which approximate orientation-bias parameter settings are optimal for KM performance; these optimal parameter settings notably vary between arrangements.
These findings will be used in Section~\ref{sec:res-comp} in our comparison between algorithms.
For the sparse diffuse arrangement (a) and the most concentrated arrangement (e), the greatest orientation-bias setting $b=6\cdot 10^{(-11)}$ performs the best in both success and treatment time.
For the sparse and slightly concentrated arrangement (b), a moderate amount of orientation-bias, $b=4\cdot 10^{(-11)}$, achieves the highest eventual success though it is slower than the larger $b$ settings. 
For the dense diffuse arrangement (c), all orientation-bias settings perform approximately the same.
For the clustered arrangement (d), a moderate amount of orientation-bias, $b=4\cdot 10^{(-11)}$, achieves the highest eventual success, though this improvement is minor.

\subsection{Results for Algorithm KMA}\label{sec:res-kma}
We now consider the more involved case in which agents amplify the attractive M-signal(s) by dropping payloads of chemical A once they reach a cancer site, that then diffuse through the environment.
We investigate the effect of this more involved treatment strategy on performance.
Our results ultimately show that Algorithm KMA improves treatment time significantly, but often has poor success unless the cancer is concentrated, i.e., if there is one cancer site with the majority of the total demand.

In the subsequent two subsections, we analyze the effect of individual parameters on performance in order to approximate the best parameter values, as well as better understand the global behavior of the nanobot swarm under Algorithm KMA and its driving mechanisms.

\subsubsection{Effect of Orientation-Bias Parameter on KMA Performance}
\begin{figure}
    \centering
    \begin{subfigure}{.48\textwidth}
        \centering
        \includegraphics[width=.85\textwidth]{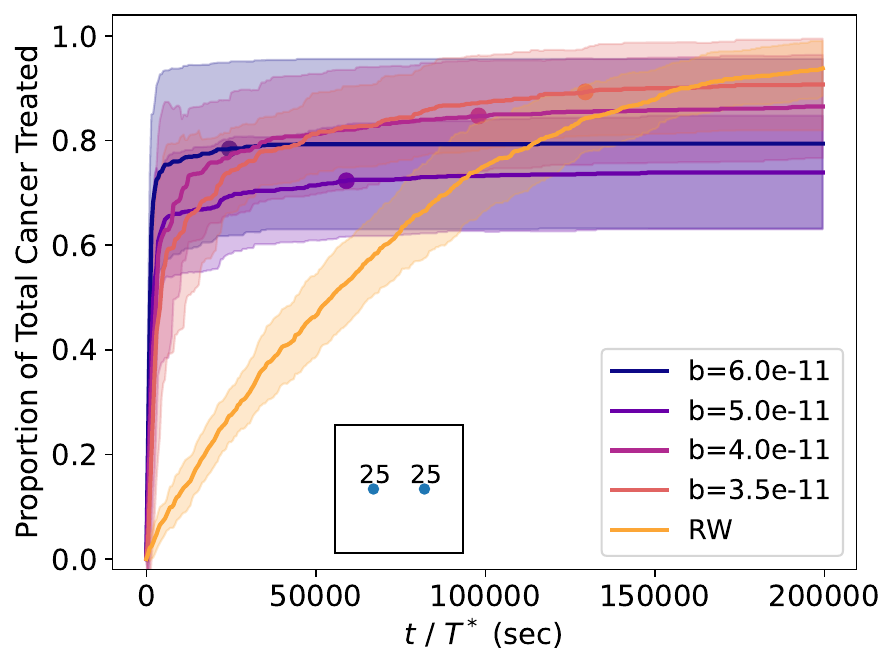}
        \caption{For $b=6\cdot 10^{(-11)}$: $(T_{\text{fin}}, S_{\text{avg}}(T_{\text{fin}}))=(24500,0.784)$; for $b=5\cdot 10^{(-11)}$: $(59000,0.723)$; for $b=4\cdot 10^{(-11)}$: $(98000,0.847)$; for $b=3.5\cdot 10^{(-11)}$: $(129500,0.893)$.}
        \label{fig:KMA-2Even}
    \end{subfigure}
    \hfill
    \begin{subfigure}{.48\textwidth}
        \centering
        \includegraphics[width=.85\textwidth]{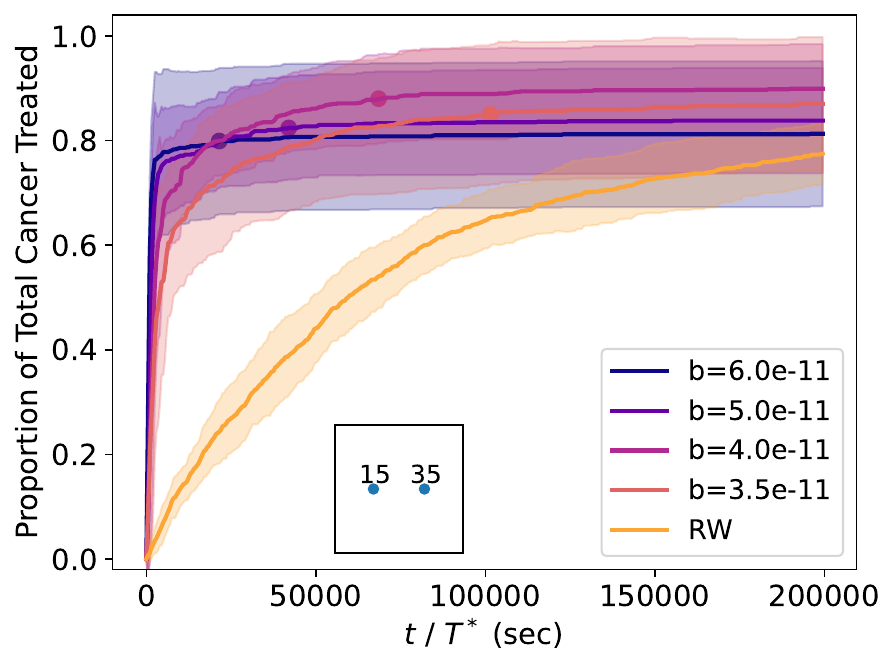}
        \caption{For $b=6\cdot 10^{(-11)}$: $(T_{\text{fin}}, S_{\text{avg}}(T_{\text{fin}}))=(21500,0.799)$; for $b=5\cdot 10^{(-11)}$: $(42000,0.824)$; for $b=4\cdot 10^{(-11)}$: $(68500,0.88)$; for $b=3.5\cdot 10^{(-11)}$: $(101500,0.85)$.}
        \label{fig:KMA-2UNeven}
    \end{subfigure}
    \hfill
    
    \centering
    \begin{subfigure}{.48\textwidth}
        \centering
        \includegraphics[width=.85\textwidth]{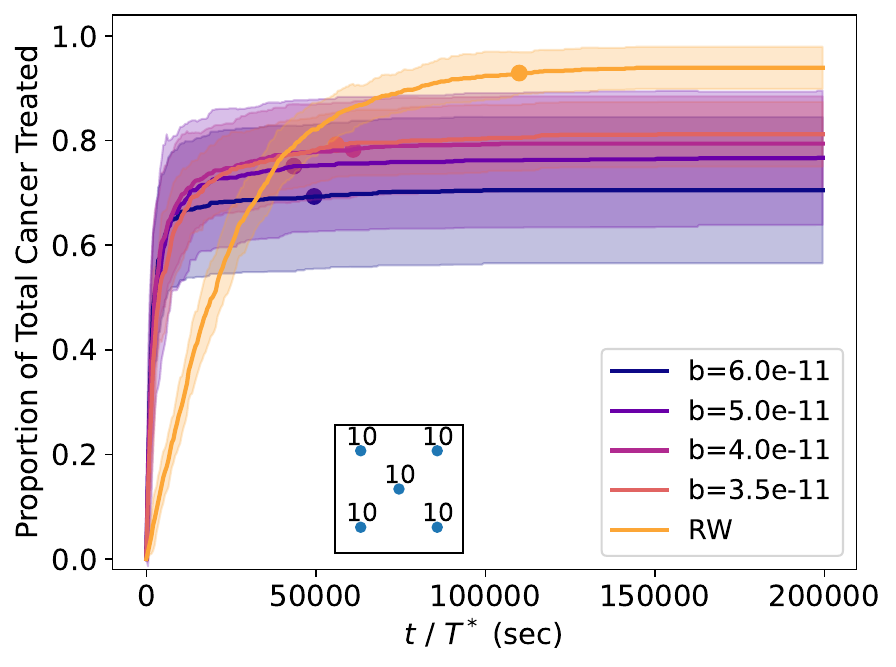}
        \caption{For $b=6\cdot 10^{(-11)}$: $(T_{\text{fin}}, S_{\text{avg}}(T_{\text{fin}}))=(49500,0.693)$; for $b=5\cdot 10^{(-11)}$: $(43500,0.751)$; for $b=4\cdot 10^{(-11)}$: $(61000,0.783)$; for $b=3.5\cdot 10^{(-11)}$: $(56000,0.792)$; for RW: $(110000,0.929)$.}
        \label{fig:KMA-spreadout}
    \end{subfigure}
    \hfill
    \begin{subfigure}{.48\textwidth}
        \centering
        \includegraphics[width=.85\textwidth]{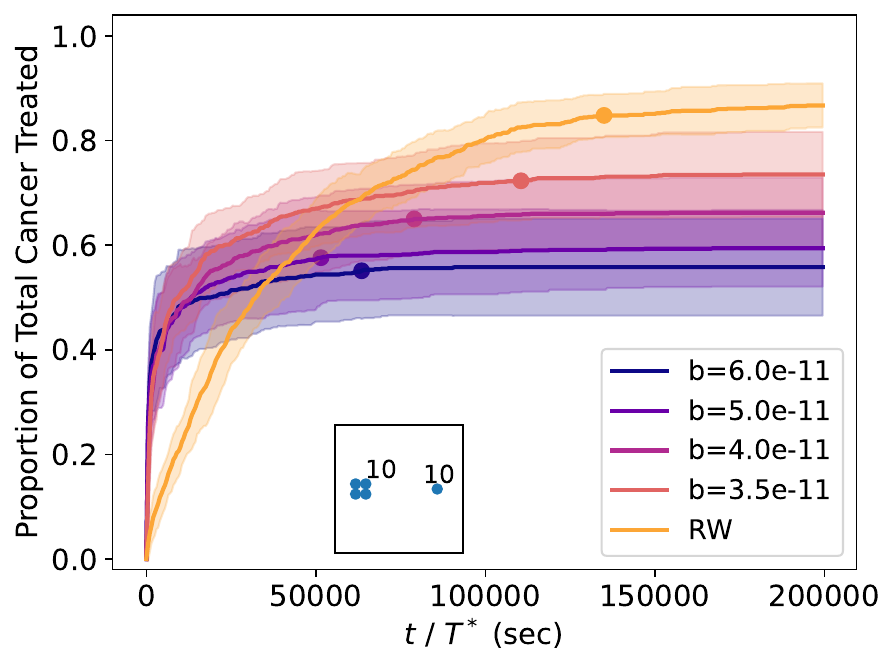}
        \caption{For $b=6\cdot 10^{(-11)}$: $(T_{\text{fin}}, S_{\text{avg}}(T_{\text{fin}}))=(63500,0.551)$; for $b=5\cdot 10^{(-11)}$: $(51500,0.576)$; for $b=4\cdot 10^{(-11)}$: $(79000,0.65)$; for $b=3.5\cdot 10^{(-11)}$: $(110500,0.723)$; for RW: $(135000,0.848)$.}
        \label{fig:KMA-clusterplusone}
    \end{subfigure}

    \centering
    \begin{subfigure}{.48\textwidth}
        \centering
        \includegraphics[width=.85\textwidth]{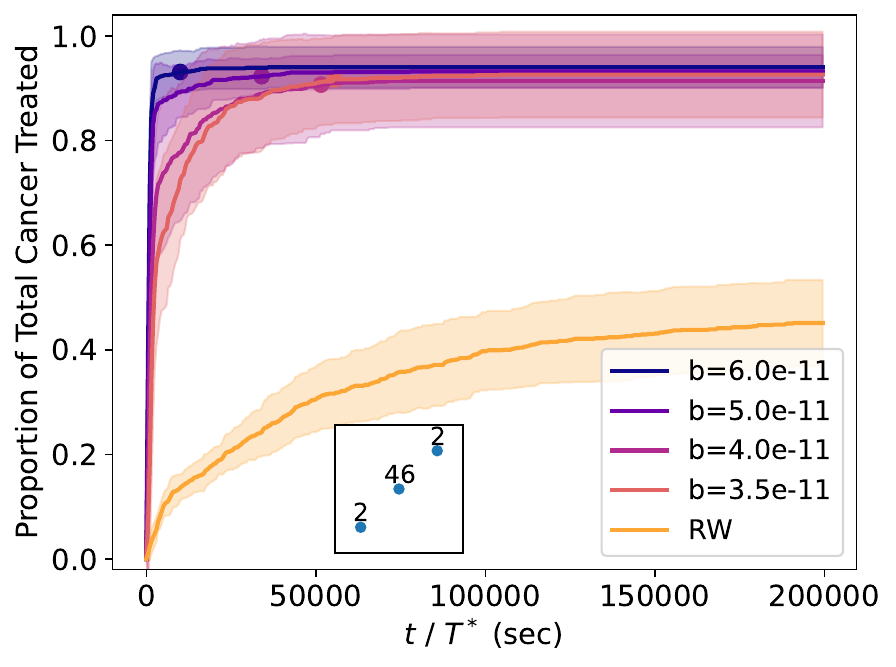}
        \caption{For $b=6\cdot 10^{(-11)}$: $(T_{\text{fin}}, S_{\text{avg}}(T_{\text{fin}}))=(10000,0.931)$; for $b=5\cdot 10^{(-11)}$: $(34000,0.923)$; for $b=4\cdot 10^{(-11)}$: $(51500,0.907)$; for $b=3.5\cdot 10^{(-11)}$: $(55500,0.917)$.}
        \label{fig:KMA-onebig}
    \end{subfigure}
    \hfill
    
    \caption{
    \textit{Alg. KMA:   }
    Simulation results for Algorithm KMA with $n=55$ agents, across varying cancer site and $\{P_{M_j}\}_j$ arrangements (depicted by small scatterplots) and for different amounts of orientation-bias. RW serves as baseline of comparison. $P_A=10, D_A=10^{(-9)}$, $\phi_{\text{max}}=0.005$, $\alpha=\epsilon=2\cdot 10^{-5}$, $m=10^{-6}$, $r_{\text{K,M}}=1$. Plotted lines show average success $S$ (at that given point in time) over $20$ trials, and shaded regions are standard deviations.
    Points on main plot have coordinates $(T_{\text{fin}}, S_{\text{avg}}(T_{\text{fin}}))$ for the given $b$ setting whose color they match with.
    For $b$ settings with no such point, $T_{\text{fin}}$ is undefined, i.e., $T_{\text{fin}} > 200000$.
    }
    \label{fig:KMA}
\end{figure}

For our chosen set of distinct cancer site and demand arrangements (as listed in Subsection~\ref{sec:arrangements}) as well as for different orientation-bias parameter $b$ values, we plot the success metric $S$ over time.
We fix the chemical A payload size $P_A=10$.
See Figure~\ref{fig:KMA}.

Our results show that Algorithm KMA has extremely fast-progressing treatment (i.e., fast treatment time) with A-payloads continually amplifying the existing attractive chemical signals, but, for many site and demand arrangements, suffers from poor success (even underperforming RW) as all agents tend to converge on a singular site.
Similar to KM, greater orientation-bias yields faster treatment time $T_{\text{fin}}$ but often also suboptimal allocation of treatment, and thus lower total success $S_{\text{avg}}(T_{\text{fin}})$.
The optimal $b$ value for overall performance varies between distinct cancer patterns.

When the first agent finds a cancer site, it drops its chemical A payload, amplifying the attractive M-signal surrounding that given site.
The other agents will then be more likely, in expectation, to find this site next with its amplified signal.
These agents will drop their chemical A payloads as well, amplifying the attractive signal even more.
This process continues, only speeding up and intensifying over time: as the attractive signal is amplified, more agents are more likely to reach the site quickly, which results in the attractive signal for that site being even more amplified by means of their A-payloads, and so on.
Essentially, Algorithm KMA can induce a sort of ``cascade effect'' in which a couple of agents (or even one) finding a particular site can spark a process in which many agents also head towards that site.
This phenomenon is only made more dramatic and rapid under greater amounts of orientation-bias (i.e., larger values of $b$).
Such a cascade effect is extremely advantageous for fast treatment time, but, when there are many separate cancer sites all needing treatment, sending a large majority of the total nanobot swarm towards a single cancer site is not ideal for success, as exhibited by the results in Figure~\ref{fig:KMA}.

Sending almost all of the agents to one of two cancer sites is better than sending almost all of the agents to one of five cancer sites, which is why success is greater for the cancer arrangements of Figures \ref{fig:KMA-2Even} and \ref{fig:KMA-2UNeven} compared to Figures \ref{fig:KMA-spreadout} and \ref{fig:KMA-clusterplusone}.

In the case of Figure \ref{fig:KMA-onebig} in which one site has almost all of the total demand, sending all of the agents to that demanding site will still yield close to perfect $100$ percent success.
Our results show that Algorithm KMA indeed achieves very high success of treatment for this arrangement, across all orientation-bias settings.
In fact, compared to the results of Algorithm KM for this same arrangement (e) (shown in Figure \ref{fig:KM-onebig}), both algorithms achieve similar levels of success but KMA is much faster, demonstrating a clear advantage in KMA over KM for arrangement (e) (as opposed to the tradeoff that is present for the other arrangements between the efficiency of KMA versus the effectiveness of KM).
Similar to the results of Algorithm KM for arrangement (e), greater orientation-bias here actually corresponds to slightly greater average success $S_{\text{avg}}(T^*)$.

\subsubsection{Effect of Chemical A Payload Size on KMA Performance}

\begin{figure}
    \centering
    \begin{subfigure}{.48\textwidth}
        \centering
        \includegraphics[width=.85\textwidth]{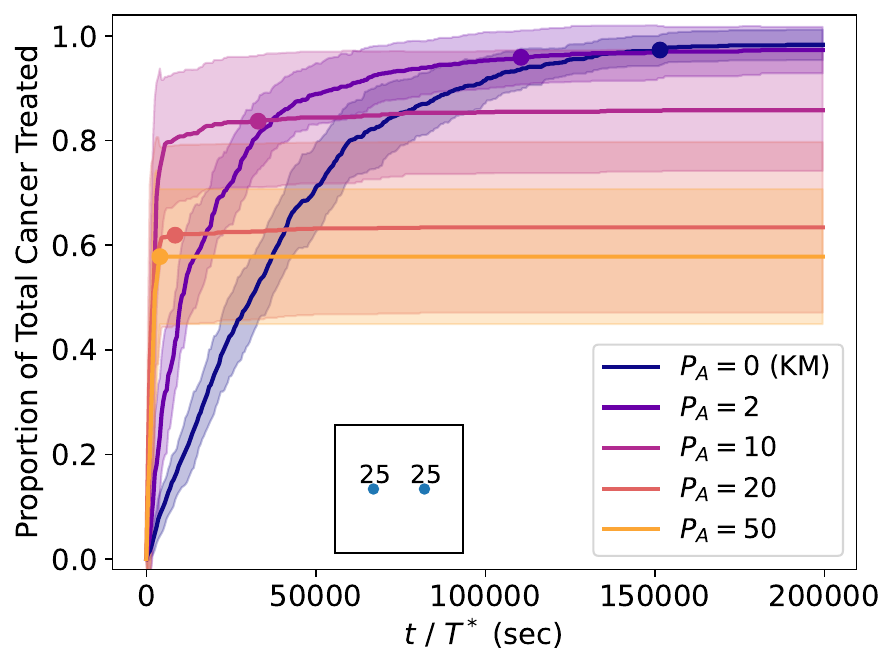}
        \caption{For $P_A=50$: $(T_{\text{fin}}, S_{\text{avg}}(T_{\text{fin}}))=(4000,0.578)$; for $P_A=20$: $(8500,0.619)$; for $P_A=10$: $(33000,0.837)$; for $P_A=2$: $(110500,0.959)$; for $P_A=0$: $(151500,0.973)$.}
        \label{fig:KMA-A-ffect-2Even}
    \end{subfigure}
    \hfill
    \begin{subfigure}{.48\textwidth}
        \centering
        \includegraphics[width=.85\textwidth]{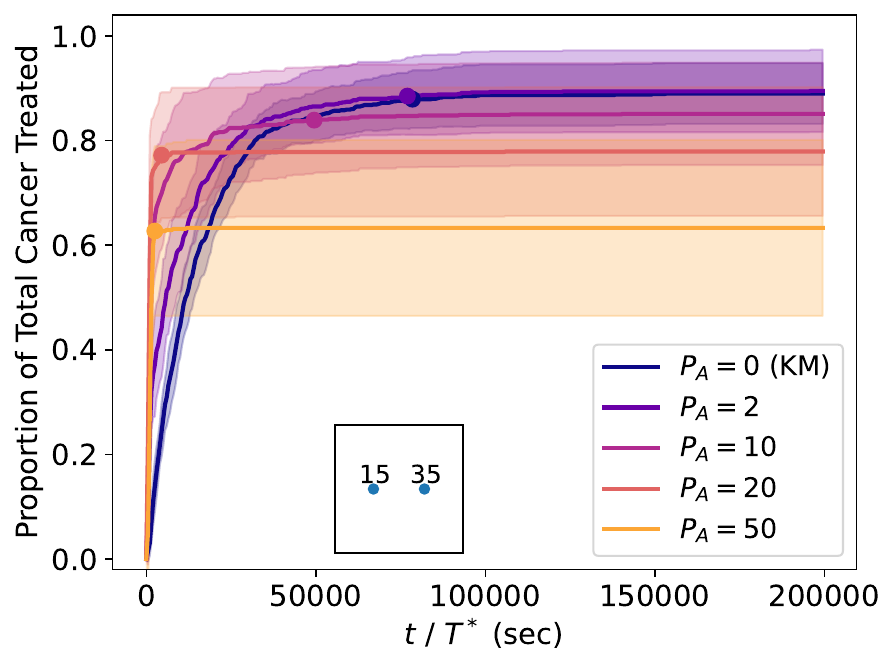}
        \caption{For $P_A=50$: $(T_{\text{fin}}, S_{\text{avg}}(T_{\text{fin}}))=(2500,0.627)$; for $P_A=20$: $(4500,0.772)$; for $P_A=10$: $(49500,0.839)$; for $P_A=2$: $(77000,0.885)$; for $P_A=0$: $(78500,0.879)$.}
        \label{fig:KMA-A-ffect-2UNeven}
    \end{subfigure}
    \hfill
    
    \centering
    \begin{subfigure}{.48\textwidth}
        \centering
        \includegraphics[width=.85\textwidth]{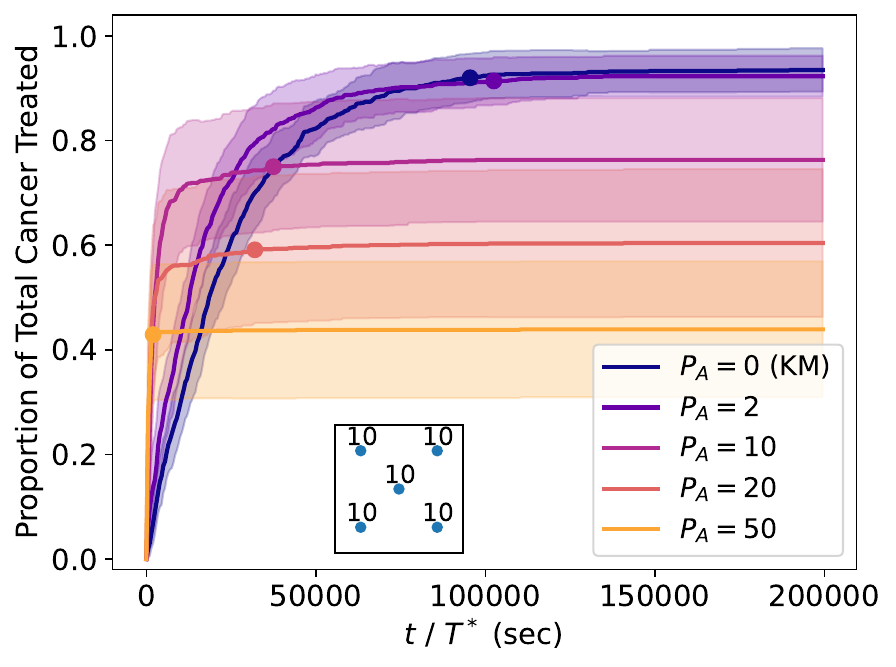}
        \caption{For $P_A=50$: $(T_{\text{fin}}, S_{\text{avg}}(T_{\text{fin}}))=(2000,0.429)$; for $P_A=20$: $(32000,0.591)$; for $P_A=10$: $(37500,0.75)$; for $P_A=2$: $(102500,0.914)$; for $P_A=0$: $(95500,0.92)$.}
        \label{fig:KMA-A-ffect-spreadout}
    \end{subfigure}
    \hfill
    \begin{subfigure}{.48\textwidth}
        \centering
        \includegraphics[width=.85\textwidth]{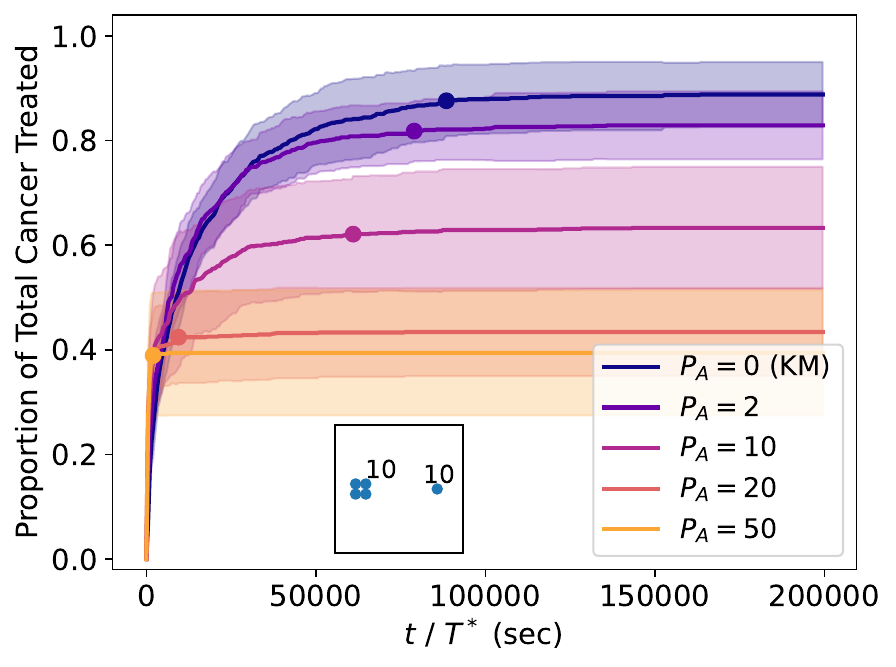}
        \caption{For $P_A=50$: $(T_{\text{fin}}, S_{\text{avg}}(T_{\text{fin}}))=(2000,0.389)$; for $P_A=20$: $(9500,0.424)$; for $P_A=10$: $(61000,0.621)$; for $P_A=2$: $(79000,0.818)$; for $P_A=0$: $(88500,0.876)$.}
        \label{fig:KMA-A-ffect-clusterplusone}
    \end{subfigure}

    \centering
    \begin{subfigure}{.48\textwidth}
        \centering
        \includegraphics[width=.85\textwidth]{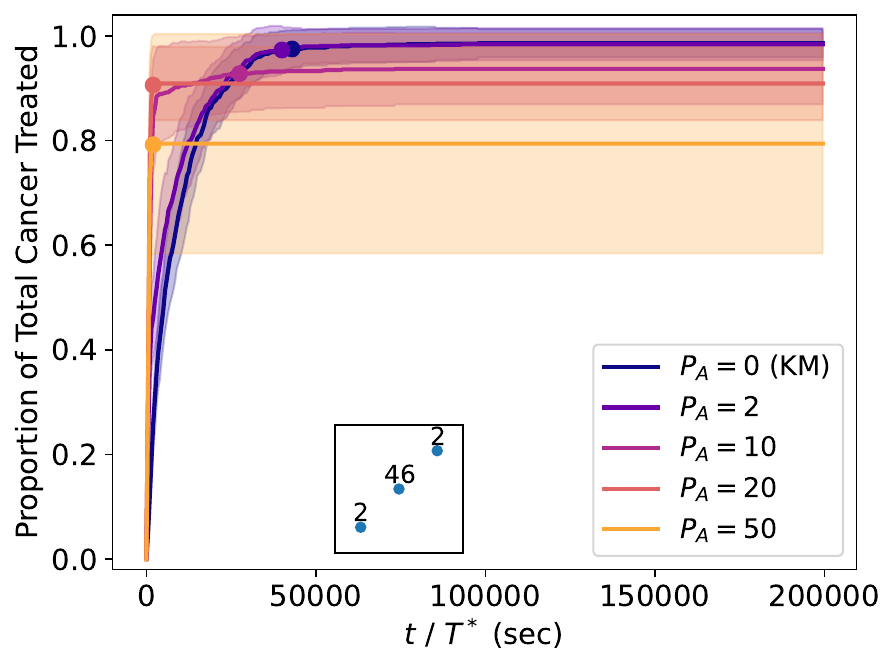}
        \caption{For $P_A=50$: $(T_{\text{fin}}, S_{\text{avg}}(T_{\text{fin}}))=(2000,0.792)$; for $P_A=20$: $(2000,0.906)$; for $P_A=10$: $(27500,0.928)$; for $P_A=2$: $(40000,0.973)$; for $P_A=0$: $(43000,0.975)$.}
        \label{fig:KMA-A-ffect-onebig}
    \end{subfigure}
    \hfill

    \caption{
    \textit{Alg. KMA:   }
    Simulation results for Algorithm KMA with $n=55$ agents across varying cancer site arrangements (depicted by the small scatterplots), investigating the effect of chemical A payload size $P_A$ on performance. $b=5\cdot 10^{(-11)}, D_A=10^{(-9)}$, $\phi_{\text{max}}=0.005$, $\alpha=\epsilon=2\cdot 10^{-5}$, $m=10^{-6}$, $r_{\text{K,M}}=1$. Plotted lines show average success $S$ (at that given point in time) over $20$ trials, and shaded regions are standard deviations.
    Points on main plot have coordinates $(T_{\text{fin}}, S_{\text{avg}}(T_{\text{fin}}))$ for the given $P_A$ setting whose color they match with.
    }
    \label{fig:KMA-A-ffect}
\end{figure}

Next, we analyze the effect of the signal strength of chemical A---or more precisely, the size of an individual chemical A payload $P_A$---on performance, across the same set of cancer site and demand arrangements (as listed in Subsection~\ref{sec:arrangements}).
Note that Algorithm KMA with $P_A=0$ is simply equivalent to Algorithm KM.
We fix the orientation-bias parameter $b=5\cdot 10^{(-11)}$.
See Figure~\ref{fig:KMA-A-ffect}.

Our results demonstrate that the stronger the A-signal(s)---i.e., the larger the value of $P_A$---the faster progressing but less successful the treatment.
These effects are significant: for the largest $P_A$ settings compared to the baseline of $P_A$ equal to zero, the treatment time $T_{\text{fin}}$ can be up to $30$ times faster and the total success $S_{\text{avg}}(T_{\text{fin}})$ can be up to $40\%$ worse.
Larger A-payloads yield stronger attractive signals, helping agents reach sites faster; however, stronger signals also produce an even more extreme cascade effect in which all of the agents converge on a single site. 
This trend is more dramatic when the cancer is more diffuse, both with respect to there being more total sites or with respect to the total demand being distributed out more evenly.

Again, sending all of the agents to one of many cancer sites is worse than sending them to one of a couple sites---for the greatest A-signal strengths, Figures \ref{fig:KMA-A-ffect-2Even} and \ref{fig:KMA-A-ffect-2UNeven} have higher success then Figures \ref{fig:KMA-A-ffect-spreadout} and \ref{fig:KMA-A-ffect-clusterplusone}.
Further, Figure \ref{fig:KMA-A-ffect-2UNeven} has greater success than Figure \ref{fig:KMA-A-ffect-2Even} as sending all of the agents to a site with demand $35$ (out of a total demand of $50$) yields a higher success metric value than sending all of the agents to a site with demand $25$ (out of a total demand of $50$).
Regarding the most extreme case of the cancer being concentrated, as shown by Figure \ref{fig:KMA-A-ffect-onebig}, the treatment is still reasonably successful (approximately $80\%$ success) even for the largest $P_A$ setting.

While the success of Algorithm KMA's treatment is generally low, it is important to note KMA will treat the site with the most demand extremely fast, thus killing the main portion of cancer.
If we were to consider multiple rounds of this cancer treatment by nanobots, then quickly killing one large cancerous cluster in the first pass of treatment could be a desirable option.
To extend this, in the case of there actually being only one, singular cancer site, Algorithm KMA would perform outstandingly both in treatment time and success.\footnote{This is demonstrated by the results for the ``active agent model'' in our single-site nanobots work \cite{ourSingleSite}.}
Returning to the case of there being multiple rounds of treatment, it is interesting to consider the remaining ``new'' cancer arrangement that is left over after one round of treatment.
For example, after one pass of KMA treatment on a concentrated arrangement similar to (e), the new arrangement of cancer still remaining is likely to be similar to the sparse diffuse arrangement of (a), with KMA mostly killing the major, most-demanding site.
Then, at this point, another algorithm (and accompanying parameter settings) which performs well for cancer patterns similar to arrangement (a), such as KM and KMAR, could be employed to ``clean up'' the final portions of cancer.

Ultimately, through the results in this section (Section~\ref{sec:res-kma}), we show which parameter settings yield the best KMA performance for each given type of cancer arrangement, respectively.
Regarding the orientation-bias parameter $b$, for very concentrated arrangements like (e), greater orientation-bias is optimal ($b=6\cdot 10^{(-11)}$), yielding both better success and faster treatment time.
For all other classes of arrangements, there is a clear trade-off between success and treatment time.
Greater $b$ values have faster treatment time but lower success; as such, moderate orientation-bias, around $b=4\cdot 10^{(-11)}$, yields intermediate performance in both metrics.
Regarding the signal strength of chemical A, or more precisely the size of an individual A-payload $P_A$, there is a similar trade-off between success and treatment time.
Across all cancer arrangements, greater $P_A$ values have faster treatment time but lower success, with a moderate $P_A$ value of around $10$ yielding intermediate performance in both metrics.
This trade-off is less dramatic for more concentrated cancer arrangements like (b) and (e), with larger $P_A$ values not worsening success so drastically.

\subsection{Results for Algorithm KMAR}\label{sec:res-kmar}

We now consider the even more involved case in which the repellent signal chemical R is introduced and used to discourage agents from planting at cancer sites that are already sufficiently treated.
Algorithm KMAR seeks to leverage the A-payloads which speed up treatment time by amplifying the attractive signal(s), while adding chemical R in order to combat the previously discussed issue with KMA's success in which too many agents treat a singular site.
Our results ultimately show that KMAR has good performance (in both success and treatment time) across all cancer site and demand arrangements, assuming well-chosen parameter values.

In each of the subsections to come, we analyze the effect of individual parameters on performance in order to tune and pin down the best values.

\subsubsection{Effect of A- Versus R-Payload Threshold on KMAR Performance}\label{sec:res-kmar-thresh}
Recall from Subsection~\ref{sec:alg-kmar} that when agents in Algorithm KMAR reach a cancer site, they drop A-payloads if the local chemical A concentration is less than $( r_{A,M} \cdot P_{M_j})$, or instead drop R-payloads if the A concentration is greater than or equal to $( r_{A,M} \cdot P_{M_j})$.
That is, the value of $r_{A,M}$ determines the threshold for whether agents drop chemical A versus chemical R upon reaching a site.

While, intuitively, we want to be able to send agents to random walk and explore elsewhere immediately once a given site is fully treated, the limitation of following diffused chemicals' gradients does not allow for this.
When agents begin dropping R-payloads at a site that has been deemed to be sufficiently treated, it takes a nonnegligible amount of time for this repellent signal to spread and impact other agents' behavior.
As such, we want to preemptively start releasing R-payloads once a site is getting close to being fully treated.
To emphasize the previous sentence, we often elect for R-payloads to be dropped \textit{before} a given site has been fully treated.
Tuning the threshold value $r_{AM}$ is exactly this problem of deciding when to have agents start releasing the repellent R-payloads instead of releasing further A-payloads.
Note that for sufficiently large $r_{A,M}$ (as $r_{A,M}$ approaches infinity), Algorithm KMAR simply becomes KMA as only A-payloads are ever dropped.
On the other extreme end, $r_{A,M}$ equal to zero results in only R-payloads ever being dropped.

\begin{figure}
    \centering
    \begin{subfigure}{.48\textwidth}
        \centering
        \includegraphics[width=.85\textwidth]{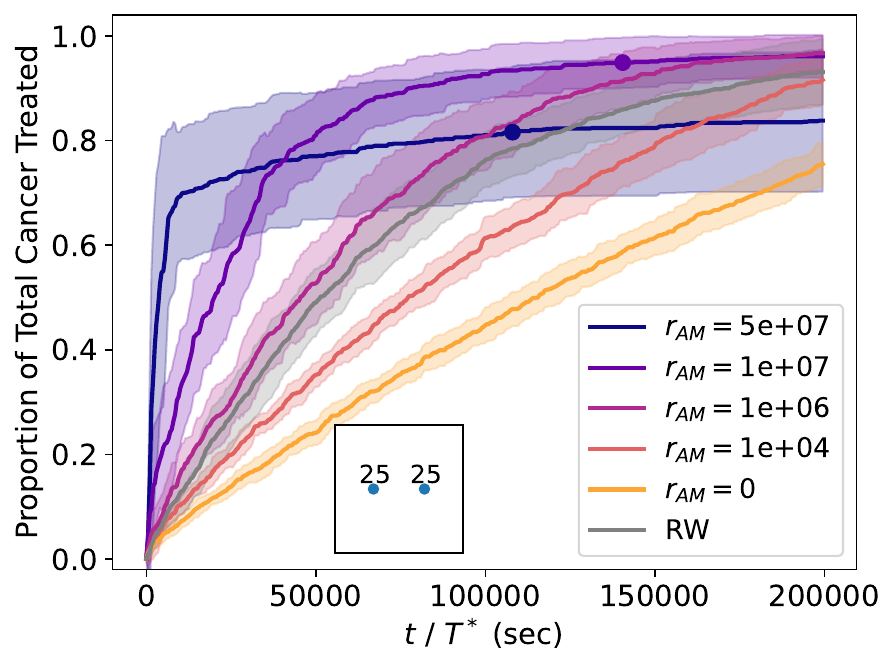}
        \caption{For $r_{AM}=5\cdot 10^7$: $(T_{\text{fin}}, S_{\text{avg}}(T_{\text{fin}}))=(108000,0.816)$; for $r_{AM}=10^7$: $(140500,0.949)$.}
    \end{subfigure}
    \hfill
    \begin{subfigure}{.48\textwidth}
        \centering
        \includegraphics[width=.85\textwidth]{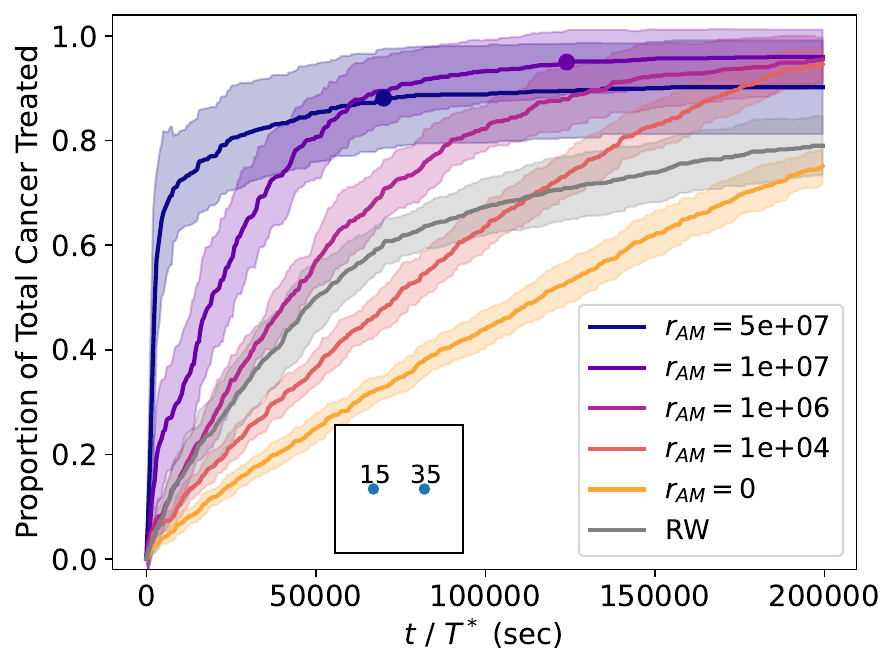}
        \caption{For $r_{AM}=5\cdot 10^7$: $(T_{\text{fin}}, S_{\text{avg}}(T_{\text{fin}}))=(70000,0.881)$; for $r_{AM}=10^7$: $(124000,0.95)$.}
    \end{subfigure}
    \hfill
    
    \centering
    \begin{subfigure}{.48\textwidth}
        \centering
        \includegraphics[width=.85\textwidth]{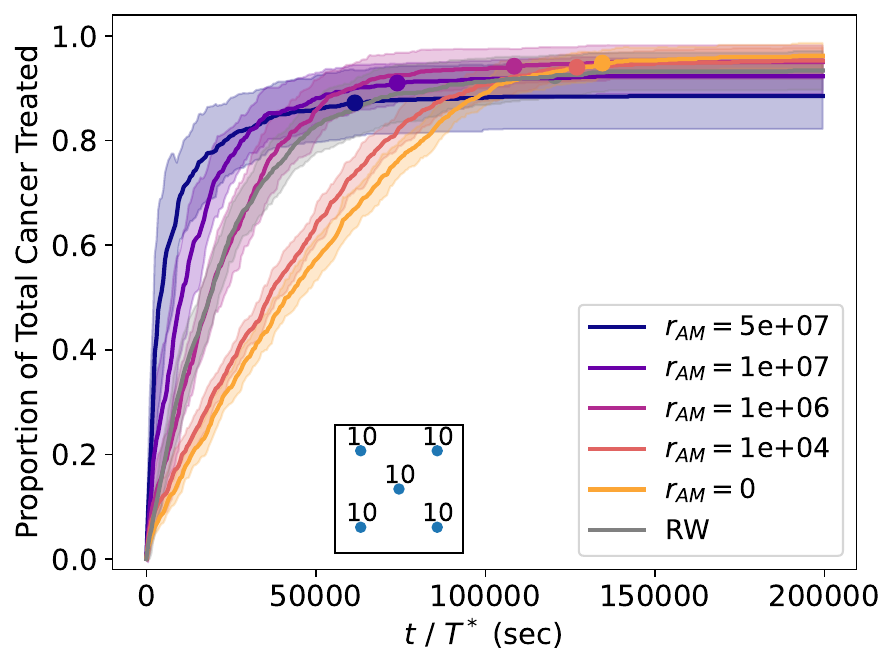}
        \caption{For $r_{AM}=5\cdot 10^7$: $(T_{\text{fin}}, S_{\text{avg}}(T_{\text{fin}}))=(61500,0.872)$; for $r_{AM}=10^7$: $(74000,0.91)$; for $r_{AM}=10^6$: $(108500,0.942)$; for $r_{AM}=10^4$: $(127000,0.94)$; for $r_{AM}=0$: $(134500,0.948)$; for RW: $(120500,0.924)$.}
    \end{subfigure}
    \hfill
    \begin{subfigure}{.48\textwidth}
        \centering
        \includegraphics[width=.85\textwidth]{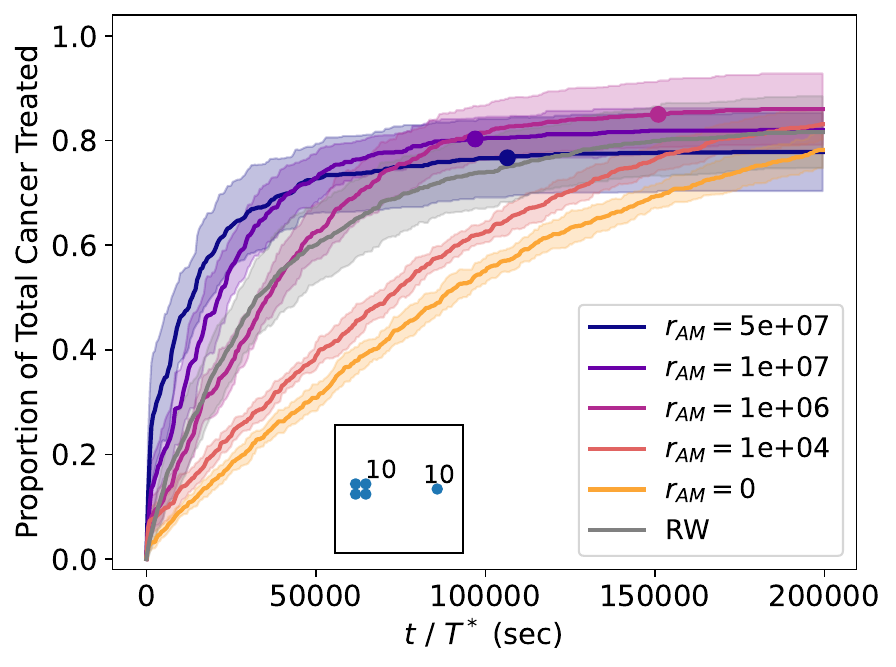}
        \caption{For $r_{AM}=5\cdot 10^7$: $(T_{\text{fin}}, S_{\text{avg}}(T_{\text{fin}}))=(106500,0.767)$; for $r_{AM}=10^7$: $(97000,0.803)$; for $r_{AM}=10^6$: $(151000,0.85)$.}
    \end{subfigure}
    \hfill

    \centering
    \begin{subfigure}{.48\textwidth}
        \centering
        \includegraphics[width=.85\textwidth]{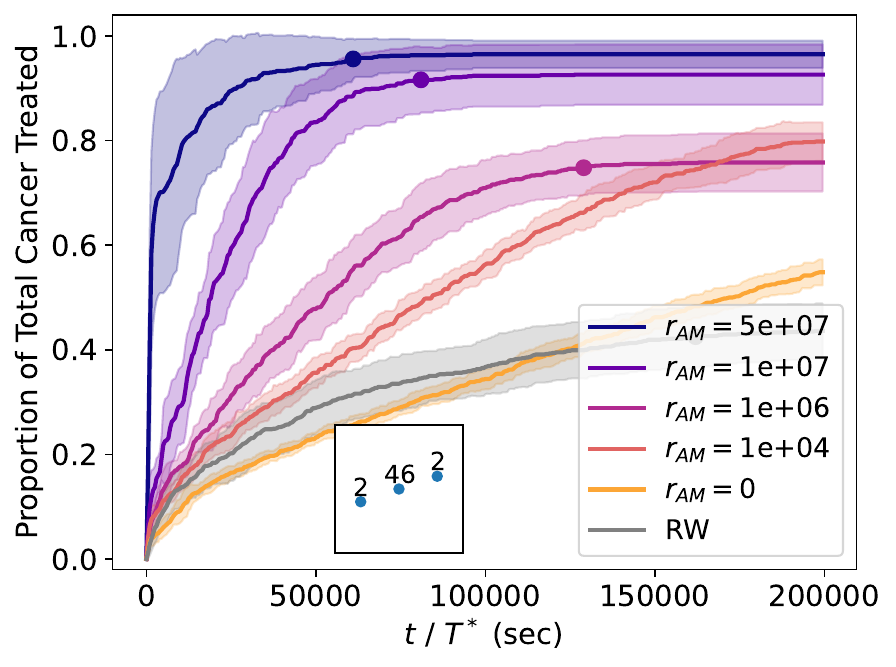}
        \caption{For $r_{AM}=5\cdot 10^7$: $(T_{\text{fin}}, S_{\text{avg}}(T_{\text{fin}}))=(61000,0.956)$; for $r_{AM}=10^7$: $(81000,0.916)$; for $r_{AM}=10^6$: $(129000,0.748)$; for RW: $(162000,0.425)$.}
    \end{subfigure}
    
    \caption{
    \textit{Alg. KMAR:   }
    Simulation results for Algorithm KMAR with $n=55$ agents across varying cancer site arrangements (depicted by small scatterplots), showing effect of threshold value for whether agents drop chemical A or R payloads, $r_{AM}$, on performance. RW serves as baseline of comparison. $b=4\cdot 10^{(-11)}, P_A=10, P_R=50, D_A=10^{(-9)}, D_R=10^{(-10)}$, $\phi_{\text{max}}=0.005$, $\alpha=\epsilon=2\cdot 10^{-5}$, $m=10^{-6}$, $r_{KM}=1$. Plotted lines show average success $S$ (at that given point in time) over $20$ trials, and shaded regions are standard deviations.
    Points on main plot have coordinates $(T_{\text{fin}}, S_{\text{avg}}(T_{\text{fin}}))$ for the given $r_{AM}$ setting whose color they match with. For $r_{AM}$ settings with no such point, $T_{\text{fin}}$ is undefined, i.e., $T_{\text{fin}}>200000$.
    }
    \label{fig:KMAR_thresheffect}
\end{figure}

For our chosen set of cancer site and demand arrangements (as listed in Subsection~\ref{sec:arrangements}), we vary $r_{AM}$ and measure the success metric $S$ over time.
See Figure~\ref{fig:KMAR_thresheffect}.
We fix the chemical A payload size $P_A=10$, the chemical R payload size $P_R=50$, and the orientation-bias parameter $b=4\cdot 10^{(-11)}$.

Our results show that for large $r_{AM}$, the treatment time is fast, but success can often be suboptimal as not enough repellent chemical R is released in order to discourage agents from wasting treatment on already treated sites. 
For small $r_{AM}$, the treatment time is extremely slow as too much repelling chemical R is released, resulting in agents aimlessly random walking around forever.
However, there is some middle-ground of $r_{AM}$ values (which can vary between distinct site and demand arrangements) for which Algorithm KMAR performs with both fast treatment time and high success. 
Ultimately, we fix a ``good'' such $r_{AM}$ value for the next set of KMAR experiments.

Across all site and demand arrangements, the results show that the treatment time $T_{\text{fin}}$ decreases as $r_{A,M}$ increases.
As $r_{A,M}$ increases, there are more total chemical A payloads being dropped by agents as well as fewer R-payloads; consequently, agents are more biased to move towards cancer sites (rather than away from them) and thus reach a final site faster, speeding up treatment time.
The effect of $r_{A,M}$ on treatment time is significant, as demonstrated by the fact that the smallest $r_{A,M}$ setting(s) has (have) even slower progressing treatment than the simple random walk RW, while the largest $r_{A,M}$ settings have significantly faster progressing treatment than RW.

Regarding the effect of $r_{A,M}$ on success, there are some differences between cancer arrangements.
For the more diffuse cancer arrangements (a)--(d), the maximum achieved success decreases once $r_{A,M}$ takes on too large a value.
More specifically, for arrangements (a) and (b), the $r_{A,M}=10^7$ setting achieves greater success $S_{\text{avg}}(T_{\text{fin}})$ than the $r_{A,M}=5\cdot 10^7$ setting; the precise declines in average achieved success for (a) and (b) are $13$ and $7$ percent, respectively.
Though not captured within the $200000$ runtime cutoff of our experiments, it appears that the $r_{A,M}=10^4$ and $r_{A,M}=10^6$ settings are on pace to surpass the larger $r_{A,M}$ settings in success as time progresses beyond $200000$.
For arrangement (c), the settings of $r_{A,M}=0,10^4,10^6$ achieve greater success than the $r_{A,M}=10^7$ setting, which achieves greater success than the $r_{A,M}=5\cdot 10^7$ setting, though these discrepancies in total achieved success are less dramatic than those seen for arrangements (a) and (b).
Similarly, for arrangement (d), the $r_{A,M}=10^6$ setting achieves greater success than the $r_{A,M}=10^7$ setting, which achieves greater success than the $r_{A,M}=5\cdot 10^7$ setting.

On the other hand, for the most concentrated cancer arrangement (e), larger values of $r_{A,M}$ actually correspond to higher success of treatment, generally, with the largest value of $r_{A,M}$ tested yielding the highest success across all points in time. 
Since the major site here has almost all of the demand, using more chemical A than chemical R (by increasing the threshold $r_{A,M}$) improves success by appropriately ensuring that almost all of the agents reach the major site.
From $r_{A,M}=10^6$ to $r_{A,M}=10^7$ the average success $S_{\text{avg}}(T_{\text{fin}})$ improves by approximately $17$ percent, and from $r_{A,M}=10^7$ to $r_{A,M}=5 \cdot 10^7$ the average success (at ``termination'', or time $T_{\text{fin}}$) improves by another four percent.
Furthermore, from $r_{A,M}=0$ to $r_{A,M}=10^6$, the average success at the maximum runtime cutoff of $t=200000$ improves by nearly twenty percent (we compare $S_{\text{avg}}(200000)$ here instead of $S_{\text{avg}}(T_{\text{fin}})$ since $T_{\text{fin}}$ is undefined for $r_{A,M}=0$).
The only slight exception to the correspondence between greater $r_{A,M}$ values and faster treatment times is for the $r_{A,M}=10^4$ setting, which slightly surpasses $r_{A,M}=10^6$ in average success from approximately $t=170000$ on.
However, $r_{A,M}=10^6$ does indeed have greater average success than $r_{A,M}=10^4$ for all clearance times less than the $t=170000$ mark, which is arguably already an unreasonably long clearance time of nearly two full days.

Motivated by the results presented in this section, we fix $r_{A,M}=10^7=10^6 \cdot r_{K,M} \cdot P_A$ for the remaining KMAR experiments to come in Subsections \ref{sec:res-kmar-b} and \ref{sec:res-kmar-PR}.
While the $r_{A,M}$ value producing the highest success is slightly different between cancer arrangements, fixing $r_{A,M}$ allows us to isolate the impact of the other given parameters that we will be testing.
Further, even for the arrangements for which $r_{A,M}=10^7$ does not produce the highest success $S_{\text{avg}}(T_{\text{fin}})$ value, the discrepancy is never more than four percent.

\subsubsection{Effect of Orientation-Bias Parameter on KMAR Performance}\label{sec:res-kmar-b}
\begin{figure}
    \centering
    \begin{subfigure}{.48\textwidth}
        \centering
        \includegraphics[width=.85\textwidth]{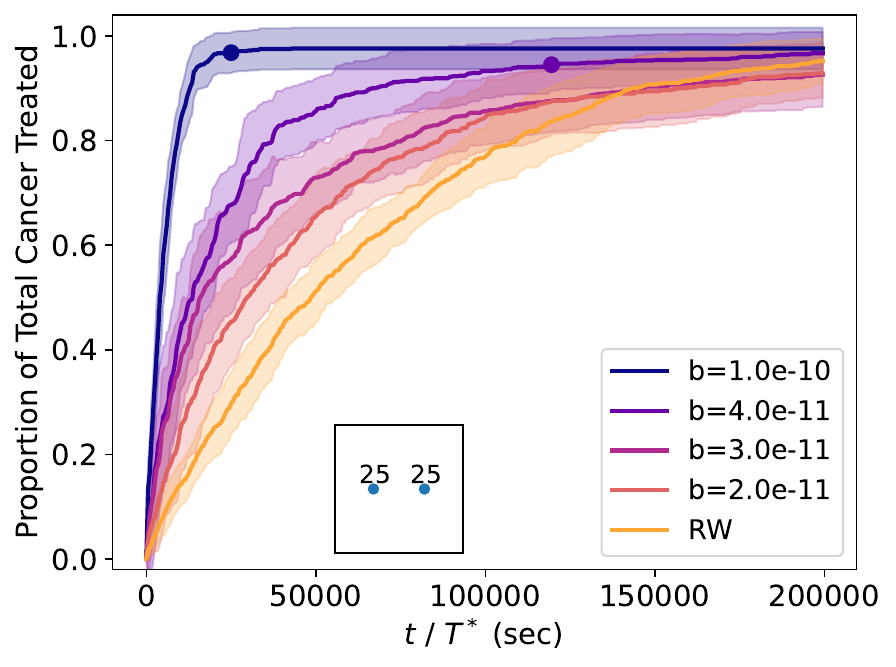}
        \caption{For $b=10^{(-10)}$: $(T_{\text{fin}}, S_{\text{avg}}(T_{\text{fin}})) = (25000,0.968)$; $b=4\cdot 10^{(-11)}$: $(119500,0.945)$.}
        \label{fig:KMAR-2Even}
    \end{subfigure}
    \hfill
    \begin{subfigure}{.48\textwidth}
        \centering
        \includegraphics[width=.85\textwidth]{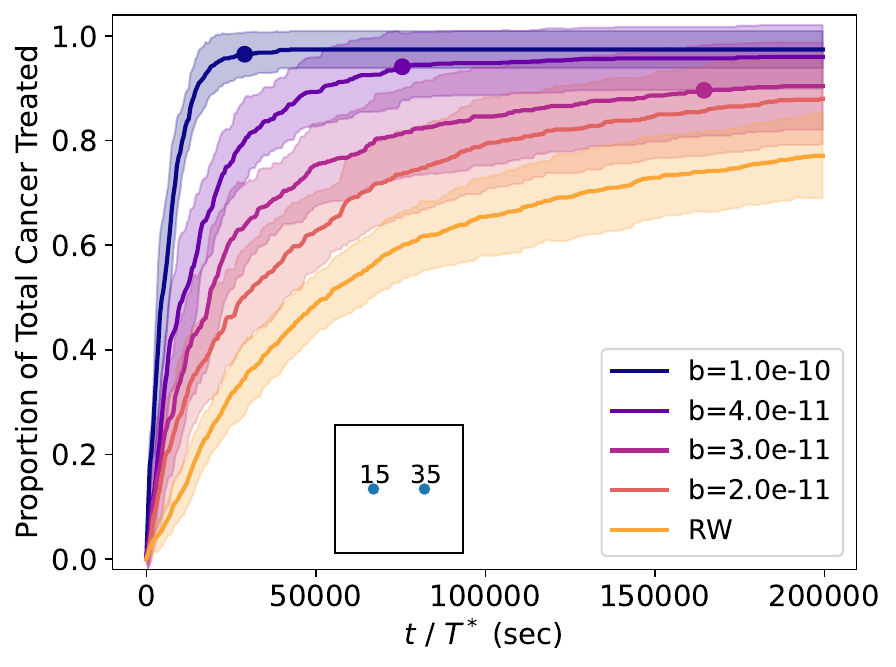}
        \caption{For $b=10^{(-10)}$: $(T_{\text{fin}}, S_{\text{avg}}(T_{\text{fin}})) = (29000,0.965)$; $b=4\cdot 10^{(-11)}$: $(75500,0.941)$; $b=3\cdot 10^{(-11)}$: $(164500,0.896)$.}
        \label{fig:KMAR-2UNeven}
    \end{subfigure}
    \hfill
    
    \centering
    \begin{subfigure}{.48\textwidth}
        \centering
        \includegraphics[width=.85\textwidth]{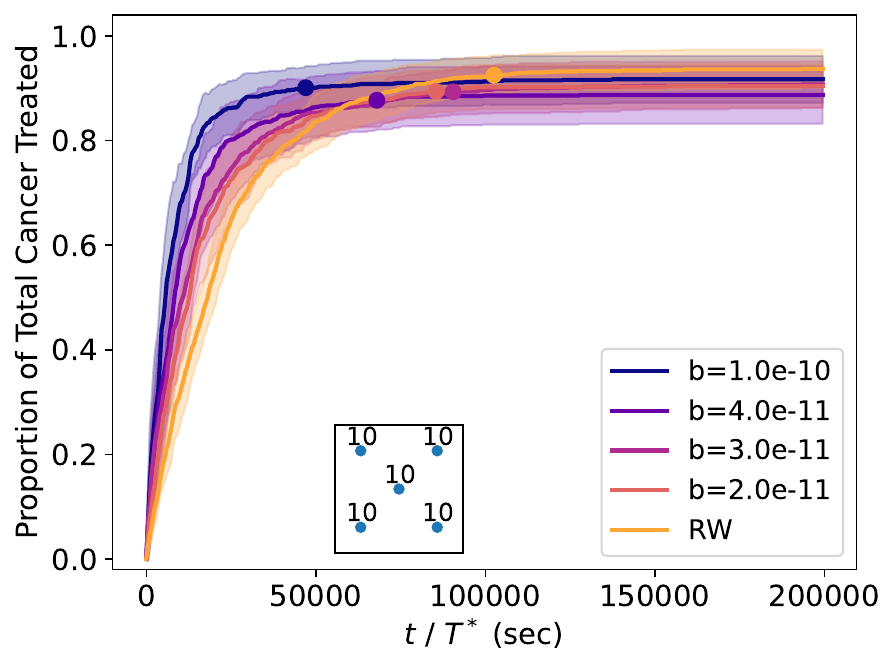}
        \caption{For $b=10^{(-10)}$: $(T_{\text{fin}}, S_{\text{avg}}(T_{\text{fin}})) = (47000,0.901)$; for $b=4\cdot 10^{(-11)}$: $(68000,0.877)$; for $b=3\cdot 10^{(-11)}$: $(90500,0.893)$; for $b=2\cdot 10^{(-11)}$: $(85500,0.894)$; for RW: $(102500,0.925)$.}
        \label{fig:KMAR-spreadout}
    \end{subfigure}
    \hfill
    \begin{subfigure}{.48\textwidth}
        \centering
        \includegraphics[width=.85\textwidth]{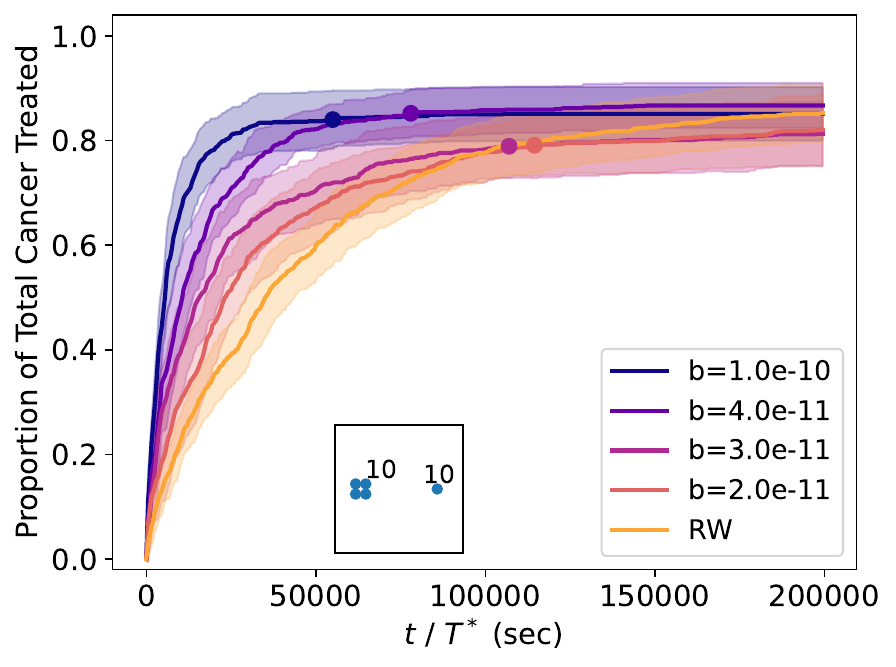}
        \caption{For $b=10^{(-10)}$: $(T_{\text{fin}}, S_{\text{avg}}(T_{\text{fin}})) = (55000,0.84)$; for $b=4\cdot 10^{(-11)}$: $(78000,0.852)$; for $b=3\cdot 10^{(-11)}$: $(107000,0.789)$; for $b=2\cdot 10^{(-11)}$: $(114500,0.791)$.}
        \label{fig:KMAR-clusterplusone}
    \end{subfigure}
    \hfill

    \centering
    \begin{subfigure}{.48\textwidth}
        \centering
        \includegraphics[width=.85\textwidth]{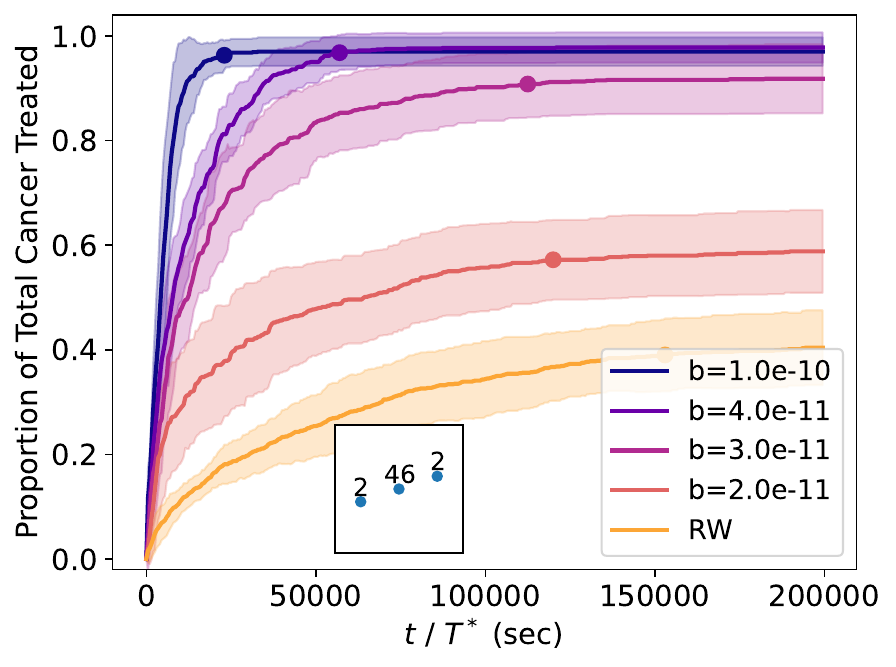}
        \caption{For $b=10^{(-10)}$: $(T_{\text{fin}}, S_{\text{avg}}(T_{\text{fin}})) = (23000,0.963)$; for $b=4\cdot 10^{(-11)}$: $(57000,0.968)$; for $b=3\cdot 10^{(-11)}$: $(112500,0.908)$; for $b=2\cdot 10^{(-11)}$: $(120000,0.572)$; for RW: $(153000,0.39)$.}
        \label{fig:KMAR-onebig}
    \end{subfigure}
    
    \caption{
    \textit{Alg. KMAR:   }
    Simulation results for Algorithm KMAR with $n=55$ agents, across varying cancer site and $\{P_{M_j}\}_j$ arrangements (depicted by the small scatterplots) and for different amounts of orientation-bias. RW serves as a baseline of comparison. $r_{AM}=10^7, P_A=10, P_R=50, D_A=D_R=10^{(-9)}$, $\phi_{\text{max}}=0.005$, $\alpha=\epsilon=2\cdot 10^{-5}$, $m=10^{-6}$, $r_{KM}=1$. Plotted lines show average success $S$ (at that given point in time) over $20$ trials, and shaded regions are standard deviations.
    Points on main plot have coordinates $(T_{\text{fin}}, S_{\text{avg}}(T_{\text{fin}}))$ for the given $b$ setting whose color they match with. For $b$ settings with no such point, $T_{\text{fin}}$ is undefined, i.e., $T_{\text{fin}}>200000$.
    }
    \label{fig:KMAR}
\end{figure}

Next, we investigate the effect of the orientation-bias parameter $b$ on the performance of Algorithm KMAR, still across the same site and demand arrangements (as listed in Subsection~\ref{sec:arrangements}).
See Figure~\ref{fig:KMAR}.
We fix $r_{A,M}=10^7$, $P_A=10$, and $P_R=50$, based on the experiments in Subsection~\ref{sec:res-kmar-thresh}.

Our results show that generally, greater orientation-bias yields better performing treatment by KMAR both in treatment time and success.
Under greater orientation-bias, agents in KMAR follow the A-signals more closely to reach sites faster as well as follow the R-signals more closely to find untreated sites faster, appropriately when tasked to do each.

Recall from the movement model defined in Subsection~\ref{sec:bots-model} (specifically Equation~\ref{eqn:beta}) that the orientation-bias parameter $b$ determines how closely (in expectation) agents follow \textit{all} of the various chemical gradients.
That is, under greater orientation-bias, agents ascend the chemical M- and A-gradients more closely as well as descend the R-gradient more closely; $b$ effects the extremeness of both attractive and repellent agent motion, indiscriminately.
Consequently, the effect of $b$ on KMAR treatment time is not obvious a priori: while greater orientation-bias results in agents following the M- and A-signals more closely, which contributes to reducing treatment time, greater orientation-bias also results in agents following the R-signals more closely, which contributes to increasing treatment time.
However, the results in Figure~\ref{fig:KMAR} show that the net effect of greater orientation-bias is in fact faster treatment times.

Recall that initially, only A-payloads are dropped at every cancer site.
Then, once the chemical A concentration reaches $(r_{A,M} \cdot P_{M_j})$ at some site $j$, R-payloads begin to be dropped there by agents.
As such, we can loosely consider two phases of KMAR treatment: the first phase is the time for which the A concentration is less than $(r_{A,M} \cdot P_{M_j})$ for every cancer site j, and the second phase is the remaining time, in which R-payloads begin to be dropped.
Notably, this first phase is simply Algorithm KMA.
It is possible that the impact of the orientation-bias parameter on the first phase is more drastic than its impact on the second phase such that the decrease in the runtime of the first phase caused by greater orientation-bias dominates the increase in the runtime of the second phase caused by greater orientation-bias.
This would explain the ultimate decrease in total treatment time that we observe under greater orientation-bias.
An alternative explanation for the correlation we see in Figure~\ref{fig:KMAR} between increasing orientation-bias and decreasing treatment times is that greater orientation-bias actually speeds up both of the aforementioned two phases.
The speedup of the first phase is easy to see, but regarding the second phase, if agents follow the R-signals extremely closely, this could cause them to quickly find another untreated site instead of aimlessly random walking by being repelled strongly enough in the direction of another site, coincidentally.
These points remain as conjectures for now.

Importantly, the A- versus R-payload threshold value $r_{A,M}$ and relative strengths of A and R $(P_R / P_A)$ are fixed here; further simulated experiments are needed in order to test whether the negative correlation between $b$ and $T_{\text{fin}}$ is a general property of KMAR, or rather dependent on the specific values of $r_{A,M}$ and/or $(P_R / P_A)$.
For example, if the threshold $r_{A,M}$ were lowered, increased orientation-bias might instead correspond to slower treatment times if the repellent signals from the extra R-payloads start to dominate the attractive M- and A-signals under high orientation-bias, but this remains to be seen.

Regarding the correlation between greater orientation-bias and greater success, this can be viewed as evidence for our chosen fixed values for $r_{A,M}$ as well as $(P_R / P_A)$ being fairly optimal.
Assuming favorable settings of $r_{A,M}$ and $(P_R / P_A)$, increasing the amount of orientation-bias results in---to put it simply---agents doing what we want them to do more often, or more deterministically, thus yielding greater success (as is indeed observed in the results).  
That is, under greater orientation-bias, agents follow the M- and A-signals---which have appropriate relative strengths by our assumption of a favorable $(P_R/P_A)$ setting---more reliably (note that we are not discussing speed here), to successfully reach the cancer sites still in need of treatment.
Also, under greater orientation-bias, agents follow the R-signals more reliably---and at the correct time(s) by our assumption of a favorable $r_{A,M}$ setting---to successfully avoid already-treated sites and find other sites elsewhere.
Algorithm KMAR is the most sophisticated of our presented algorithms, involving the most nuanced agent behavior.
If we properly leverage these extra capabilities to design a successful treatment strategy, then increasing the amount of orientation-bias is just reducing noise compared to otherwise optimal agent behavior.

It is worth noting that the degree of correspondence between orientation-bias and success varies slightly among the different cancer site and demand arrangements.
For example, for Figures \ref{fig:KMAR-clusterplusone} and \ref{fig:KMAR-onebig}, the two largest $b$ settings yield approximately the same success rate $S_{\text{avg}}(T_{\text{fin}})$, while for the other arrangements, there is indeed a nonegligible improvement in success between these two $b$ values.
There are also certain anomalies such as for arrangement (e) in which the $b=2\cdot 10^{(-11)}$ and $b=3\cdot 10^{(-11)}$ settings have similar treatment times (within $10000$ seconds) but very different success rates (nearly $35$ percent).
Generally, for the arrangements with fewer total cancer sites, (a), (b), and (e), the effect of $b$ on success $S$ is more dramatic; the difference in success across $b$ settings is the least drastic for arrangement (c).

While our results demonstrate that increasing the amount of orientation-bias improves overall performance for KMAR, such a choice to vary $b$ may be limited in implementation by the constraints of nanotechnologies.

\subsubsection{Effect of Relative Strengths of Chemicals A and R on KMAR Performance}\label{sec:res-kmar-PR}

\begin{figure}
    \centering
    \begin{subfigure}{.48\textwidth}
        \centering
        \includegraphics[width=.85\textwidth]{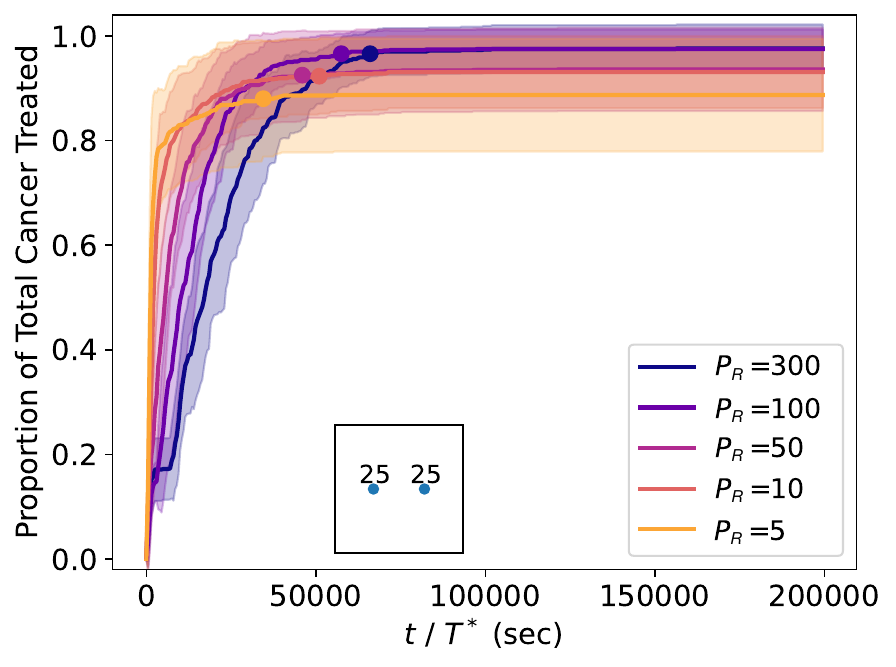}
        \caption{For $P_R=5$: $(T_{\text{fin}}, S_{\text{avg}}(T_{\text{fin}}))=(34500,0.88)$; for $P_R=10$: $(51000,0.923)$; for $P_R=50$: $(46000,0.925)$; for $P_R=100$: $(57500,0.966)$; for $P_R=300$: $(66000,0.966)$.}
        \label{fig:KMAR-Reffect-2Even}
    \end{subfigure}
    \hfill
    \begin{subfigure}{.48\textwidth}
        \centering
        \includegraphics[width=.85\textwidth]{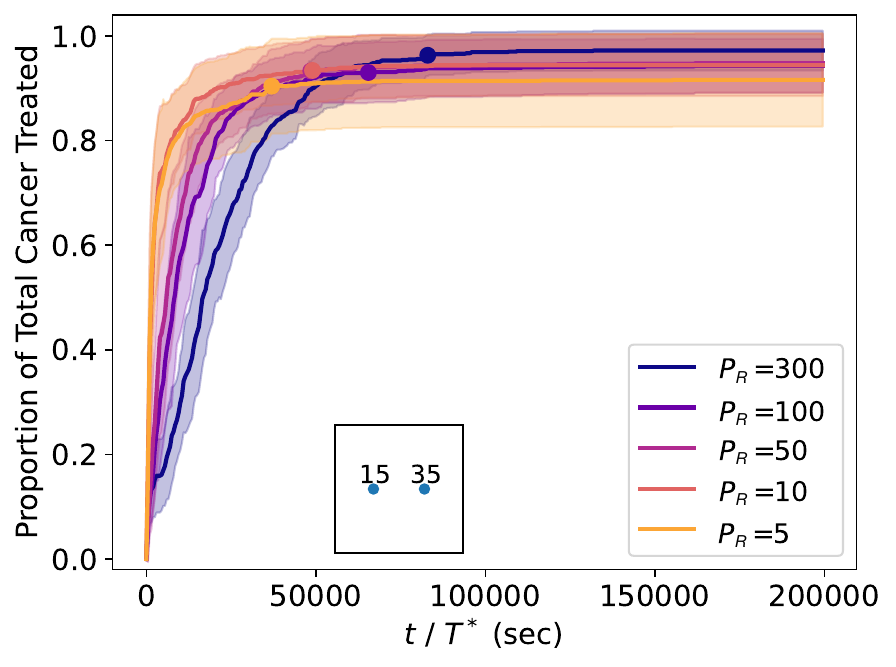}
        \caption{For $P_R=5$: $(T_{\text{fin}}, S_{\text{avg}}(T_{\text{fin}}))=(37000,0.904)$; for $P_R=10$: $(49000,0.934)$; for $P_R=50$: $(48500,0.933)$; for $P_R=100$: $(65500,0.93)$; for $P_R=300$: $(83000,0.963)$.}
        \label{fig:KMAR-Reffect-2UNeven}
    \end{subfigure}
    \hfill
    
    \centering
    \begin{subfigure}{.48\textwidth}
        \centering
        \includegraphics[width=.85\textwidth]{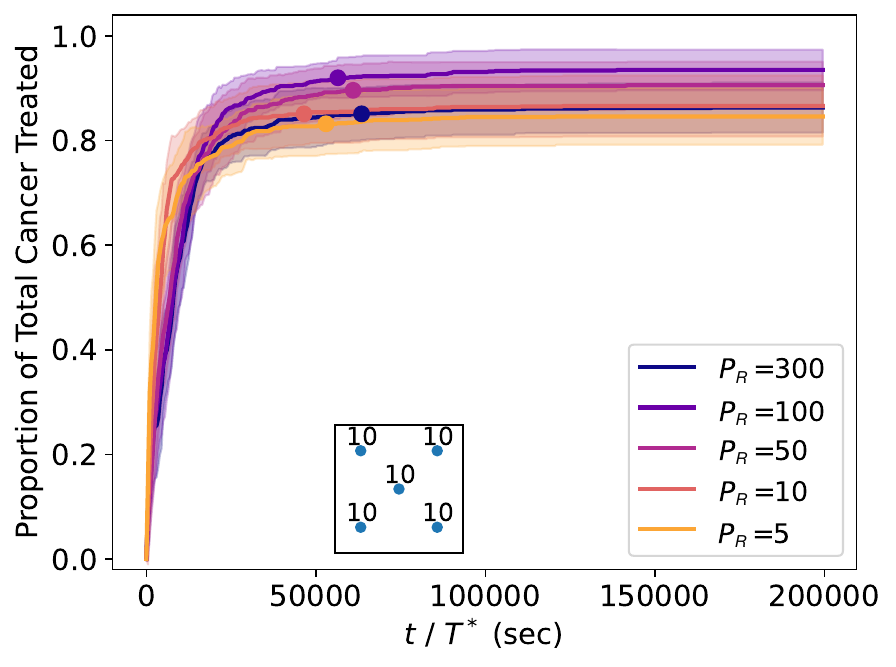}
        \caption{For $P_R=5$: $(T_{\text{fin}}, S_{\text{avg}}(T_{\text{fin}}))=(53000,0.832)$; for $P_R=10$: $(46500,0.851)$; for $P_R=50$: $(61000,0.896)$; for $P_R=100$: $(56500,0.92)$; for $P_R=300$: $(63500,0.851)$.}
        \label{fig:KMAR-Reffect-spreadout}
    \end{subfigure}
    \hfill
    \begin{subfigure}{.48\textwidth}
        \centering
        \includegraphics[width=.85\textwidth]{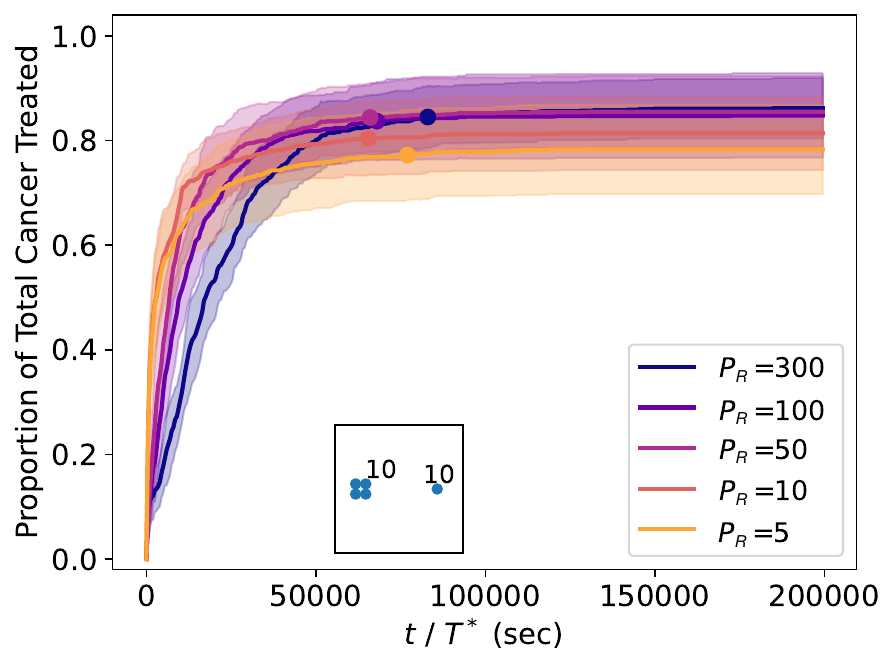}
        \caption{For $P_R=5$: $(T_{\text{fin}}, S_{\text{avg}}(T_{\text{fin}}))=(77000,0.772)$; for $P_R=10$: $(65500,0.804)$; for $P_R=50$: $(66000,0.844)$; for $P_R=100$: $(68000,0.837)$; for $P_R=300$: $(83000,0.845)$.}
        \label{fig:KMAR-Reffect-clusterplusone}
    \end{subfigure}

    \centering
    \begin{subfigure}{.48\textwidth}
        \centering
        \includegraphics[width=.85\textwidth]{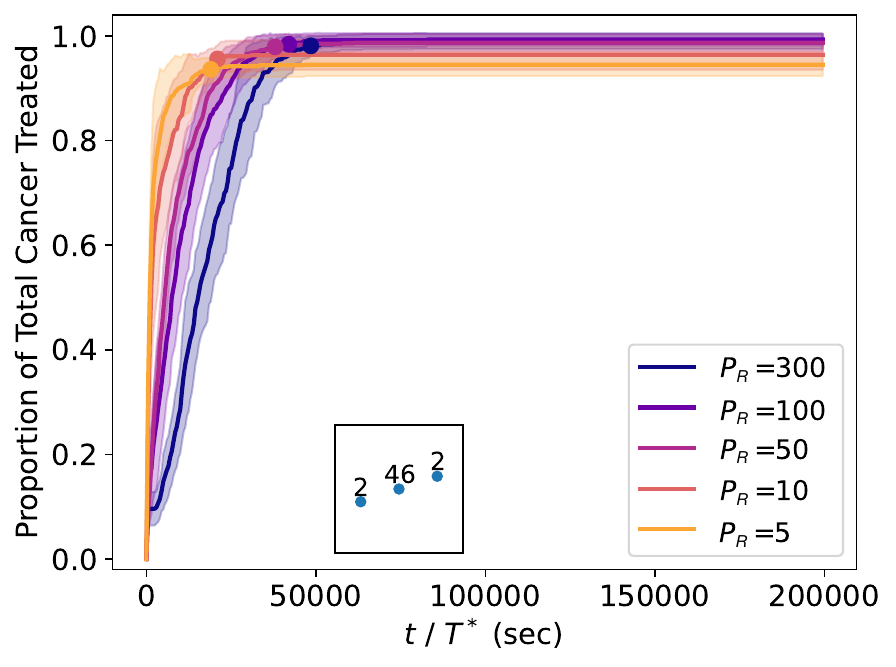}
        \caption{For $P_R=5$: $(T_{\text{fin}}, S_{\text{avg}}(T_{\text{fin}}))=(19000,0.936)$; for $P_R=10$: $(21000,0.956)$; for $P_R=50$: $(38000,0.979)$; for $P_R=100$: $(42000,0.984)$; for $P_R=300$: $(48500,0.981)$.}
        \label{fig:KMAR-Reffect-onebig}
    \end{subfigure}
    
    \caption{
    \textit{Alg. KMAR:   }
    Simulation results for Algorithm KMAR with $n=55$ agents across varying cancer site arrangements (depicted by the small scatterplots), showing the effect of the relative strengths of chemicals A and R ($P_R/P_A$) on performance. $b=6\cdot 10^{(-11)}, P_A=10, D_A=D_R=10^{(-9)}, r_{AM}=10^7$, $\phi_{\text{max}}=0.005$, $\alpha=\epsilon=2\cdot 10^{-5}$, $m=10^{-6}$, $r_{KM}=1$. Plotted lines show average success $S$ (at that given point in time) over $20$ trials, and shaded regions are standard deviations.
    Points on main plot have coordinates $(T_{\text{fin}}, S_{\text{avg}}(T_{\text{fin}}))$ for the given $P_R$ setting whose color they match with. For $P_R$ settings with no such point, $T_{\text{fin}}$ is undefined, i.e., $T_{\text{fin}}>200000$.
    }
    \label{fig:KMAR-Reffect}
\end{figure}

Lastly, we analyze the effect of the relative strengths of the chemical A and chemical R signals on KMAR performance by varying the size of chemical R payloads $P_R$ for a fixed A-payload size $P_A$, i.e., varying $(P_R/P_A)$.
These results are carried out for the same set of distinct cancer arrangements (as listed in Subsection~\ref{sec:arrangements}).
See Figure~\ref{fig:KMAR-Reffect}.
We fix $P_A=10$, $b=6\cdot 10^{(-11)}$, and $r_{AM}=10^7$, based on the experiments in the previous Subsections~\ref{sec:res-kmar-thresh} and \ref{sec:res-kmar-b}.

Our results show that as the strength of chemical R increases relative to A, the treatment time generally becomes slower as agents are more strongly and reliably repelled away from already treated sites and forced to spend time searching the rest of the space for other untreated sites.
However, this additional random walking and searching reduces the chance of any of the cancer sites being missed by the nanobot swarm's detection and treatment; indeed, our results show that greater strengths of chemical R relative to A correspond with higher success rates.

The settings which achieve the greatest success here have R-signals that are stronger than the A-signals, i.e., $(P_R/P_A)$ which is greater than one. 
Specifically, the largest  $(P_R/P_A)$ values we tested, which typically yield the highest success rates, were ten and thirty, i.e., R-payloads which are ten and thirty times larger than their A-payload counterparts, respectively.
Since R-payloads are not dropped until after multiple A-payloads have already been dropped, the R-signals must overcome and dominate the A-signals (as well as the persistent M-signals) in order to actually influence agent motion. 
Thus, it is intuitive that extremely strong R-payloads (relative to A) are needed in order to quickly shift agent behavior from being attracted to the given site to being repelled away from it, as is desired and needed for high success of treatment.

The effect of $(P_R/P_A)$ on success is slightly more dramatic for the most diffuse cancer arrangements (a) and (c), and the least dramatic for the most concentrated cancer arrangement (e), as evidenced by the differing ranges of $S_{\text{avg}}(T_{\text{fin}})$ values.
The effect of $(P_R/P_A)$ on treatment time is most dramatic for arrangement (b), and the least dramatic for the arrangements with the most total cancer sites (c) and (d), as evidenced by the differing ranges of $T_{\text{fin}}$ values.

The correspondence between larger $(P_R/P_A)$ values and slower treatment times is less strictly followed for arrangements (c) and (d), with larger $(P_R/P_A)$ settings sometimes performing faster than certain smaller $(P_R/P_A)$ settings.
However, as was just stated above, the total range of $T_{\text{fin}}$ values for these two particular arrangements is smaller than the range for the other arrangements.
Thus, these anomalies (observed in the average behavior of just $20$ trials) are less likely to be indicative of an actually interesting phenomenon.

The only exception to the correspondence between larger $(P_R/P_A)$ values and higher success rates is the dense diffuse cancer arrangement (c) shown in Figure~\ref{fig:KMAR-Reffect-spreadout}.
For arrangement (c), too large a value of $P_R$ (relative to $P_A$) actually corresponds with lower achieved success.
Once $P_R$ reaches $300$, i.e., $(P_R/P_A)$ equal to thirty, the average achieved success value $S_{\text{avg}}(T_{\text{fin}})$ drops off by about seven percent from the next closest $(P_R/P_A)$ setting of $100$.
The correspondence between larger $(P_R/P_A)$ and higher success is otherwise followed before this point of $P_R$ reaching $300$.
This drop in success for too large of a $(P_R/P_A)$ value is possibly a result of agents getting ``stuck'' at the corners of the bounded space, repelled by an overly strong R-signal from the given nearby cancer site, since in this dense arrangement, there are sites closer to the boundaries of the space. 
If this is the case, this exception to the correspondence between larger $(P_R/P_A)$ values and higher success rates is mostly a distracting anomaly not caused by actual general mechanics of the algorithm.

For the other arrangements, it is left to be seen as to whether there is an extreme point after which the chemical R-signals become too strong and cause the achieved success to drop.
Recall that, to account for the dynamic nature of the diffusing chemicals A and R, $r_{A,M}$ is often set such that agents start releasing R-payloads slightly before a given site is fully treated.
For such $r_{A,M}$ settings, it is possible that too strong of R-signals could result in agents being sent away from sites too early, before they are fully treated, yielding suboptimal success

Our results in Figure~\ref{fig:KMAR-Reffect} demonstrate which $P_A$ and $P_R$ settings produce the best treatment times and success rates, respectively, for each given site and demand arrangement.
In particular, increasing $P_R$ proves to boost the success of treatment for most arrangements.
However, in the implementation of actual nanobots, there is a limit to increasing the size of chemical payloads as a result of the extremely small size, and even smaller carrying capacity, of an individual nanobot.
This presents a problem for algorithm optimization in practice.
With that said, having more total nanobot agents, i.e., increasing $n$, can actually mimic increasing $P_A$ or $P_R$.
For example, ten agents dropping payloads each of size one over the span of one minute is roughly equivalent to one agent dropping a payload of size ten over that same span of time.
Let us assume that for a specific setting it holds true that for $n$ total agents exactly one agent reaches a certain cancer site every minute in expectation, and that for $kn$ total agents ten agents reach that site every minute in expectation.
Then, we indeed have an equivalence achieved between increasing the number of agents by a multiplicative factor of $k$ and increasing the given payload size $P_{A/R}$ by a multiplicative factor of ten.

Ultimately, through the results in this Section~\ref{sec:res-kmar}, we show that moderately high threshold values $r_{A,M}$ (approximately $10^7$), high orientation-bias parameter $b$ values, and relative strengths of chemicals A and R in which R is much stronger by approximately $10$ to $30$ times, are the parameter settings which yield the best overall performance for Algorithm KMAR across all cancer arrangements.
The only exceptions to this are for the most concentrated arrangement (e) for which an even higher threshold value ($r_{A,M}$ equal to approximately $5\cdot 10^7$) produces the best results, and for the dense diffuse arrangement (c) for which too strong of chemical R signals relative to A can be detrimental---$(P_R/P_A)$ approximately equal to between $5$ and $10$ is instead optimal here.
While many experiments across many parameter settings were carried out for KMAR, we still acknowledge the inexhaustive and rudimentary nature of our grid search across the parameter domain $(r_{A,M} \times b \times (P_R/P_A))$ for the optimal KMAR performance.

\subsection{Comparing Algorithms KM vs. KMA vs. KMAR}\label{sec:res-comp}
\begin{figure}
    \centering
    \begin{subfigure}{.48\textwidth}
        \centering
        \includegraphics[width=.85\textwidth]{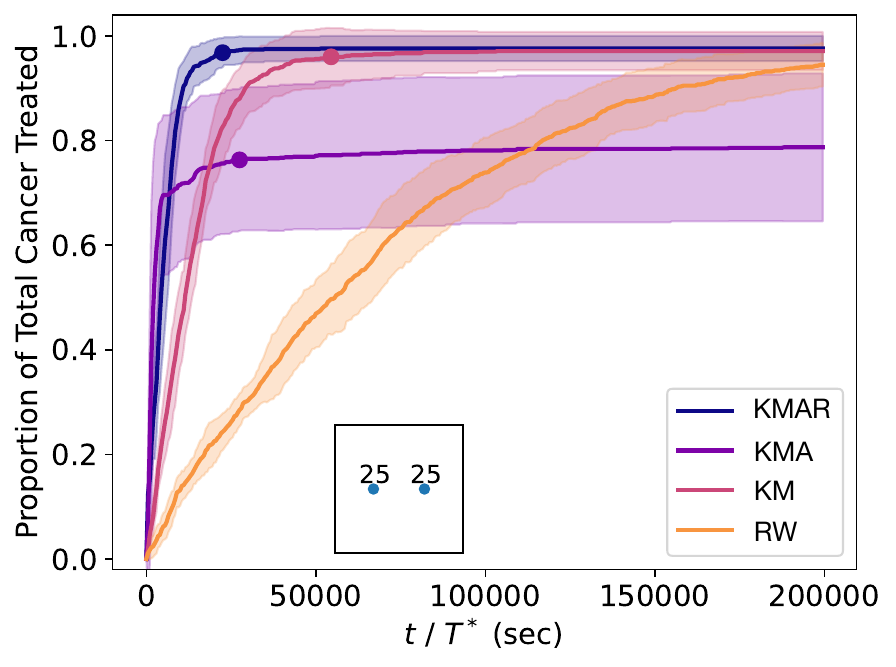}
        \caption{
        For KMAR: $b=10^{(-10)}$; $(T_{\text{fin}}, S_{\text{avg}}(T_{\text{fin}}))= (22500,0.968)$. 
        For KMA: $b=5\cdot 10^{(-11)}$; $(27500,0.763)$. 
        For KM: $b=6\cdot 10^{(-11)}$; $(54500,0.96)$.
        }
    \end{subfigure}
    \hfill
    \begin{subfigure}{.48\textwidth}
        \centering
        \includegraphics[width=.85\textwidth]{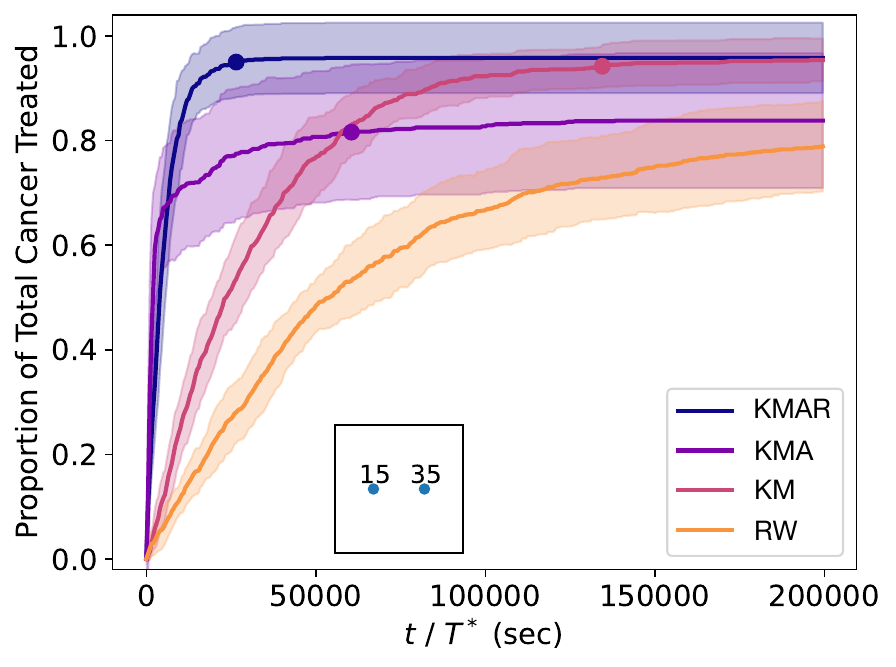}
        \caption{
        For KMAR: $b=10^{(-10)}$; $(T_{\text{fin}}, S_{\text{avg}}(T_{\text{fin}}))= (26500,0.95)$. 
        For KMA: $b=4\cdot 10^{(-11)}$; $(60500,0.816)$. 
        For KM: $b=4\cdot 10^{(-11)}$; $(134500,0.942)$.
        }
    \end{subfigure}
    \hfill
    
    \centering
    \begin{subfigure}{.48\textwidth}
        \centering
        \includegraphics[width=.85\textwidth]{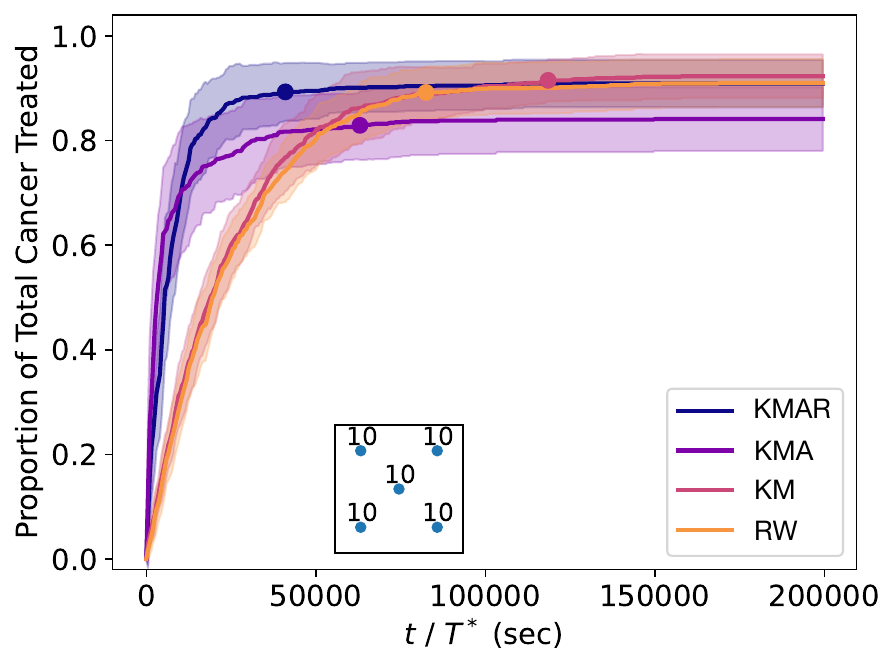}
        \caption{
        For KMAR: $b=10^{(-10)}$; $(T_{\text{fin}}, S_{\text{avg}}(T_{\text{fin}}))= (41000,0.893)$. 
        For KMA: $b=3.5\cdot 10^{(-11)}$; $(63000,0.829)$. 
        For KM: $b=4\cdot 10^{(-11)}$; $(118500,0.915)$.
        For RW: $(82500,0.892)$.
        }
    \end{subfigure}
    \hfill
    \begin{subfigure}{.48\textwidth}
        \centering
        \includegraphics[width=.85\textwidth]{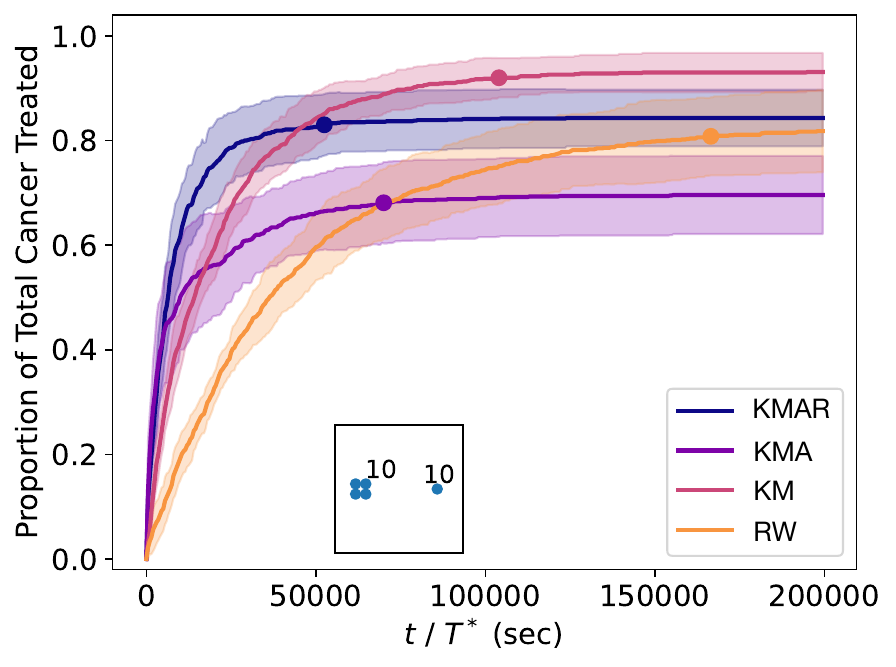}
        \caption{
        For KMAR: $b=10^{(-10)}$; $(T_{\text{fin}}, S_{\text{avg}}(T_{\text{fin}}))= (52500,0.83)$. 
        For KMA: $b=4\cdot 10^{(-11)}$; $(70000,0.681)$. 
        For KM: $b=4\cdot 10^{(-11)}$; $(104000,0.92)$.
        For RW: $(166500,0.808)$.
        }
    \end{subfigure}

    \centering
    \begin{subfigure}{.48\textwidth}
        \centering
        \includegraphics[width=.85\textwidth]{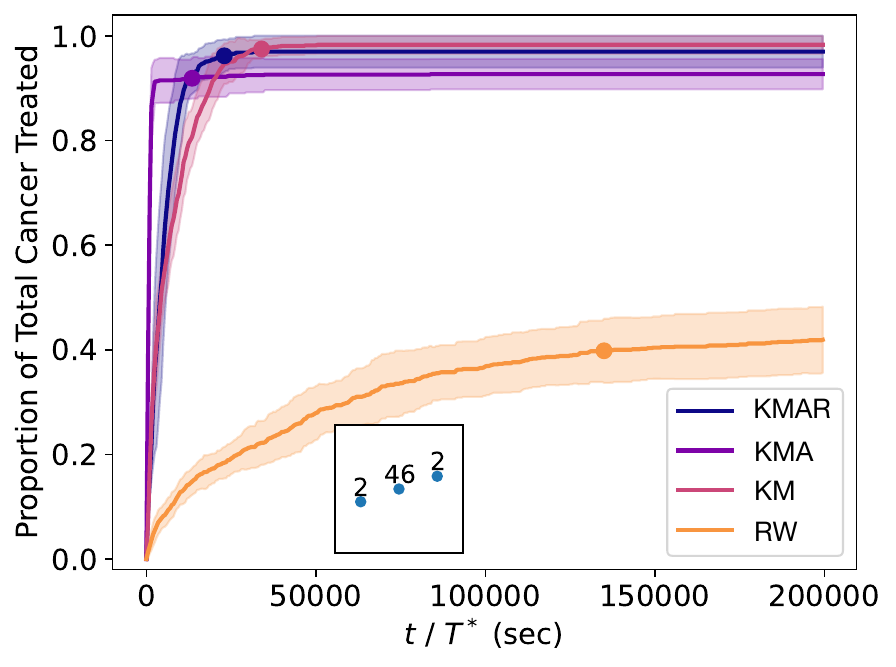}
        \caption{
        For KMAR: $b=10^{(-10)}$; $(T_{\text{fin}}, S_{\text{avg}}(T_{\text{fin}}))= (23000,0.962)$. 
        For KMA: $b=6\cdot 10^{(-11)}$; $(13500,0.919)$. 
        For KM: $b=6\cdot 10^{(-11)}$; $(34000,0.975)$.
        For RW: $(135000,0.398)$.
        }
    \end{subfigure}
    \hfill
    
    \caption{
    \textit{All Algs.:   }
    Simulation results comparing performance between all of the different algorithms (KM, KMA, KMAR, and RW), across the same set of cancer site and demand arrangements.
    For each algorithm and site arrangement (pair), respectively, a specific, well-performing parameter setting was chosen.
    $P_A=10, P_R=50, D_A=D_R=10^{(-9)}, r_{A,M}=10^7$, $\phi_{\text{max}}=0.005$, $\alpha=\epsilon=2\cdot 10^{-5}$, $m=10^{-6}$, $r_{KM}=1$. Plotted lines show average success $S$ (at that given point in time) over $20$ trials, and shaded regions are standard deviations.
    Points on main plot have coordinates $(T_{\text{fin}}, S_{\text{avg}}(T_{\text{fin}}))$ for the given algorithm whose color they match with. For algorithms with no such point, $T_{\text{fin}}$ is undefined, i.e., $T_{\text{fin}}>200000$.
    }
    \label{fig:compareAlgs}
\end{figure}

We fix a single specific parameter setting for each algorithm, respectively, and proceed to compare their performance.
For Algorithms KM and KMAR, we choose the parameter setting which yields the highest success.
For Algorithm KMA, there is a tradeoff when varying parameter values in which faster treatment times correspond with lower success rates and vice versa; as such, we choose a parameter setting for KMA here which is both moderately fast as well as moderately successful.
For each distinct algorithm, including the simple random walk RW, we plot the success metric $S$ over time ($t/T^*$), again across the same set of cancer arrangements (as listed in Subsection~\ref{sec:arrangements}).
See Figure~\ref{fig:compareAlgs}.

Overall, across all site and demand arrangements, our results show that Algorithms KM and KMAR both generally achieve the highest success, though KMAR has faster treatment time.
Algorithm KMA has fast treatment time, but outside of very concentrated cancer arrangements, it has lower success compared to all other algorithms, at least for sufficiently long clearance time.
For dense site arrangements, RW performs just as well as Algorithms KM and KMAR. 

For sparse cancer arrangements that are either diffuse or only moderately concentrated, like arrangements (a) and (b), the trend in relative performance between algorithms is similar.
KMAR and KM both achieve the same, highest eventual success rate $S_{\text{avg}}(T_{\text{fin}})$, though KMAR is faster.
KMA has a lower final success rate than KMAR and KM, but, because of its fast progressing treatment (still not faster than KMAR), it outperforms RW and KM below certain thresholds of clearance times.
For arrangement (b), the more  concentrated cancer arrangement between (a) and (b), KMA's relative performance is better than it is for (a), having higher success than KM for clearance times less than $50000$ seconds (approximately $14$ hours) and higher success than RW for all of the clearance times we tested (less than $200000$ seconds).
By comparison, for arrangement (a), KMA outperforms KM in success only for clearance times less than around $20000$ seconds, and outperforms RW only for clearance times less than $110000$ seconds.  
Further, for arrangement (b), KMA achieves an eventual success rate $S_{\text{avg}}(T_{\text{fin}})$ of nearly $82$ percent, compared to $76$ percent for arrangement (a).
The improvement in treatment time of KMAR over KM is also significantly more dramatic for the more unbalanced arrangement (b).
For an arrangement like (b) in which there is a site with dominant demand, KMAR's and KMA's use of amplifying A-payloads is shown to be particularly beneficial.

For dense diffuse arrangements like (c), KMA achieves lower eventual success than the rest of the algorithms, but for clearance times less than $50000$ seconds (approximately $14$ hours), it has greater success than RW and KM because of its fast treatment speed.
While RW, KM, and KMAR all achieve around the same final success rate $S_{\text{avg}}(T_{\text{fin}})$, KMAR has a significantly faster treatment time so, similar to KMA, KMAR outperforms RW and KM for all clearance times less than approximately $75000$ seconds ($21$ hours). 

For arrangement (d) in which there is a small cluster of cancer sites plus one faraway outlier site, KM actually outperforms KMAR in achieved success $S_{\text{avg}}(T_{\text{fin}})$ by nine percent, though KM's treatment time $T_{\text{fin}}$ is approximately twice as long as KMAR.
KMA has the lowest eventual success, but for clearance times less than around $75000$ seconds (approximately $21$ hours), it has greater success than RW.

For the most concentrated cancer arrangement (e), all of the algorithms (KM, KMA, and KMAR) perform very well in both success and treatment time.
Although KM achieves the greatest success and KMA has the fastest treatment time, the three algorithms' levels of performance are all very close to each other, with the range of $S_{\text{avg}}(T_{\text{fin}})$ values being less than six percent and the range of $T_{\text{fin}}$ values only being approximately $20000$ (seconds).
This most concentrated cancer arrangement is the arrangement for which our three algorithms demonstrate the largest improvement in performance over the simple random walk RW, with RW achieving an average success rate which is over $52$ percent lower than any of the other algorithms as well as an average treatment time which is nearly a full order of magnitude slower.
Concentrated cancer arrangements yield the strongest chemical signals in magnitude (surrounding the major site); our algorithms that follow these chemical signals are thus able to significantly outperform the simple random walk, which is blind to any chemical signals.

Although KM achieves high success across all arrangements, it is arguably less reasonable to assume to have choice over the strength of M-signals compared to having choice over the strength of A- and R-signals, as the former is a feature of the natural environment while the latter is under the control of nanobot designers. 
Assuming extremely weak M-signals (or low $b$ values, equivalently), for example, KM would likely have extremely slow progressing treatment, to the point at which KMAR and even KMA would significantly outperform KM in achieved success for finite clearance times.
This point helps to illustrate the true benefit of KMA's fast treatment time, for all arrangements.

Despite the improvements of KMA over KM in treatment time, and the improvements in KMAR over both KM and KMA in both success and treatment time, we acknowledge that each more involved algorithm (KM to KMA, to KMAR, in that order) is more speculative than the previous regarding individual nanobot capabilities.

\section{Discussion}\label{sec:discuss}
We have considered the problem of cancer detection and treatment by nanobots in the situation in which there are multiple, distinct cancer sites spread about the environment.
Because there are multiple cancer sites, the treatment, which is administered via drug payloads being dropped by agents once at the sites, must be allocated according to the demand of each respective site in order to maximize the success of the treatment.
These demands are unknown, at least directly, to the fully distributed and minimally capable nanobots.
However, we assumed that there exist endogenous chemical signals, or gradients, surrounding each cancer site, where the strength of each signal is proportional to the demand, or amount of treatment the respective site requires.

We have presented a mathematical model of the nanobot agents as well as of the colloidal environment through which they navigate.
This includes a model of agent movement based upon actual nanoparticles. 
We then presented three incrementally sophisticated algorithms, KM, KMA, and KMAR, which describe which artificial chemical payloads agents carry onboard (beyond the cancer-treating drug) and when agents drop those payloads.
These additional chemical payloads diffuse throughout the space, forming gradients that agents can noisily follow, similar to the aforementioned natural signals.

We considered a set of distinct cancer site and demand arrangements and presented simulation results for all of the algorithms across these arrangements, investigating the effect of the arrangement, as well as other parameters, on performance.
For performance, we considered both the success of the treatment, i.e., how much cancer is ultimately killed, as well as the speed of the treatment, since the lifespan of the nanobots (before dissolving, etc.) is finite.

While the distinct cancer site and demand arrangements tested in our experiments are inexhaustive, each is representative of a larger class of common cancer patterns.
The optimal algorithm, as well as the optimal settings of specific parameters, vary between arrangements; ``optimality'' also varies once a specific clearance time $T^*$ is fixed.
Our results shed light on which algorithm and accompanying parameter settings is ideal when given a site and demand arrangement as input.
If initial imaging allows for the cancer site and demand arrangement to be estimated prior to treatment, then, informed by experiments like those presented in our results, one could select a specific algorithm and set of parameter values in an effort to optimize the success of treatment.

For more diffuse arrangements (like arrangements (a), (b), and (c)), our results show that Algorithm KMAR, with high orientation-bias, moderately high $r_{A,M}$, and chemical R stronger than chemical A, has the best performance compared to all other algorithms in both treatment time and success (though KM with moderate orientation-bias matches its level of success assuming a sufficiently long clearance time).
For slightly more concentrated arrangements like (d), KMAR, again with high orientation-bias, moderately high $r_{A,M}$, and chemical R stronger than chemical A, is the fastest algorithm, though KM with moderate orientation-bias has higher success; to get a more precise sense of this tradeoff, KMAR is approximately twice as fast as KM but achieves $10$ percent lower success.
For the most concentrated cancer arrangements like (e), KMA with high orientation-bias is the fastest algorithm while KM, also with high orientation-bias, has the highest success rate, though all algorithms (KM, KMA, and KMAR) perform well here with very similar treatment times and success values.
It is also worth mentioning that for dense diffuse cancer arrangements like (c), RW also achieves the highest level of success, approximately, alongside KM and KMAR; RW here also has reasonable treatment time which is only twice as long as KMAR.

While KMAR, our most sophisticated algorithm, does not always have the absolute best overall performance in both metrics compared to the other algorithms, such as for arrangements (d) and (e), the fact that it always has reasonably close to optimal performance across all types of cancer arrangements is crucial in demonstrating its \textit{adaptability} as an algorithm.
For example, if the initial cancer arrangement is completely unknown, then KMAR's consistently good performance across arrangements---its adaptibility---is extremely useful.
In contrast, while KMA, for example, performs well for concentrated cancer arrangements, if this information about the arrangement is unknown, than blindly administering KMA may be unsuccessful.

While KM achieves high success across all arrangements, for weak chemical M signals (or equivalently, weak orientation-bias), KM's treatment progresses too slowly for it to yield high success for any reasonable clearance times.
Importantly, in practice, the strength of M-signals is a predetermined feature of the natural environment, i.e., we are unable to simply set the orientation-bias parameter $b$ as desired.

While in this work, we focused on only a single pass of treatment, it is reasonable to imagine several repeated passes of treatment by nanobots given the nontoxicity promised by their precise, selective drug delivery. 
In this case, after the first pass of treatment, further imaging could be carried out to estimate the new site and demand arrangement, following which a new algorithm and set of parameter values could be selected to optimize treatment success once again.
This process would continue with each subsequent round of treatment.

For ``multi-pass'' treatment, the success of the process is now given by the total proportion of cancerous matter treated after \textit{all} rounds of treatment have finished.
Maximizing the total success after all rounds may or may not coincide with maximizing the success of each round, individually.
For example, a slightly less effective (in total cancerous mass treated) first round of treatment could yield a remaining cancer arrangement that actually results in greater overall success after the second round of treatment.
Although for ``single-pass'' treatment KMA has the lowest success across all algorithms (excluding RW), it has potential usefulness here in the multi-pass treatment scenario, as KMA quickly and reliably kills the main cancer site, via an algorithm with lower nanobot capabilities requirements than KMAR. 

To conclude on the topic of algorithmic sophistication, or equivalently nanobot capabilities, while our results suggest that KMAR is usually the best performing algorithm, in a case where KM or KMA perform close to as well as KMAR, KM or KMA might be better choices in application over KMAR because of their simplicity (which could correspond to lower cost, safer treatment, etc.).

\subsection{Future Work}
Future work could vary further parameters of the model such as the total number of nanobot agents $n$, the size of the total space of activity $\phi_{\text{max}}$, and the diffusion coefficients for chemicals A and R, $D_A$ and $D_R$, respectively.
More cancer site and demand arrangements, beyond the inexhaustive five considered here, could be tested in experiments.
Future work could also work towards a more general and quantitative approach to studying the impact of different cancer patterns on algorithms' treatment success.
For example, one could formulate quantitative metrics for site density and demand distribution, and then carry out experiments in which for a single trial of a given cancer pattern metric values, a site and demand arrangement is generated randomly.

In the model presented in this paper, the naturally existing chemical M signals are completely persistent and time-constant; however, one could consider a model in which the M-signals decrease in magnitude during the time of treatment in accordance with the given cancer site being progressively killed, and/or decrease passively via diffusion.

Future work could include studying the impact of different types of noise on agent behavior, such as imperfect accuracy in detecting the presence of a nearby cancer site and agent motion that is influenced by some external force(s) such as the blood flow.
The three-dimensional setting also remains to be investigated. 

Future work could also consider different initial arrangements of agents at time $t=0$, beyond uniformly random initial positions. 
For example, consider the sparse site arrangements (a) and (b) and imagine if all of the agents started at the same position which happened to be near a single particular site (and far away from any other other sites). 
Then Algorithm KM, for example, would be very unsuccessful in its treatment, as most of the agents would converge upon the site to which they are closest, in expectation, leaving all other sites completely untreated. 
In this scenario, the more sophisticated mechanisms of Algorithm KMAR are likely to be even more crucial to achieve optimal allocation. 

As alluded to in our earlier discussion, the case of multiple, repeated passes or rounds of treatment remains to be directly investigated in simulation.
This would involve more careful study of the site and demand arrangement that is left over after each round of treatment; this ``output'' arrangement becomes the input for the next round of treatment.
For simplicity, we have restricted continuous space to a bounded region as well as ignored the effect of agents that never reach a cancer site within the given clearance time.
It is reasonable to assume that an agent's chemical payloads are released upon the agent's dissolution at clearance time; this includes the toxic cancer-treating drug.
Thus, future work could more thoroughly study this complication in which toxic drug payloads are released across the entire space including noncancerous areas (i.e., directly factor it into treatment success), with ``lost'' or wandering agents dissolving at clearance time.

While difficult given the dynamic and interdependent nature of the system, it is possible that analytical results for the performance of different algorithms can be derived.
Such analytical results could provide performance estimates for many arrangements and parameter settings, removing the need for exhaustive simulations.

Finally, we can explore the feasibility of our nanobot model for applications, particularly its more speculative additions to the simpler KM, via experiments on actual nanoparticles and nanorobots.

\paragraph{Acknowledgments}
Special thanks to Sabrina Drammis and Mien Brabeeba Wang for their helpful suggestions given throughout the completion of this work.
As referenced earlier, this work is a direct extension on the previous work \cite{ourSingleSite} done in collaboration with Claudia Contini, Cristina Gava, and Frederik Mallmann-Trenn and thus builds upon many of the early ideas and discussions we all shared regarding the multi-site problem.

This work was supported by NSF Award 2003830.

\bibliographystyle{plain}
\bibliography{bibliography}

@article{claudiavesicles,
    author = {Joseph, Adrian and Contini, Claudia and Cecchin, Denis and Nyberg, Sophie and Ruiz-Perez, Lorena and Gaitzsch, Jens and Fullstone, Gavin and Tian, Xiaohe and Azizi, Juzaili and Preston, Jane and Volpe, Giorgio and Battaglia, Giuseppe},
    title = {Chemotactic synthetic vesicles: Design and applications in blood-brain barrier crossing},
    journal = {Science Advances},
    year = {2017}
}

@misc{diffusion,
    author={Junping Shi},
    title={Diffusion of point source and biological dispersal, {N}otes},
    year={2006},
    howpublished={William \& Mary, Math 490-01 Partial Differential Equations and Mathematical Biology}
}

@article{
ourSingleSite,
author = {Noble Harasha  and Cristina Gava  and Nancy Lynch  and Claudia Contini  and Frederik Mallmann-Trenn },
title = {Modeling feasible locomotion of nanobots for cancer detection and treatment},
journal = {Proceedings of the National Academy of Sciences},
volume = {122},
number = {48},
pages = {e2510036122},
year = {2025},
doi = {10.1073/pnas.2510036122},
URL = {https://www.pnas.org/doi/abs/10.1073/pnas.2510036122}
}

@article{posNEGchemotax,
    title={Positive and negative chemotaxis of enzyme-coated liposome motors},
    author={Somasundar, Ambika and Ghosh, Subhadip and Mohajerani, Farzad and Massenburg, Lynnicia N. and Yang, Tinglu and Cremer, Paul S. and Velegol, Darrell and Sen, Ayusman},
    journal={Nature Nanotechnology},
    volume={14},
    pages={1129–-1134},
    year={2019}
}

@article{NEGchemotaxEx,
    author = {Popescu, Mihail N. and Uspal, William E. and Bechinger, Clemens and Fischer, Peer},
    title = {Chemotaxis of Active Janus Nanoparticles},
    journal = {Nano Letters},
    volume = {18},
    number = {9},
    pages = {5345--5349},
    year = {2018}
}

@inproceedings{grace,
author = {Cai, Grace and Harasha, Noble and Lynch, Nancy},
title = {A Comparison of New Swarm Task Allocation Algorithms in Unknown Environments with Varying Task Density},
year = {2023},
publisher = {International Foundation for Autonomous Agents and Multiagent Systems},
booktitle = {Proceedings of the 2023 International Conference on Autonomous Agents and Multiagent Systems},
pages = {525--533},
numpages = {9},
location = {London, United Kingdom},
series = {AAMAS '23}
}

@misc{adithya,
      title={Swarm Algorithms for Dynamic Task Allocation in Unknown Environments}, 
      author={Balachandran, Adithya and Harasha, Noble and Lynch, Nancy},
      year={2024},
      eprint={2409.09550},
      archivePrefix={arXiv},
      primaryClass={cs.MA},
      url={https://arxiv.org/abs/2409.09550}
}

@book{crandall1996nanotechnology,
  title={Nanotechnology: molecular speculations on global abundance},
  author={Crandall, BC},
  year={1996},
  publisher={Mit Press}
}

@article{brigger2012nanoparticles,
  title={Nanoparticles in cancer therapy and diagnosis},
  author={Brigger, Ir{\`e}ne and Dubernet, Catherine and Couvreur, Patrick},
  journal={Advanced drug delivery reviews},
  volume={64},
  pages={24--36},
  year={2012},
  publisher={Elsevier}
}

@article{golestanian2005propulsion,
  title={Propulsion of a molecular machine by asymmetric distribution of reaction products},
  author={Golestanian, Ramin and Liverpool, Tanniemola B and Ajdari, Armand},
  journal={Physical review letters},
  volume={94},
  number={22},
  pages={220801},
  year={2005},
  publisher={APS}
}

@article{howse2007self,
  title={Self-motile colloidal particles: from directed propulsion to random walk},
  author={Howse, Jonathan R and Jones, Richard AL and Ryan, Anthony J and Gough, Tim and Vafabakhsh, Reza and Golestanian, Ramin},
  journal={Physical review letters},
  volume={99},
  number={4},
  pages={048102},
  year={2007},
  publisher={APS}
}

@article{kostarelos2010nanorobots,
  title={Nanorobots for medicine: how close are we?},
  author={Kostarelos, Kostas},
  journal={Nanomedicine},
  volume={5},
  number={3},
  pages={341--342},
  year={2010},
  publisher={Future Medicine}
}

@article{martel2009flagellated,
  title={Flagellated magnetotactic bacteria as controlled MRI-trackable propulsion and steering systems for medical nanorobots operating in the human microvasculature},
  author={Martel, Sylvain and Mohammadi, Mahmood and Felfoul, Ouajdi and Lu, Zhao and Pouponneau, Pierre},
  journal={The International journal of robotics research},
  volume={28},
  number={4},
  pages={571--582},
  year={2009},
  publisher={SAGE Publications Sage UK: London, England}
}

@article{gwisai2022magnetic,
  title={Magnetic torque--driven living microrobots for increased tumor infiltration},
  author={Gwisai, Tinotenda and Mirkhani, Nima and Christiansen, Michael G and Nguyen, Thuy Trinh and Ling, V and Schuerle, S},
  journal={Science Robotics},
  volume={7},
  number={71},
  pages={eabo0665},
  year={2022},
  publisher={American Association for the Advancement of Science}
}

@inproceedings{gomez2021markov,
  title={Markov model for the flow of nanobots in the human circulatory system},
  author={G{\'o}mez, Jorge Torres and Wendt, Regine and Kuestner, Anke and Pitke, Ketki and Stratmann, Lukas and Dressler, Falko},
  booktitle={Proceedings of the Eight Annual ACM International Conference on Nanoscale Computing and Communication},
  pages={1--7},
  year={2021}
}

@book{freitas1999nanomedicine,
  title={Nanomedicine, volume I: basic capabilities},
  author={Freitas, Robert A},
  volume={1},
  year={1999},
  publisher={Landes Bioscience Georgetown, TX}
}

@article{zhang2023advanced,
  title={Advanced medical micro-robotics for early diagnosis and therapeutic interventions},
  author={Zhang, Dandan and Gorochowski, Thomas E and Marucci, Lucia and Lee, Hyun-Taek and Gil, Bruno and Li, Bing and Hauert, Sabine and Yeatman, Eric},
  journal={Frontiers in Robotics and AI},
  volume={9},
  pages={1086043},
  year={2023},
  publisher={Frontiers}
}

@article{sanchez_lm,
author = {Xu, Dandan and Hu, Jing and Pan, Xi and Sánchez, Samuel and Yan, Xiaohui and Ma, Xing},
title = {Enzyme-Powered Liquid Metal Nanobots Endowed with Multiple Biomedical Functions},
journal = {ACS Nano},
volume = {15},
number = {7},
pages = {11543--11554},
year = {2021},
doi = {10.1021/acsnano.1c01573},
URL = {https://doi.org/10.1021/acsnano.1c01573},
eprint = {https://doi.org/10.1021/acsnano.1c01573}
}

@article{sanchez_catalase,
author = {Serra-Casablancas, Meritxell and Di Carlo, Valerio and Esporrín-Ubieto, David and Prado-Morales, Carles and Bakenecker, Anna C. and Sánchez, Samuel},
title = {Catalase-Powered Nanobots for Overcoming the Mucus Barrier},
journal = {ACS Nano},
volume = {18},
number = {26},
pages = {16701--16714},
year = {2024},
doi = {10.1021/acsnano.4c01760},
URL = {https://doi.org/10.1021/acsnano.4c01760},
eprint = {https://doi.org/10.1021/acsnano.4c01760}
}

@article{
doi:10.1073/pnas.0610298104,
author = {Dmitri Simberg  and Tasmia Duza  and Ji Ho Park  and Markus Essler  and Jan Pilch  and Lianglin Zhang  and Austin M. Derfus  and Meng Yang  and Robert M. Hoffman  and Sangeeta Bhatia  and Michael J. Sailor  and Erkki Ruoslahti },
title = {Biomimetic amplification of nanoparticle homing to tumors},
journal = {Proceedings of the National Academy of Sciences},
volume = {104},
number = {3},
pages = {932-936},
year = {2007},
doi = {10.1073/pnas.0610298104},
URL = {https://www.pnas.org/doi/abs/10.1073/pnas.0610298104},
eprint = {https://www.pnas.org/doi/pdf/10.1073/pnas.0610298104}}

@article{doi.org/10.1038/nmat3049,
author = {Geoffrey von Maltzahn and Ji-Ho Park and Kevin Y. Lin and Neetu Singh and Christian Schwöppe and Rolf Mesters and Wolfgang E. Berdel and Erkki Ruoslahti and Michael J. Sailor and Sangeeta N. Bhatia},
title = {Nanoparticles that communicate in vivo to amplify tumour targeting},
journal = {Nature Materials},
volume = {10},
number = {7}, 
pages = {1476-4660},
year = {2011},
doi = {10.1038/nmat3049},
URL = {https://doi.org/10.1038/nmat3049}
}

@article{
strano,
author = {Ge Zhang  and Sungyun Yang  and Jing Fan Yang  and David Gonzalez-Medrano  and Marc Z. Miskin  and Volodymyr B. Koman  and Yuwen Zeng  and Sylvia Xin Li  and Matthias Kuehne  and Albert Tianxiang Liu  and Allan M. Brooks  and Mahesh Kumar  and Michael S. Strano },
title = {High energy density picoliter-scale zinc-air microbatteries for colloidal robotics},
journal = {Science Robotics},
volume = {9},
number = {93},
pages = {eade4642},
year = {2024},
doi = {10.1126/scirobotics.ade4642},
URL = {https://www.science.org/doi/abs/10.1126/scirobotics.ade4642}
}

\appendix
\section{Appendix}

\subsection{Lookup Table of Notation}\label{sec:appendix-table}
Table~\ref{fig:paramtable} is provided as a sort of glossary for the various notation and abbreviations used throughout the paper.

\begin{table}
    \centering
    \begin{tabular}{ |p{2.1cm}||p{13cm}|  }
     \hline
     Notation & Description\\
     \hline
     \hline
     $n$ & number of nanobot agents \\
     \hline
     $x_i^{(t)}, \theta_i^{(t)}$ & location and orientation of agent $i$ at time $t$, respectively \\
     \hline
     $c$ & number of cancer sites \\
     \hline
     $y_j$ & location of cancer site $j$ \\
     \hline
     $\epsilon$ & cancer site detection distance \\
     \hline
     $\phi_{\text{max}}$ & dimensions of bounded space \\
     \hline
     chemical K & cancer-treating drug \\
     \hline 
     chemical M & natural, persistent attractive signal chemical \\
     \hline
     chemical A & artificial, dissipating ``attractive'' signal chemical \\
     \hline
     chemical R & artificial, dissipating ``repellent'' signal chemical \\
     \hline
     $P_{M_j}$ & strength of M-signal surrounding cancer site $j$, proportional to amount of treatment needed (`demand') \\
     \hline
     $\gamma_M(x)$ & concentration of chemical M at location $x$ \\
     \hline
     $m$ & parameter in chemical M concentration function \\
     \hline 
     $r_{K,M}$ & minimum ratio of chemical K to M needed to fully treat a cancer site \\
     \hline
     $K_j^{(t)}$ & number of K-payloads dropped at site $j$ by time $t$ \\
     \hline
     $A_j^{(t)}, R_j^{(t)}$ & sets of timesteps when A- and R-payloads, respectively, were dropped at site $j$ before time $t$ \\
     \hline
     $\gamma_A^{(t)}(x), \gamma_R^{(t)}(x)$ & concentrations of chemicals A and R, respectively, at location $x$ at time $t$ \\
     \hline
     $P_A, P_R$ & sizes of individual A- and R-payloads, respectively \\
     \hline
     $D_A, D_R$ & diffusion coefficients for chemicals A and R, respectively \\
     \hline
     $\alpha$ & displacement per timestep in agent motion \\
     \hline
     $b$ & orientation-bias parameter \\
     \hline 
     $T^*$ & clearance time \\
     \hline
     $S(t)$ & effectiveness metric; proportion of cancer treated by time $t$ \\
     \hline
     $T_{\text{fin}}$ & efficiency metric; time for treatment to stabilize and mostly finish \\
     \hline
     $\delta, D$ & parameters for $T_{\text{fin}}$ calculations \\
     \hline
     KM & algorithm involving K- and M-payloads \\
     \hline
     KMA & algorithm involving K-, M-, and A-payloads \\
     \hline
     KMAB & algorithm involving K-, M-, A-, and R-payloads \\
     \hline
     $r_{A,M}$ & for KMAR; minimum ratio of chemical A to M to begin dropping R-payloads \\
     \hline
     RW & simple random walk in 2D \\
     \hline
    \end{tabular}
    
    \caption{Summary table of main notation used.}
    \label{fig:paramtable}
\end{table}

\subsection{``Nanobot Black Holes''}
The following section represents an interesting mathematical nuance which we posit could be worth investigating for its relevance to this work and beyond to more general applications.

Recall the chemical concentration functions $\gamma_M(\cdot)$ and $\gamma_A(\cdot)$ for the attractive signal chemicals M and A, respectively (see Equations \ref{eqn:M} and \ref{eqn:A}, respectively).
Recall that chemical M is persistent and fixed over time, while chemicals A and R diffuse and dissipate over time.
Regardless, for all chemicals, we model and calculate the total chemical concentration (of the given chemical) at a given point in space by summing the contributions over all nearby cancer sites' individual signals.
This summation yields the total or global chemical landscape ($\gamma_M(\cdot)$, $\gamma_A(\cdot)$) that agents are actually following and ascending (noisily).

Ideally, or intuitively, each local maximum of the given total/global chemical gradient should correspond to a distinct cancer site (via a bijection) such that agents converging upon these local maxima in expectation (which is what the movement model of noisy gradient ascent for attractive signal chemicals indeed entails) corresponds to agents converging to find and treat cancer sites efficiently.
However, under certain conditions, this process of summing over all cancer sites' individual signals can yield local maxima in the total/global chemical gradient which do \textit{not} correspond to actual cancer site locations.
If this happens, then agents would be converging towards ``bogus points'' that are not actually cancer sites needing treatment, significantly hurting efficiency of our treatment strategy.

\begin{figure}
    \centering
    
    \begin{subfigure}{.45\textwidth}
        \centering
        \includegraphics[width=\textwidth]{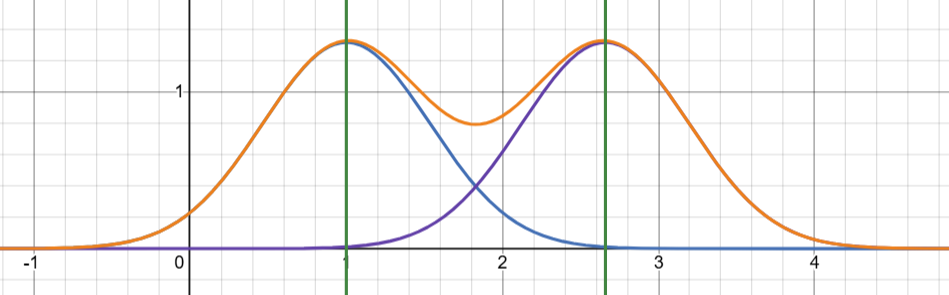}
        \caption{Favorable case.}
        \label{fig:blackholeA}
    \end{subfigure}
    \hfill
    \begin{subfigure}{.45\textwidth}
        \centering
        \includegraphics[width=\textwidth]{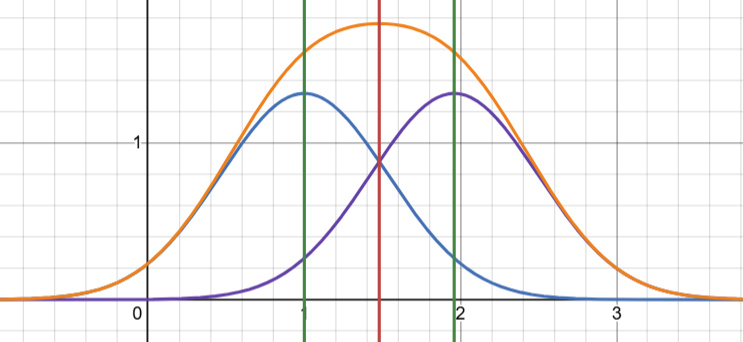}
        \caption{Unfavorable case, first possibility. If cancer sites too close, black hole arises. Individual M-signal gradients have same shape as in (a).}
        \label{fig:blackholeB1}
    \end{subfigure}
    \hfill
    \begin{subfigure}{.45\textwidth}
        \centering
        \includegraphics[width=\textwidth]{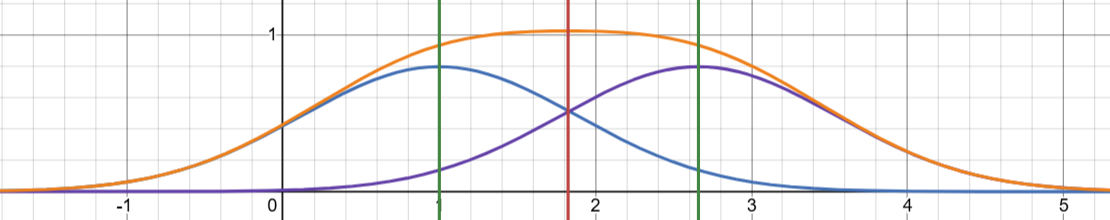}
        \caption{Unfavorable case, second possibility. If individual M-signal gradients too uniform/flat, black hole arises. Cancer sites have same locations as in (a), respectively.}
        \label{fig:blackholeB2}
    \end{subfigure}
    \hfill
    
    \caption{Example depiction of ``nanobot black hole'' between two cancer sites ($c=2$). X-axis is $x$ and Y-axis is concentration of chemical M at location $(x,0)$. Blue and purple curves are individual chemical M signals of each cancer site, respectively, and orange curve is their summation, i.e. $\gamma_M((x,0))$. Agents are noisily ascending the orange curve. Green vertical lines represent locations of cancer sites, and red line (if present) represents location of a spurious/bogus local maximum of $\gamma_M((x,0))$, i.e., a ``black hole''.}
    \label{fig:blackhole}
\end{figure}

As an example, consider two cancer sites and only consider chemical M.
See Figure~\ref{fig:blackhole} for a depiction.
Again, chemical M is an \textit{attractive} signal chemical, so agents are going to be (noisily) ascending its global gradient.
If the two cancer sites are too close to each other (see Figure~\ref{fig:blackholeB1}), or if the individual cancer sites' chemical M signals are too ``spread out'' (see Figure~\ref{fig:blackholeB2}), then the total/global chemical M gradient $\gamma_M(\cdot)$ will only have a single local maximum located at a bogus point in between the two cancer sites, instead of having two distinct local maxima each located at the two cancer sites, respectively.
In the bad case of $\gamma_M(\cdot)$ only having one bogus local maximum, agents will be converging in expectation towards this bogus point, instead of converging towards the cancer sites and administering their treatment more efficiently.

Work needs to be done to better characterize and analyze the precise conditions under which these ``black holes'', or spurious/bogus local maxima, are (or are not) able to occur.
Further, even if black holes are present, enough noise in agent motion could ensure that agents explore enough to still find the correct nearby cancer sites and not get completely stuck moving towards/around the black hole; consequently, this ``black hole'' phenomenon may not be fatal to performance in practice.

Note this this phenomenon occurring for chemical R (a repellent signal chemical) has a different effect on resulting agent motion, one that is a bit less direct but still potentially harmful to performance.
Here, agents would be moving away from some bogus point in expectation, in turn potentially sending them towards a nearby site that we were trying to send them away from.

\begin{figure}
    \centering
    \includegraphics[width=0.45\linewidth]{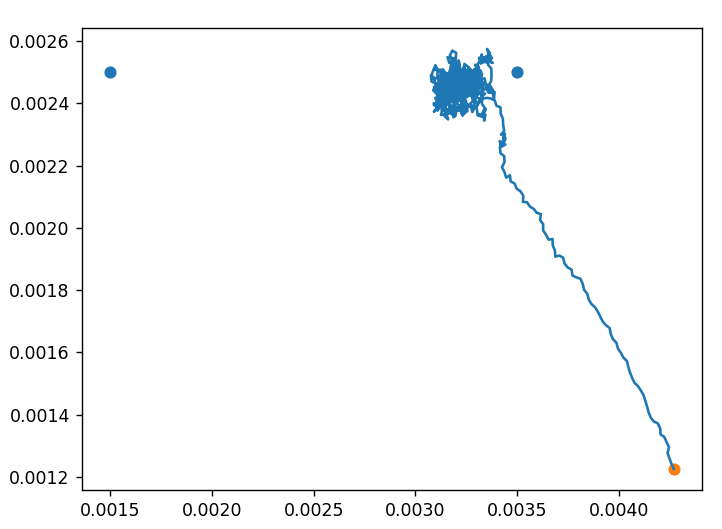}
    \caption{Example depiction of single agent  converging to bogus point right next to cancer site, even when summation indeed preserves unique local maxima corresponding to each cancer site. Two blue points are cancer site locations, orange point is agent's initial location. Blue trajectory is agent's movement over time.}
    \label{fig:grayhole}
\end{figure}

Even when the above described phenomenon doesn't occur, i.e., when there are still unique local maxima corresponding to each cancer site (via a bijection), the location of the local maximum corresponding to a given cancer site will still be slightly shifted away from the precise cancer site location by other nearby local maxima, which can cause similar ``black hole'' issues under certain conditions (see Figure~\ref{fig:grayhole}).

In the simulations for this paper, we set the values of model parameters such that all of these issues are avoided in practice (e.g., sufficiently small $D_A$ and $D_R$, sufficiently small $m$, sufficiently large $\epsilon$, sufficiently small $b$).

\end{document}